\documentclass[iop, letterpaper]{emulateapj}
\usepackage{apjfonts}
\usepackage{color}

\newcommand{\msun}{\mbox{$M_{\sun}$}}

\newcommand{\teff}{\mbox{$T_{\rm eff}$}}
\newcommand{\logg}{\mbox{$\log g$}}
\newcommand{\feh}{\mbox{$\rm{[Fe/H]}$}}
\newcommand{\kep}{\mbox{\it Kepler}}

\shorttitle{Close-out {\it Kepler} star properties catalog for Q1-Q17 targets}
\shortauthors{MATHUR ET AL.}

\begin{document}

\title{Revised Stellar Properties of {\it Kepler} Targets for the Q1--17 (DR25) Transit Detection Run}

\author{Savita Mathur\altaffilmark{1}, Daniel Huber\altaffilmark{2,3,4}, Natalie M. Batalha\altaffilmark{5}, David R. Ciardi\altaffilmark{6},  Fabienne A. Bastien\altaffilmark{7, 8, 9}, Allyson Bieryla\altaffilmark{10}, Lars A. Buchhave\altaffilmark{11, 12}, William D. Cochran\altaffilmark{13}, Michael Endl\altaffilmark{13}, Gilbert A. Esquerdo\altaffilmark{10}, Elise Furlan\altaffilmark{6}, Andrew Howard\altaffilmark{14}, Steve B. Howell\altaffilmark{5}, Howard Isaacson\altaffilmark{14}, David W. Latham\altaffilmark{10}, Phillip J. MacQueen\altaffilmark{13}, David R. Silva\altaffilmark{15}}

\altaffiltext{1}{Space Science Institute, 4750 Walnut street Suite\#205, Boulder, CO 80301, USA}
\altaffiltext{2}{Sydney Institute for Astronomy, School of Physics, University of Sydney, NSW 2006, Australia}
\altaffiltext{3}{SETI Institute, 189 Bernardo Avenue, Mountain View, CA 94043, USA}
\altaffiltext{4}{Stellar Astrophysics Centre, Department of Physics and Astronomy, Aarhus University, Ny Munkegade 120, DK-8000 Aarhus C, Denmark}
\altaffiltext{5}{NASA Ames Research Center, Moffett Field, CA 94035 USA}
\altaffiltext{6}{IPAC, Mail Code 100-22, Caltech, 1200 E. California Blvd., Pasadena, CA 91125 USA}
\altaffiltext{7}{Department of Astronomy and Astrophysics, 525 Davey Lab, The Pennsylvania State University, University Park, PA 16803, USA}
\altaffiltext{8}{Center for Exoplanets and Habitable Worlds, The Pennsylvania State University, 525 Davey Laboratory, University Park, PA 16802, USA}
\altaffiltext{9}{Hubble Fellow}
\altaffiltext{10}{Harvard-Smithsonian Center for Astrophysics, 60 Garden Street, Cambridge, MA 02138, USA}
\altaffiltext{11}{Niels Bohr Institute, University of Copenhagen, Copenhagen DK-2100, Denmark}
\altaffiltext{12}{Centre for Star and Planet Formation, Natural History Museum of Denmark, University of Copenhagen, DK-1350 Copenhagen, Denmark}
\altaffiltext{13}{The University of Texas, Austin TX 78712, USA}
\altaffiltext{14}{Astronomy Department, University of California Berkeley, CA, US}
\altaffiltext{15}{National Optical Astronomy Observatory, 950 N. Cherry Ave., Tucson, AZ 85719 USA}

\begin{abstract}


The determination of exoplanet properties and occurrence rates using {\it Kepler} data critically depends on our knowledge of the fundamental properties (such as temperature, radius and mass) of the observed stars. We present revised stellar properties for 197,096 {\it Kepler} targets observed between Quarters 1--17 (Q1--17), which were used for the final transiting planet search run by the {\it Kepler} Mission (Data Release 25, DR25). Similar to the Q1--16 catalog by Huber et al.\ the classifications are based on conditioning published atmospheric parameters on a grid of Dartmouth isochrones, with significant improvements in the adopted methodology and over 29,000 new sources for temperatures, surface gravities or metallicities. In addition to fundamental stellar properties the new catalog also includes distances and extinctions, and we provide posterior samples for each stellar parameter of each star. Typical uncertainties are $\sim$\,27\% in radius, $\sim$\,17\% in mass, and $\sim$\,51\% in density, which is somewhat smaller than previous catalogs due to the larger number of improved \logg\ constraints and the inclusion of isochrone weighting when deriving stellar posterior distributions. On average, the catalog includes a significantly larger number of evolved solar-type stars, with an increase of 43.5\% in the number of subgiants.
We discuss the overall changes of radii and masses of {\it Kepler} targets as a function of spectral type, with particular focus on exoplanet host stars.

\end{abstract}


\section{INTRODUCTION}

Since the launch of the NASA {\it Kepler} mission \citep{2010Sci...327..977B,2010ApJ...713L..79K} in 2009, a tremendous number of discoveries in exoplanet science have been made possible thanks to the near-continuous, high-precision photometric data collected for over four years. 
To date 4,706 planet-candidates have been identified, over 49\,\% of which have been  confirmed or validated \citep{2014ApJ...784...45R,2016ApJ...822...86M}
This large number of detections allowed statistical studies of planet occurrence rates \citep[e.g.][]{2012ApJS..201...15H,2013ApJ...766...81F,2015ApJ...807...45D,2015ApJ...809....8B, 2015ApJ...799..180S} as well as numerous individual discoveries such as \kep's first rocky exoplanet, Kepler-10b \citep{2011ApJ...729...27B}, circumbinary planets \citep[e.g.][]{2012ApJ...758...87O,2014ApJ...784...14K,2015ApJ...809...26W}, or the detection of planets in or near the habitable zone \citep[e.g.][]{2013ApJ...773...98B,2013Natur.494..452B,2013Sci...340..587B,2015ApJ...800...99T,2016ApJ...830....1K}.

Stellar astrophysics also benefited from the exquisite data of {\it Kepler} with a large number of breakthrough discoveries, such as the asteroseismic measurement of the internal rotation \citep{2012Natur.481...55B, 2012ApJ...756...19D,2012A&A...548A..10M, 2014A&A...564A..27D} and magnetic fields \citep{2015Sci...350..423F, 2016Natur.529..364S} of subgiants and red giants, the detection of surface rotation and its relation to ages of solar-like stars \citep[e.g.][]{2014A&A...572A..34G,2014ApJS..211...24M, 2016MNRAS.456..119C, 2016Natur.529..181V}, as well as the measurement of magnetic activity of main-sequence stars \citep[e.g.][]{2014A&A...562A.124M,2015ApJ...807..109A,2016A&A...589A.118S}. Asteroseismic data of red giants are now also being used to perform galactic archeology by combining them with high-resolution spectroscopy \citep[e.g.][]{2014ApJS..215...19P,2014MNRAS.445.2758R,2015MNRAS.451.2230M}.

Since the transit technique measures planet properties only relative to the host star, it is crucial to characterize the parameters of the host stars to derive precise parameters of the planets. Before the launch of the mission, the {\it Kepler} Input Catalog \citep[KIC][]{2011AJ....142..112B} was constructed based on broadband photometry, with the primary purpose to select targets for observations \citep{2010ApJ...713L.109B} and provide an initial classification of planet candidates. In order to improve the KIC, \citet{2014ApJS..211....2H} presented revised stellar properties for 196,468 {\it Kepler} targets, which were used for the Q1-16 Transit Planet Search and Data Validation run \citep{2014ApJS..211....6T}. The catalog was based on atmospheric properties (temperature $T_{\rm eff}$, surface gravity $\log g$, and metallicity [Fe/H]) published in the literature using a variety of methods (asteroseismology, spectroscopy, exoplanet transits, photometry), which were then homogeneously fitted to a grid of Dartmouth (DSEP) isochrones \citep{2008ApJS..178...89D}. The catalog was updated in early 2015 for a Q1-17 transit detection run \citep[Data Release 24\footnote{\url{ http://exoplanetarchive.ipac.caltech.edu/docs/KeplerStellar\_Q1\_17\_documentation.pdf}}, DR24, ][]{Huber2014} based on the latest classifications of {\it Kepler} targets in the literature and using the same methodology as \citet{2014ApJS..211....2H}. {   We discarded the stars observed only in Q0 as the transit search pipeline does not investigate the data from the commissioning phase for planets.} { However we note that 180 stars with only Q0 data have slipped into the catalog during the input data consolidation.}


\begin{figure*}[htbp]
\begin{center}
\includegraphics[width=5.8cm, trim=2cm 0.5cm 2cm 0]{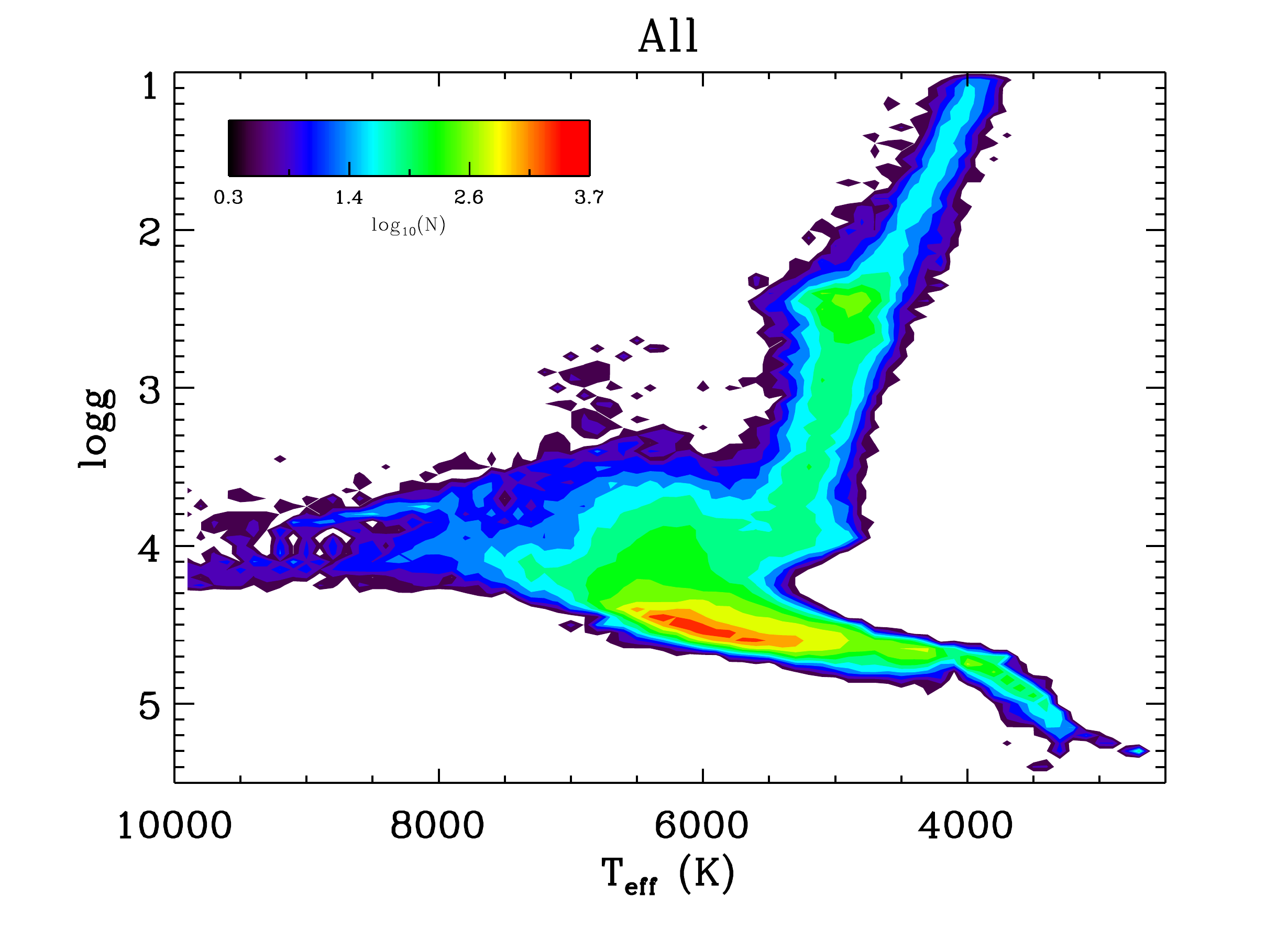}
\includegraphics[width=5.8cm, trim=2cm 0.5cm 2cm 0]{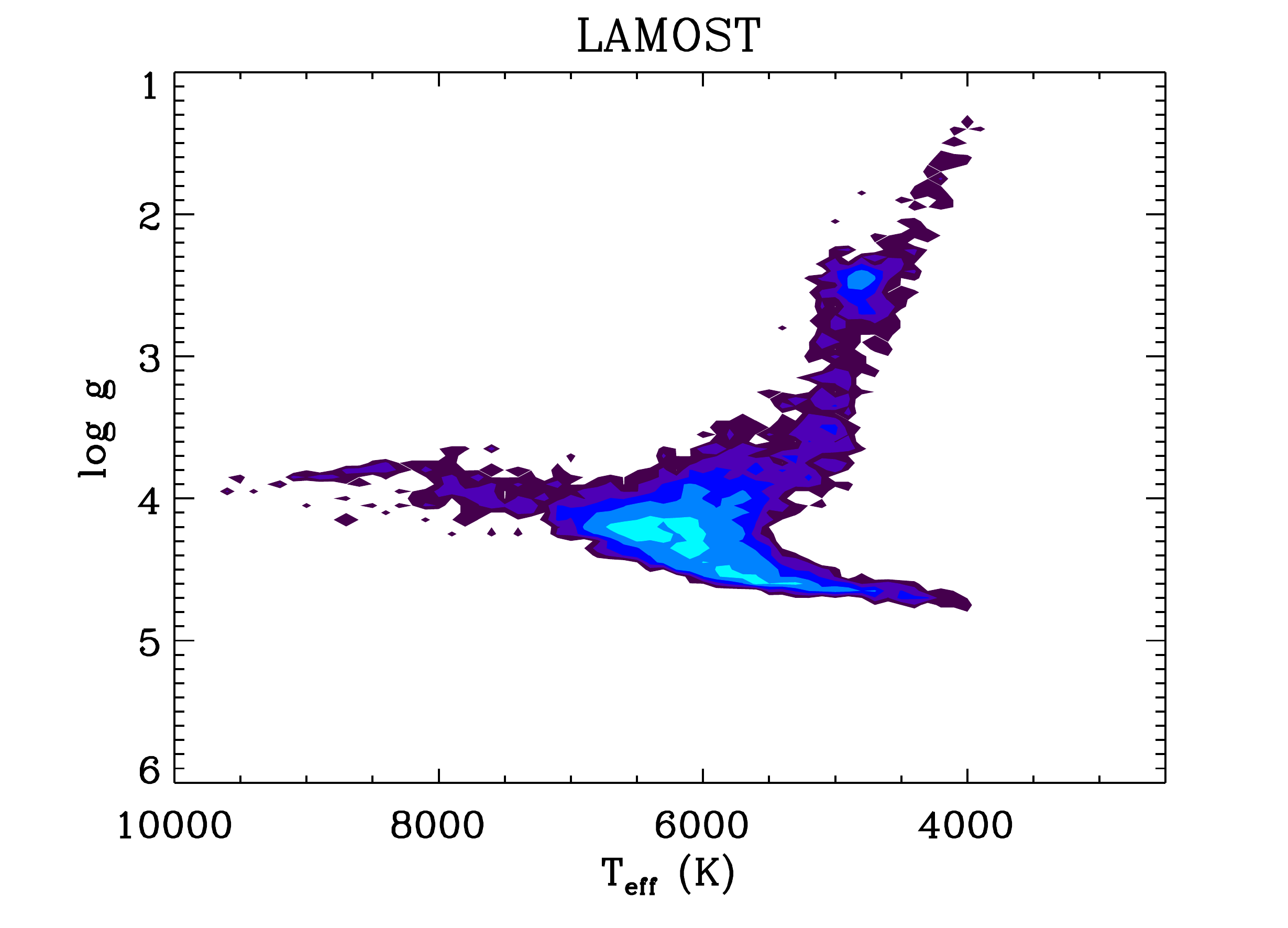}
\includegraphics[width=5.8cm, trim=2cm 0.5cm 2cm 0]{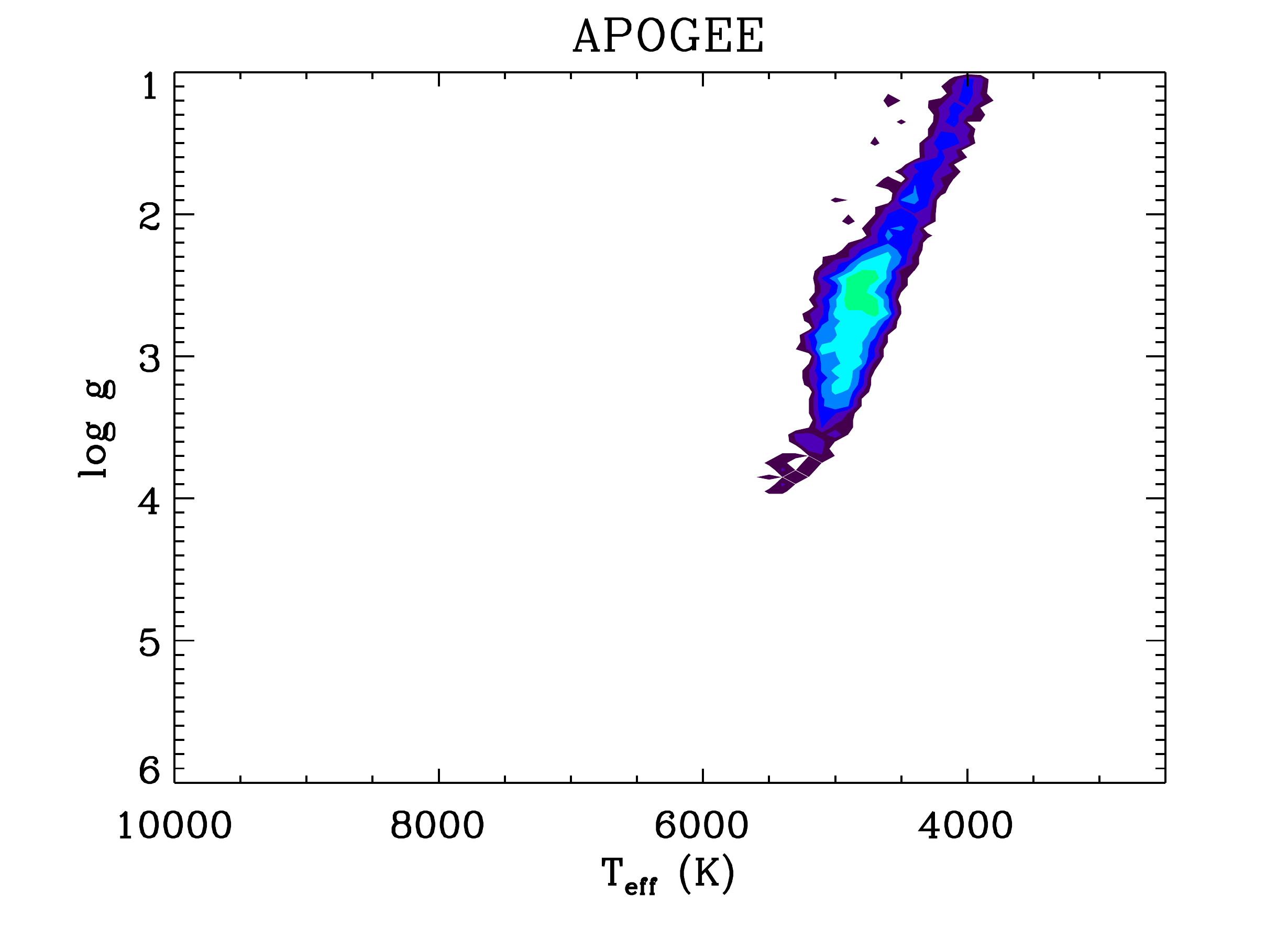}
\includegraphics[width=5.8cm, trim=2cm 0.5cm 2cm 0]{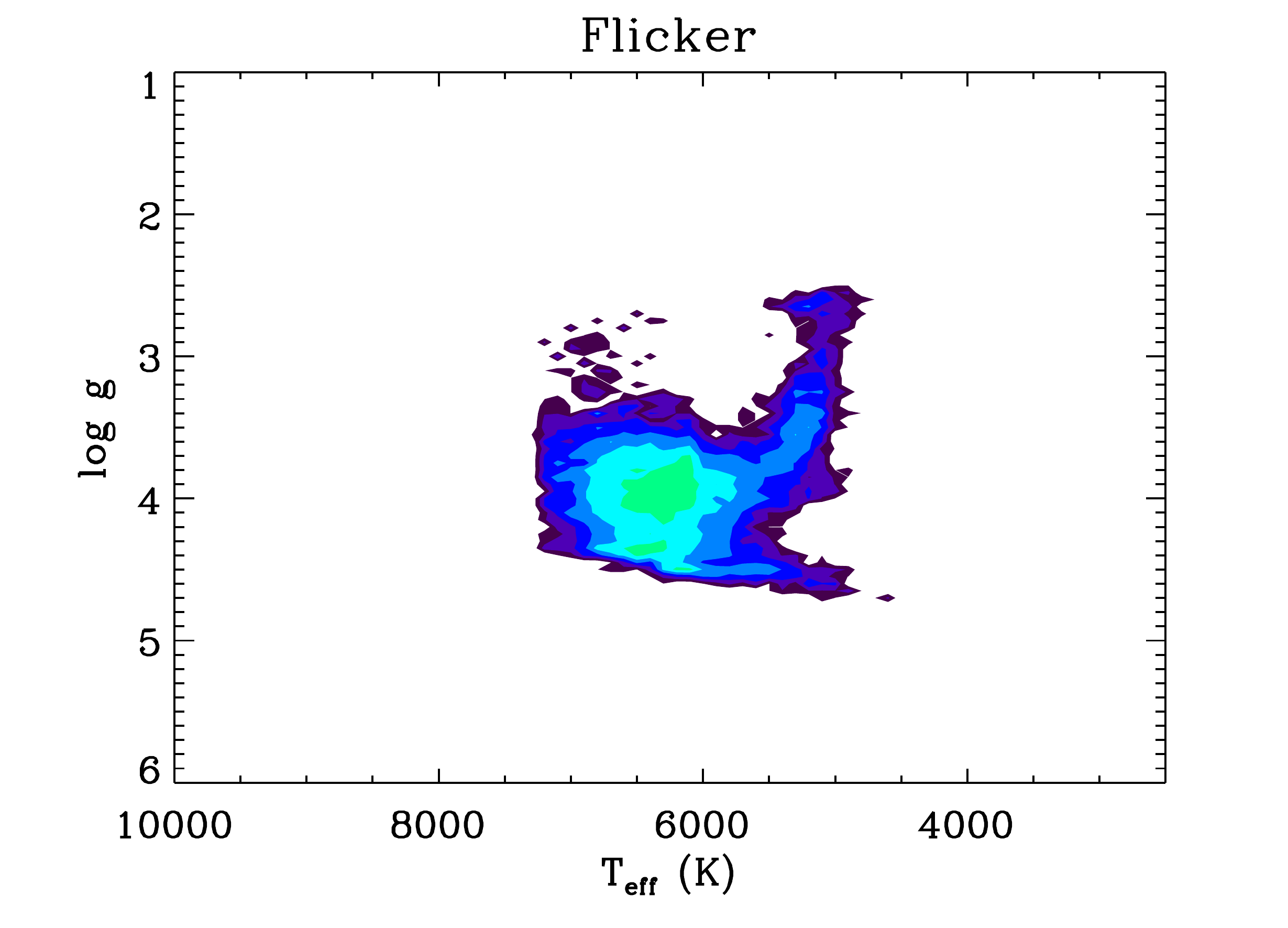}
\includegraphics[width=5.8cm, trim=2cm 0.5cm 2cm 0]{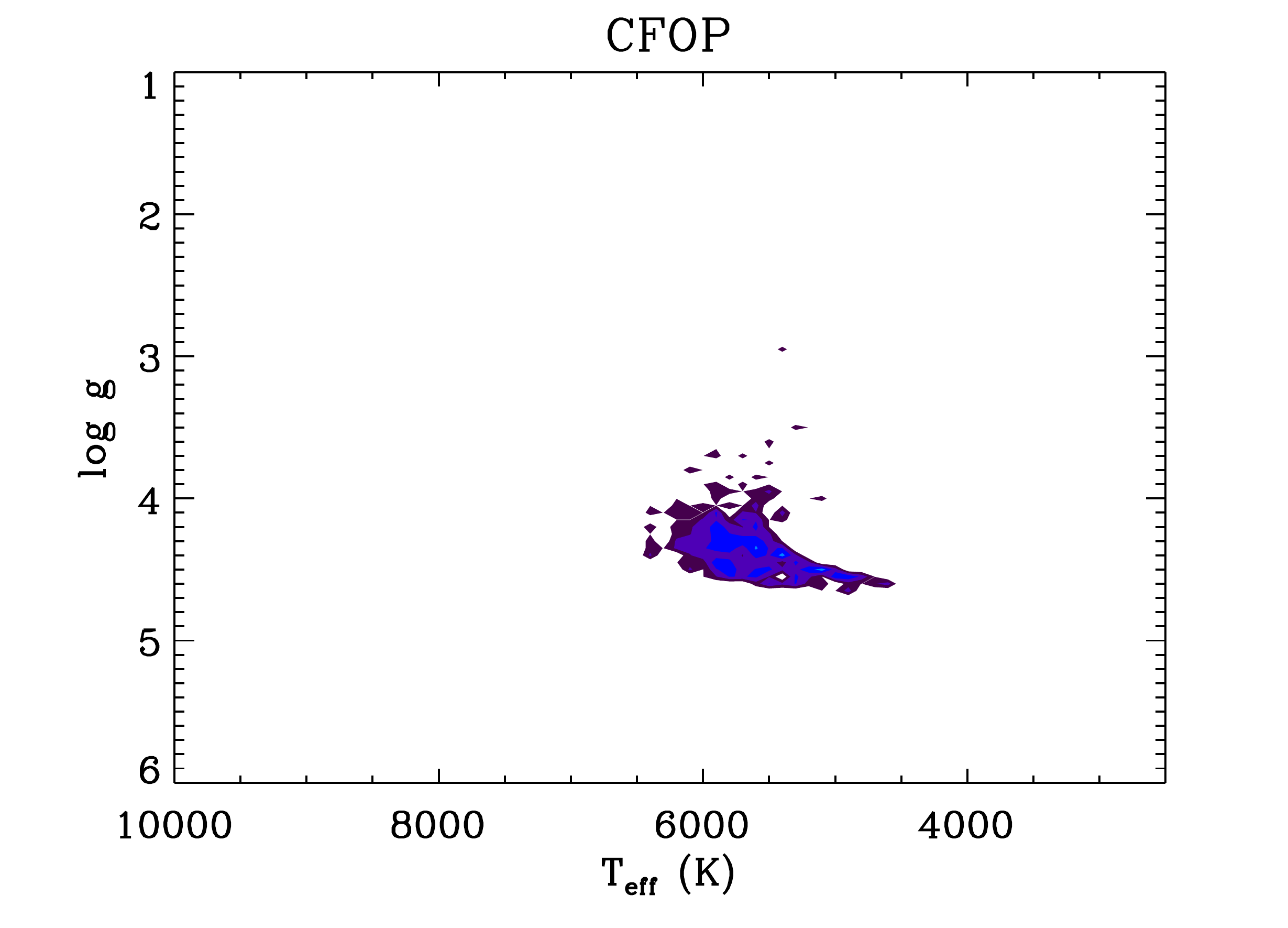}
\includegraphics[width=5.8cm, trim=2cm 0.5cm 2cm 0]{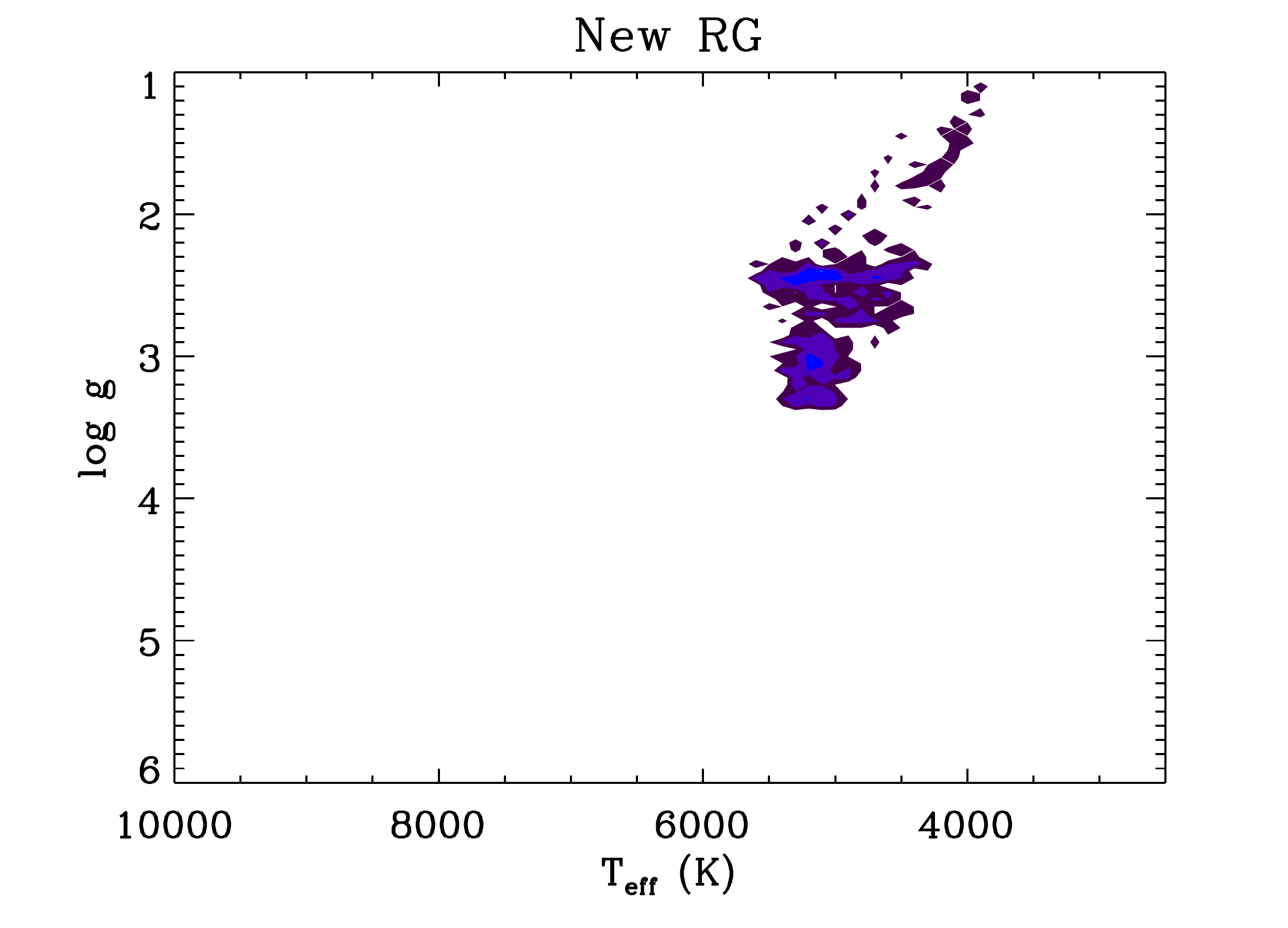}
\caption{Input surface gravity and effective temperature for the full catalog (top left panel) and for the five largest sources of new input values (see legend on the top of each panel). Color-coding denotes the logarithmic number density as shown in the color bar in the top left panel.}
\label{Fig1}
\end{center}
\end{figure*}

In this work we present another major update of the \kep\ stellar properties catalog for 197,096 {\it Kepler} targets. The catalog was developed to support the final transit detection run (Data Release 25, hereafter DR25) prior to the close-out of the {\it Kepler} mission. Initial plans for the catalog included a homogeneous reclassification based on broadband colors only (i.e.\ without relying on classifications from the KIC, see Section 9 in H14).  {   However, the limited sensitivity of available broadband colors and the complexity of constructing priors that accurately reproduce the {\it Kepler} target selection function made such a classification scheme unfeasible for the delivery of the catalog.} Thus, similar to previous versions, the updated catalog presented here is based on the consolidation of atmospheric properties (temperature \teff, surface gravity \logg, metallicity \feh) that were either published in the literature or provided by the {\it Kepler} community follow-up program \citep[CFOP,][]{2010arXiv1001.0352G}, with input values taken from different methods such as asteroseismology, spectroscopy, {   Flicker}, and photometry. 

\section{Consolidation of Input Values}

\subsection{Inputs in previous catalogs}

The stellar properties in the KIC were derived from Sloan {\it griz} and 2MASS {\it JHK} broadband photometry as well as an intermediate-band filter $D51$ that has some sensitivity to surface gravity. More details on the methodology used to build the KIC can be found in \cite{2011AJ....142..112B}. Several studies have showed a few shortcomings with the KIC. For instance, \cite{2012ApJS..199...30P} used KIC $griz$ photometry for more than 120,000 dwarfs to derive temperatures from color-temperature relations, and found that the KIC effective temperatures are underestimated by up to 200\,K. Moreover, several studies have shown that the KIC surface gravities appear to be overestimated for solar-type stars, based on comparisons to asteroseismology \citep{2011ApJ...738L..28V}, spectroscopy \citep{2013ApJ...771..107E} and surface gravities derived from stellar granulation \citep{2014ApJ...788L...9B}.

In the Q1-16 catalog, H14 consolidated literature values for temperature, surface gravity, and metallicity from asteroseismology, transits, spectroscopy, photometry, and the KIC to derive the fundamental properties of {\it Kepler} targets by fitting isochrones to these observables. However, several shortcomings remained in that catalog. For instance, 70\% of {   all {\it Kepler} target's} $\log g$ and [Fe/H] values were still based on the KIC, a number of targets without KIC stellar parameters remained unclassified, and the adopted methodology to infer stellar properties did not use priors for inferring posterior distributions. The motivation for this updated catalog was to overcome some of these shortcomings, in particular in order to have the most homogeneous catalog possible with the most up to date observables available for all {\it Kepler} targets.


\subsection{New Input Values}

The main new input values for the DR25 stellar properties catalog can be summarized as follows:


\begin{enumerate}
\item For 6,383 stars we used the effective temperatures available from Data Release 1 {   \citep{2015RAA....15.1095L}} of the Large Sky Area Multi-Object Fiber Spectroscopic Telescope (LAMOST, Xinglong observatory, China) survey \citep{2012RAA....12..723Z}. The classifications are based on medium resolution (R\,$\sim$1,800) spectra and cover a large number of stars in the {\it Kepler} field. {   There is a specific project between LAMOST and the {\it Kepler} field \citep{2015ApJS..220...19D} but the delivery of the stellar parameters \citep{2016A&A...594A..39F} was provided outside the timeframe of our catalog. The comparison of the DR25 and the LAMOST-Kepler spectroscopic results showed a good agreement in general with a standard deviation of the temperature differences of 228K for dwarfs and 205K for red giants and of surface gravity of 0.26dex for dwarfs and 0.40dex  for red giants.}

\item The Apache Point Observatory for Galactic Evolution Experiment \citep[APOGEE,][]{2015arXiv150905420M} also targeted a large number of {\it Kepler} stars to obtain high-resolution (R\,$\sim$\,22,500) H-band spectra, mostly for red giant stars. We adopted the effective temperature from APOGEE for 5678 stars, surface gravities for 1544 stars, and metallicities for 5662 stars from DR12 \citep{2015ApJS..219...12A}. 

\item For 14,535 stars we adopted surface gravities estimated from the detection of granulation in the {\it Kepler} light curves \citep[the Flicker method,][]{2016ApJ...818...43B}. We limited the Flicker $\log g$ values to stars for which the reported uncertainty was smaller than 0.2 dex to ensure a higher reliability of the input values. 

\item For more than 1,000 stars, we used spectroscopic parameters (\teff, \logg, \feh) provided by the {\it Kepler} community follow-up program (CFOP) that observed around 800 planet candidate host stars and 535 solar-like stars for which solar-oscillations had been detected in the {\it Kepler} data.

\item We included a sample of ~835 stars, which were classified as dwarfs in the original KIC but were shown to be red giants based on the detection of giant-like oscillations in the \kep\ data. We adopted $\log g$ values estimated from asteroseismology in combination with revised effective temperatures for these stars \citep{2016ApJ...827...50M}. 

\item For 62 newly confirmed {\it Kepler} exoplanet hosts we adopted stellar parameters (\teff, \logg, \feh) as published in the discovery papers. 

\item We also report spectroscopic parameters (\teff, \logg, \feh) for 317 stars, which were so far unclassified but were included in either the APOGEE or LAMOST surveys. 

\item We added 311 stars that were new targets observed during Q17.
\end{enumerate}

Compared to the H14 catalog, new input values are used for 14.7\% of the stars. In this final catalog, the input $\log g$ values are taken from seismology for {   16,947 stars (8.6\% of the stars)}, from {   Flicker} for {   14,535 stars (7.4\%)}, {   from spectroscopy for 9,277 stars (4.7\%)} and from the KIC for {   143,785 stars, corresponding to $\sim$\,72.9\% of the total sample compared to $\sim$84\% for the H14 catalog. The remaining stars have their $\log g$ input values either from photometry or transit search}. For the input effective temperature, the source is spectroscopy for {   14,813 stars}, non-KIC photometry for {   151,118 stars} and the KIC for {   31,165 stars}.

\begin{figure}[htbp]
\begin{center}
\includegraphics[width=9cm]{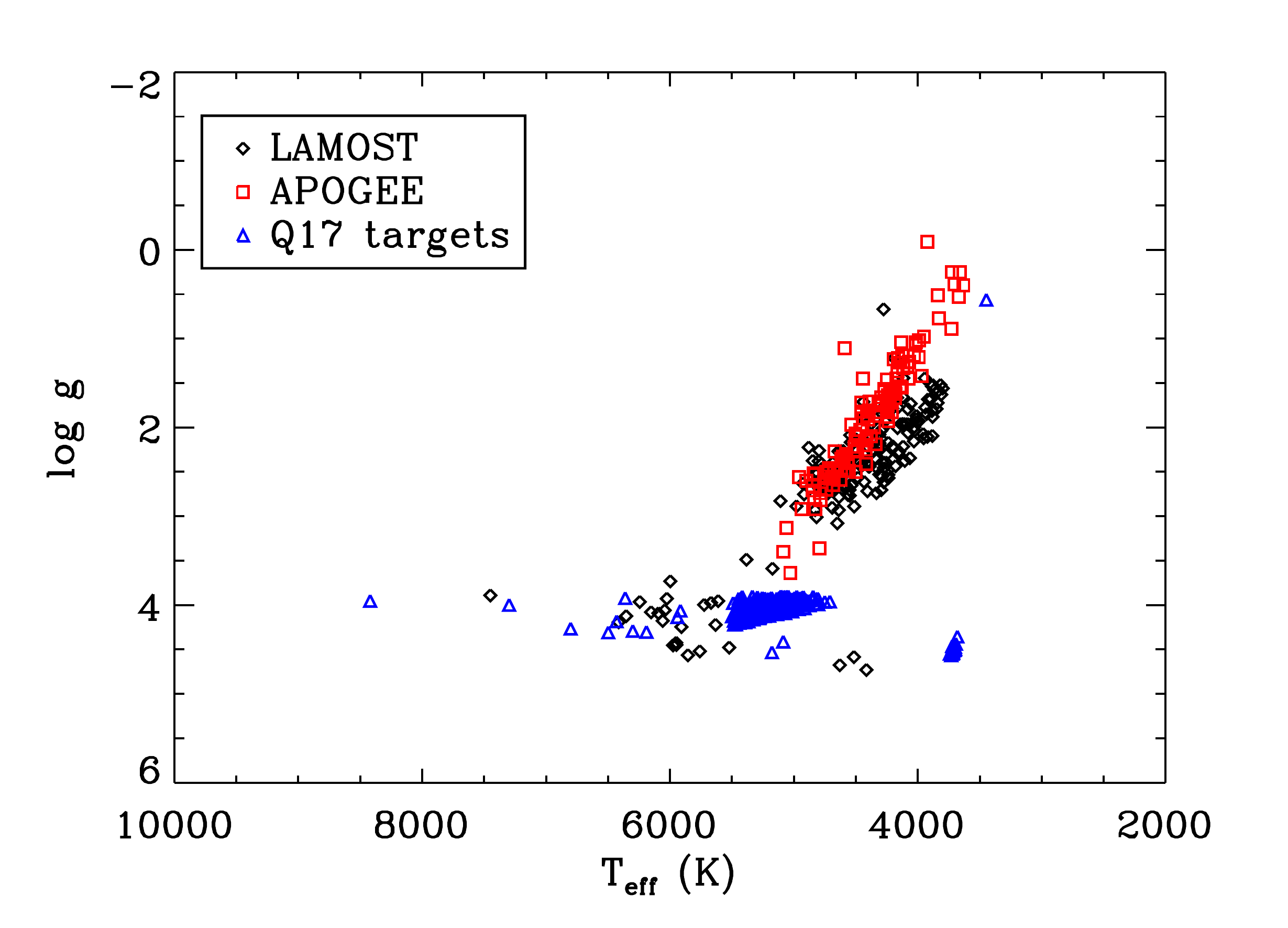}
\caption{Surface gravity versus effective temperature for targets that were newly added compared to the Q1--16 catalog. Different symbols show stars with classifications adopted from LAMOST (black diamonds), APOGEE (red squares), and new stars targeted during Q17 (blue triangles).}
\label{newstars}
\end{center}

\end{figure}

Figure~\ref{Fig1} shows an HR diagram of the largest sources of new input values, namely LAMOST, APOGEE, Flicker, CFOP, and the sample of misclassified red giants.  Figure~\ref{newstars} represents the stars that were added compared to the H14 catalog, which are either stars that remained unclassified in the Q1--16 due to a lack of 2MASS photometry or targets that were first observed in Q17. There is no overlap between the new Q17 targets and the unclassified stars with LAMOST and APOGEE spectra. Among the new 628 additional stars, {   294} are red giants and {   332} are dwarfs, the remaining stars being subgiants.



Given that some stars have input parameters from different literature sources a prioritization scheme had to be adopted. This prioritization was mostly based on the precision and accuracy of the sources used to derive the input values, as follows. For surface gravity, the highest priority was given to asteroseismology, then high-resolution (HR) spectroscopy, low-resolution (LR) spectroscopy, {   Flicker}, photometric observations, and finally the KIC. For the temperature and metallicity, the highest priority was given to high-resolution spectroscopy, low-resolution spectroscopy, photometric observations, and the KIC. In other words, priority was given to the CFOP observations and published values for confirmed planets, then APOGEE, LAMOST, and finally to the KIC. The prioritization scheme is given in Table~\ref{tab_priority}.

\begin{table}
\caption{Priority list for input surface gravity, effective temperature, and metallicity from different techniques.}
\begin{center}
\begin{tabular}{ccc}
\hline
\hline
Parameter & Priority & Input\\
\hline
$\log g$ &1 & Asteroseismology\\
 & 2 & HR spectroscopy\\
 & 3 & LR spectroscopy\\
 & 4 & Flicker\\
 & 5 & KIC\\
 \hline
 $T_{\rm eff}$ / [Fe/H] & 1& HR spectroscopy\\
  & 2 &LR spectroscopy \\
  & 3 & Photometry\\
  & 4 & KIC\\
\hline
\end{tabular}
\flushleft Notes: LR = low resolution (R < 5000); HR = high resolution (R $\ge$ 5000).
\end{center}
\label{tab_priority}
\end{table}%

Typical uncertainties associated with each observable are taken from H14 and listed in Table~\ref{tab_uncert}.  In addition, we adopted a typical uncertainty of 0.2 dex for the Flicker $\log$ g. These are the uncertainties used as inputs.

\begin{table}
\caption{Uncertainties adopted for the input parameters.}
\begin{center}
\begin{tabular}{lccc}
\hline
\hline
Method & $\sigma_{T_{\rm eff}}$ & $\sigma_{\log g}$ & $\sigma_{[Fe/H]}$\\
        & (\%) & (dex) & (dex) \\
\hline
Asteroseismology & -- & 0.03 & --\\
Transits & -- & 0.05 & --\\
Spectroscopy & 2 & 0.15 & 0.15\\
Flicker & -- & 0.20 & --\\
Photometry & 3.5 & 0.40 & 0.30\\
KIC & 3.5 & 0.40 & 0.30\\
\hline
\end{tabular}
\end{center}
\label{tab_uncert}
\end{table}%

The input values and provenances used for the full catalog are listed in Table~\ref{input}. Following H14, the provenances are comprised of three letters and a number corresponding to the reference key of Table~\ref{tab:refs} in the Appendix A. The provenances keywords are: {   AST for Asteroseismology, FLK for Flicker, KIC for Kepler Input Catalog, PHO for Photometry, SPE for Spectroscopy, and TRA for Transits.}

\begin{table*}
\begin{center}
\caption{Input values of the DR25 stellar properties catalog.}
\begin{tabular}{ccccccc}
\hline
\hline
KIC & $T_{\rm eff}$ & $\log g$ & [Fe/H] & $P_{\rm Teff}$ & $P_{\rm logg}$ & $P_{\rm Fe/H}$ \\
\hline
 757076 &  5164\,$\pm$\, 154&3.601\,$\pm$\,0.400&-0.083\,$\pm$\,0.300&PHO1&KIC0&
KIC0\\
757099 &  5521\,$\pm$\, 168&3.817\,$\pm$\,0.400&-0.208\,$\pm$\,0.300&PHO1&KIC0&
KIC0\\
757137 &  4751\,$\pm$\, 139&2.378\,$\pm$\,0.030&-0.079\,$\pm$\,0.300&PHO1&AST9&
KIC0\\
757280 &  6543\,$\pm$\, 188&4.082\,$\pm$\,0.400&-0.231\,$\pm$\,0.300&PHO1&KIC0&
KIC0\\
757450 &  5330\,$\pm$\, 106&4.500\,$\pm$\,0.050&-0.070\,$\pm$\,0.150&SPE51&TRA51
&SPE51\\
891901 &  6325\,$\pm$\, 186&4.411\,$\pm$\,0.400&-0.084\,$\pm$\,0.300&PHO1&KIC0&
KIC0\\
891916 &  5602\,$\pm$\, 165&4.591\,$\pm$\,0.400&-0.580\,$\pm$\,0.300&PHO1&KIC0&
KIC0\\
892010 &  4834\,$\pm$\, 151&2.163\,$\pm$\,0.030& 0.207\,$\pm$\,0.300&PHO1&AST9&
KIC0\\
892107 &  5086\,$\pm$\, 161&3.355\,$\pm$\,0.400&-0.085\,$\pm$\,0.300&PHO1&KIC0&
KIC0\\
892195 &  5521\,$\pm$\, 184&3.972\,$\pm$\,0.400&-0.054\,$\pm$\,0.300&PHO1&KIC0&
KIC0\\
892203 &  5945\,$\pm$\, 208&4.081\,$\pm$\,0.400&-0.118\,$\pm$\,0.300&PHO1&KIC0&
KIC0\\
892376 &  3963\,$\pm$\, 138&4.471\,$\pm$\,0.400& 0.122\,$\pm$\,0.300&KIC0&KIC0&
KIC0\\
892667 &  6604\,$\pm$\, 209&4.100\,$\pm$\,0.400&-0.256\,$\pm$\,0.300&PHO1&KIC0&
KIC0\\
892675 &  6312\,$\pm$\, 208&4.048\,$\pm$\,0.400&-0.257\,$\pm$\,0.300&PHO1&KIC0&
KIC0\\
892678 &  6136\,$\pm$\, 177&3.939\,$\pm$\,0.400&-0.260\,$\pm$\,0.300&PHO1&KIC0&
KIC0\\
 ...\\
\hline
\end{tabular}
\label{input}
\flushleft Notes: See Table~\ref{tab:refs} for the reference key for the provenances listed in the last three columns.
\end{center}
\end{table*}%


\section{Catalog Construction}

\subsection{Methodology}

We followed H14 by comparing the input \teff, \logg\ and \feh\ values to stellar evolution models in order to infer additional stellar parameters such as radii, which are required by the {\it Kepler} planet detection pipeline. For the current catalog we adopted the isochrones from the Dartmouth Stellar evolution Database \citep[DSEP,][]{2008ApJS..178...89D}, which cover a wide range in parameter space and have demonstrated good agreement with interferometric observations of low-mass dwarfs. We improved the original DSEP grid adopted by H14 by interpolating each isochrone of a given age and [Fe/H] in mass to yield a stepsize of at most 0.02\msun\ for all models with $\log g> 4.0$, which removes significant gaps in the original grid for cool dwarfs. The final grid included around 1.8 $\times 10^7$ models, and spanned from 1--15\,Gyr in steps of 0.5\,Gyr in age and -2.5--0.56\,dex in steps of 0.02\, dex in \feh. 

\begin{figure*}
\begin{center}
\includegraphics[width=5.8cm, trim=2cm 0.5cm 1cm 0.5cm]{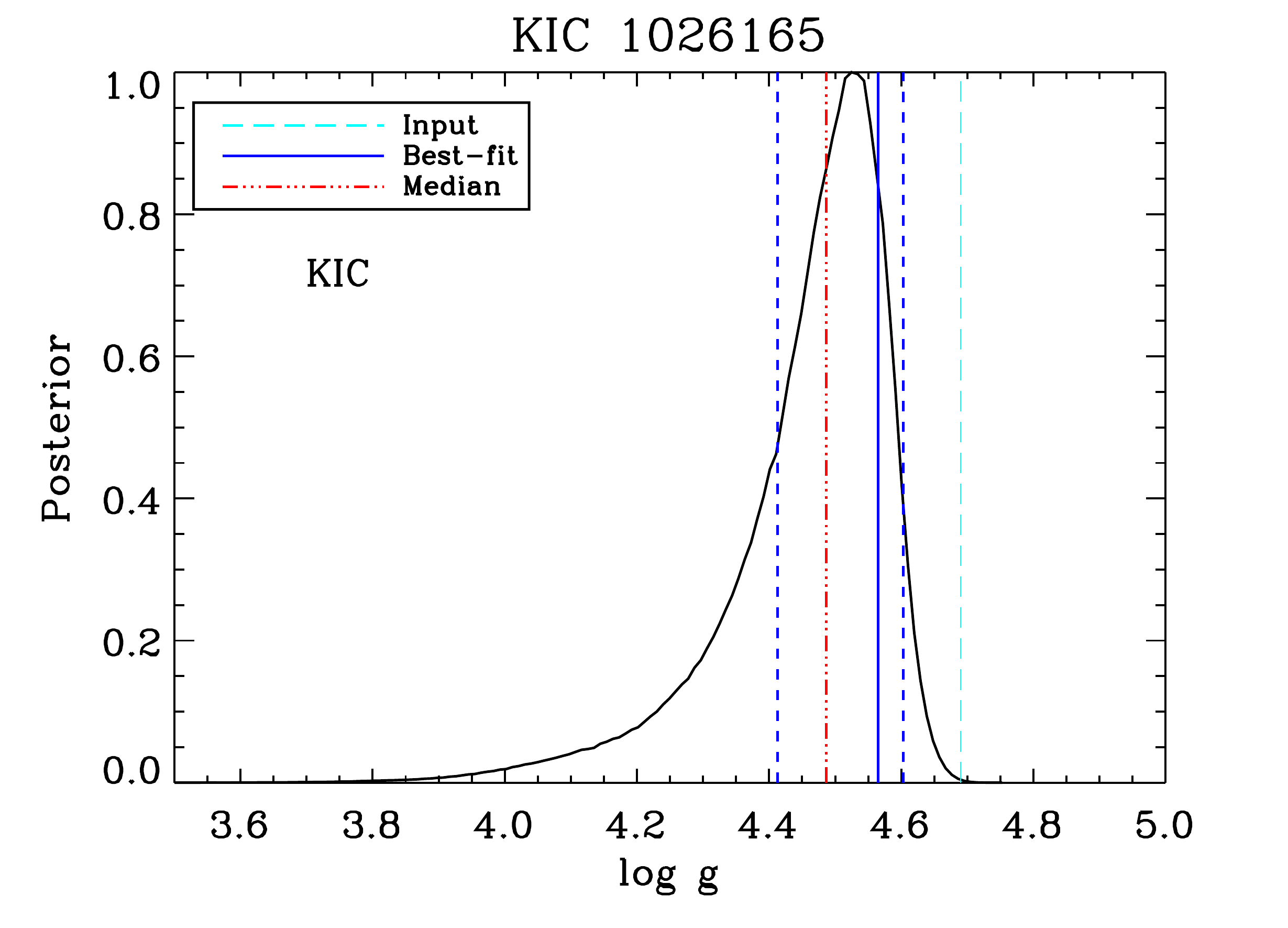}
\includegraphics[width=5.8cm, trim=2cm 0.5cm 1cm 0.5cm]{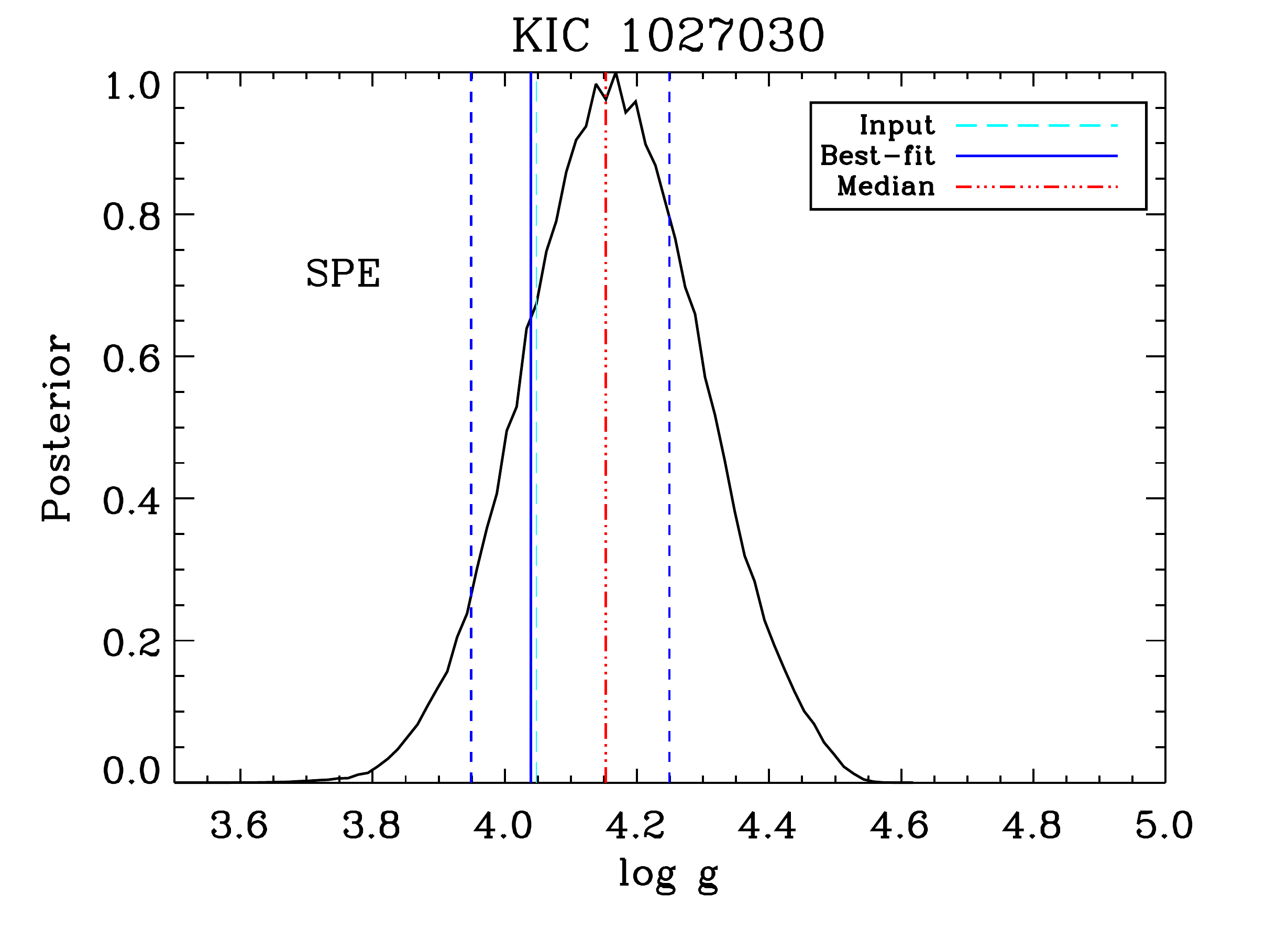}
\includegraphics[width=5.8cm, trim=2cm 0.5cm 1cm 0.5cm]{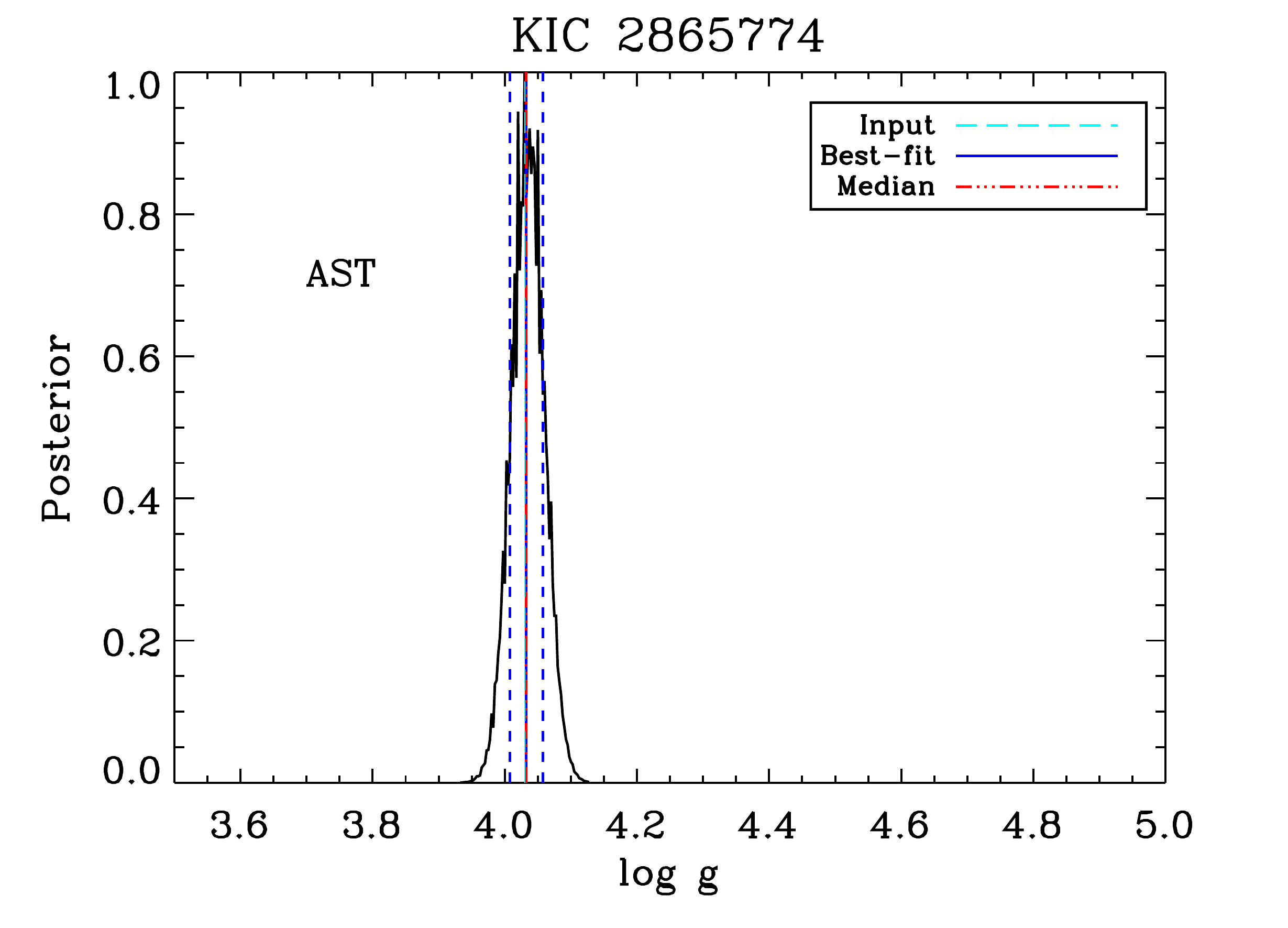}\\
\includegraphics[width=5.8cm, trim=2cm 1cm 1cm 0]{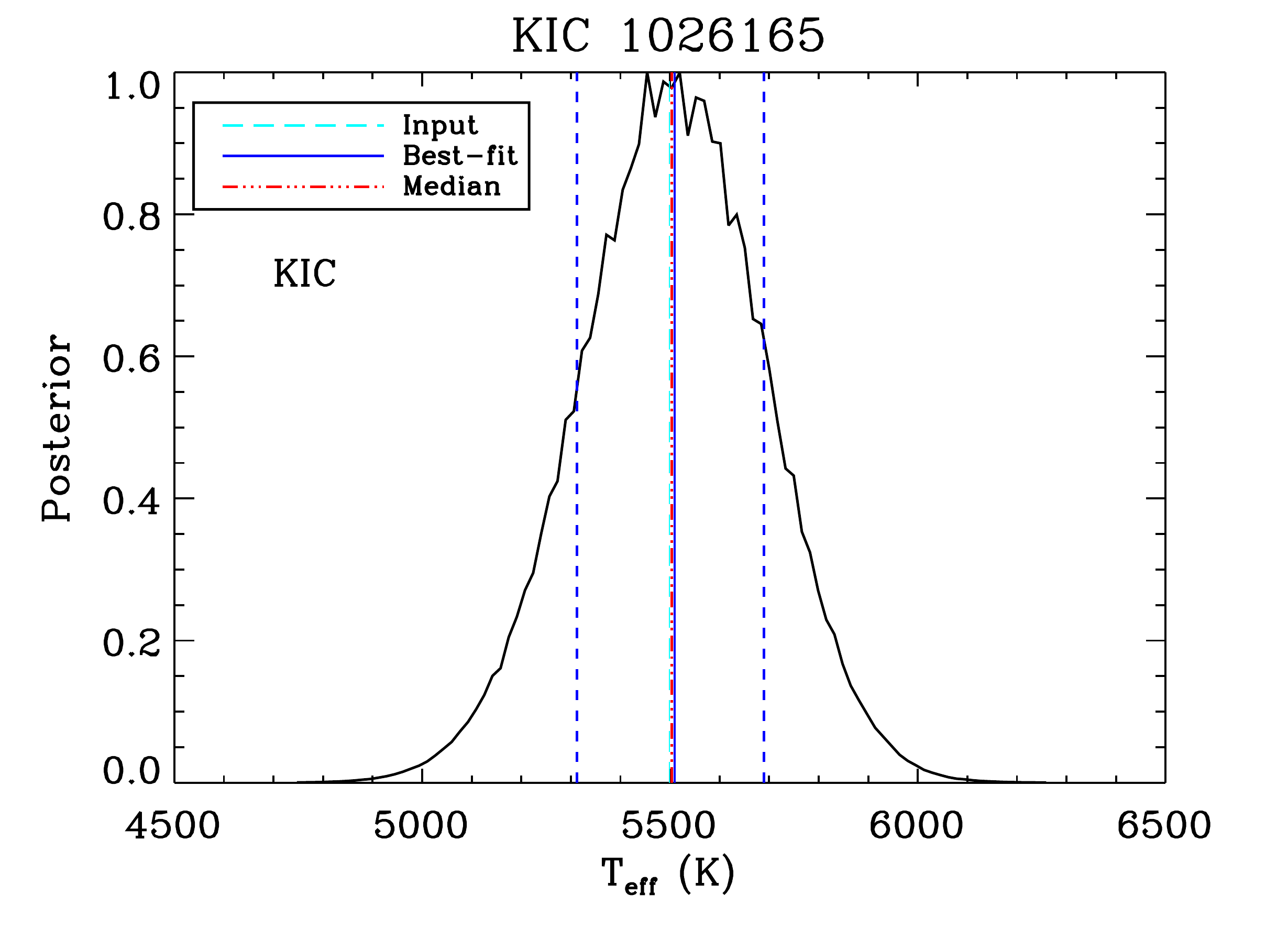}
\includegraphics[width=5.8cm, trim=2cm 1cm 1cm 0]{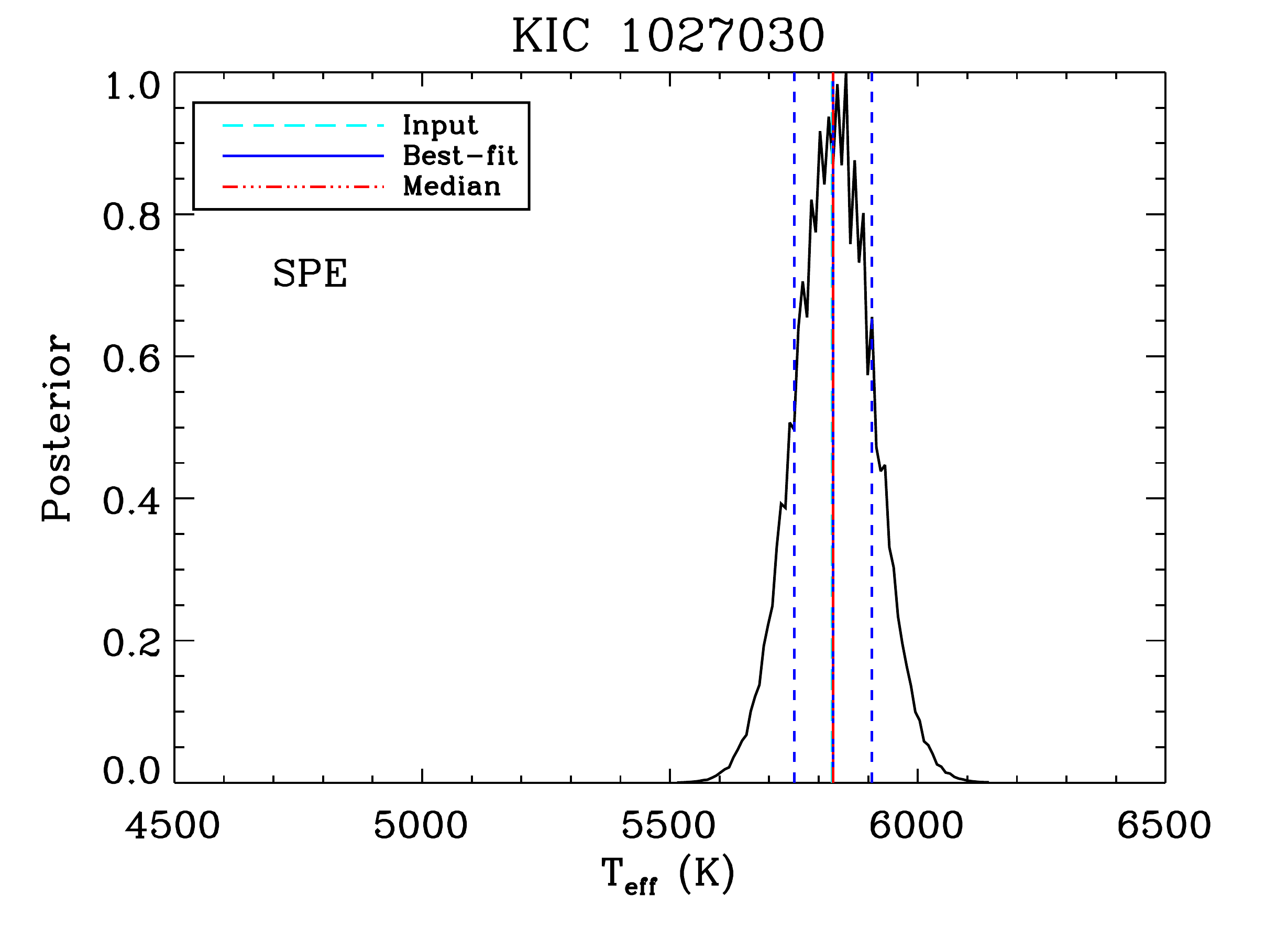}
\includegraphics[width=5.8cm, trim=2cm 1cm 1cm 0]{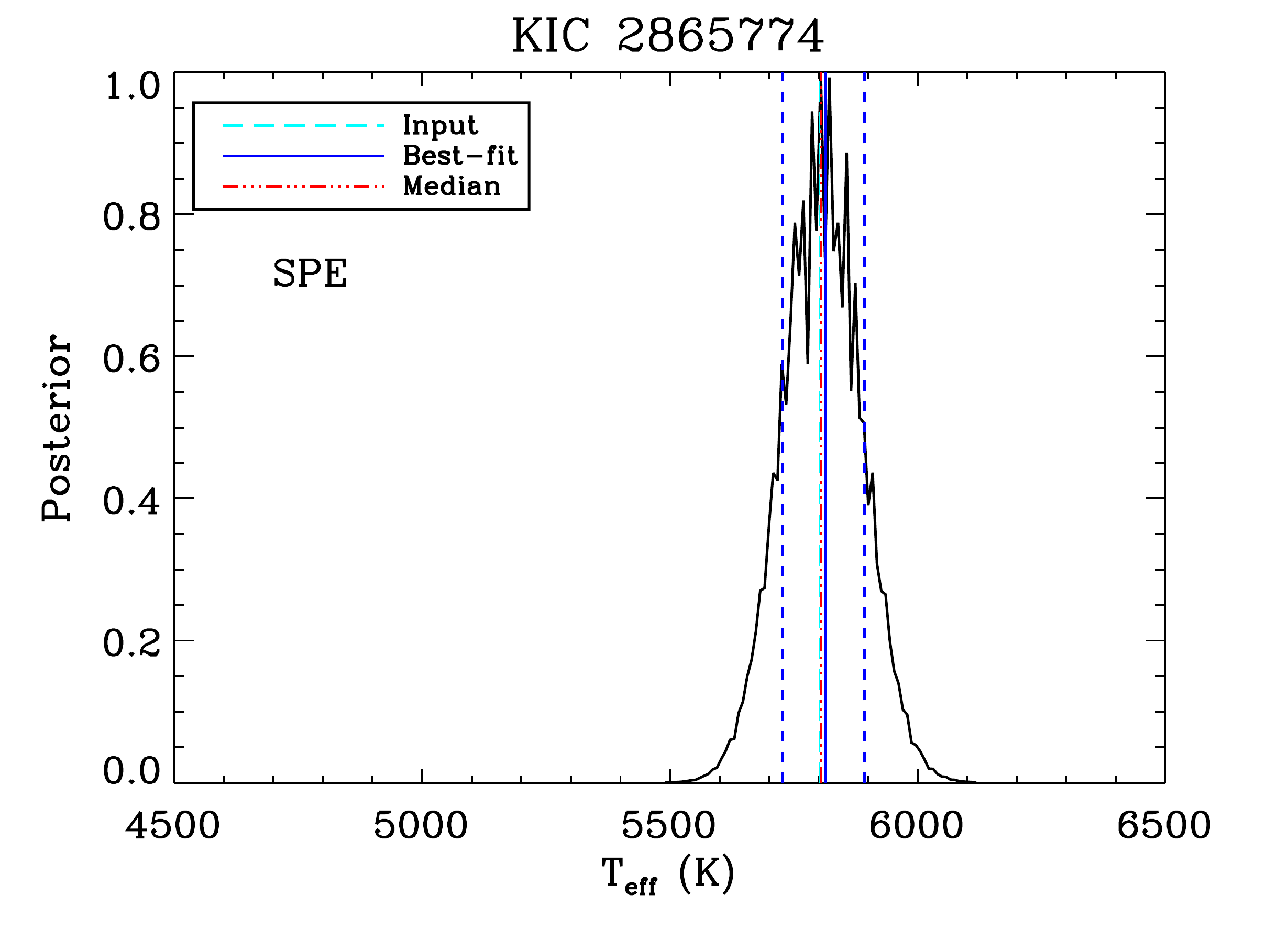}
\caption{Discrete posterior distributions for surface gravity (top panels) and effective temperature (bottom panels) for three different stars. Input values in $\log g$ were adopted from KIC (left panel), spectroscopy (middle panel), and asteroseismology (right panels). Input values in $T_{\rm eff}$ were adopted from KIC (left panel) and spectroscopy (middle and right panel). The dash cyan line marks the input value and the solid blue line is the output value with associated uncertainties (blue dashed lines). The dash-dotted red line corresponds to the median value of the distribution.}
\label{discrete_post}
\end{center}
\end{figure*}

We followed the methodology by \citet{2013MNRAS.429.3645S} to infer stellar parameters from isochrones. Given a set of input values  $x=\{\teff,\logg,\feh\}$ with Gaussian uncertainties $\sigma_x$ and a set of intrinsic parameters $y=\{\rm age,\ [Fe/H],\ mass\}$, we calculated the posterior probability of the observed star
having intrinsic parameters $y$ as:

\begin{equation}
p(y|x) = p(y)p(x|y) =  p(y)\prod_i \exp{\left(-\frac{(x_i-x_i(y))^2}{2 \sigma_{x,i}^2}\right)} \: .
\end{equation}

We adopted flat priors $p(y)$ for mass, age, and metallicity. Probability distribution functions for any given stellar parameter were then obtained by weighting $p(y|x)$ by the volume which each isochrone point encompasses in mass, age and metallicity, and summing the resulting distribution along a given stellar parameter.  The bin size was initially fixed to either an absolute value for $\log g$, [Fe/H], mass, density or a fractional step size of the best fit value for radius and distance. From this initial distribution we calculated the 1-$\sigma$ confidence interval, and then iterated the stepsize to yield at least 10 bins within a 1-$\sigma$ confidence interval. The posteriors calculated using this method are hereafter referred to as ``Discrete posteriors''. 




Figure~\ref{discrete_post} shows examples of discrete posteriors in effective temperature and surface gravity for three stars with an input $\log g$ from the KIC (top left panel), spectroscopy (top middle panel), and asteroseismology (top right panel) and an input $T_{\rm eff}$ from the KIC (bottom left panel) and spectroscopy (bottom middle and bottom right panels). The large input uncertainty in $\log g$ for the KIC yields a distribution which peaks near the main sequence (the most probable for a star with a weak $\log g$ constraint) and has a tail towards lower $\log g$ values, reflecting the uncertainty of the evolutionary state of the star. On the other hand, the smaller uncertainty of the spectroscopic and asteroseismic $\log g$ and spectroscopic $T_{\rm eff}$ values yield discrete posteriors which are considerably more narrow.

Since the {\it Kepler} pipeline requires a single value and uncertainty for each stellar parameter, a suitable summary statistic had to be chosen. We decided to report the best-fit value (calculated by maximizing Equation 1), with an uncertainty derived from the 1-$\sigma$ interval around the best fit. As shown in Figure~\ref{discrete_post}, the best-fit value does not always coincide with the mode of the posterior distribution. Adopting the best-fit was motivated by the fact that adopting the mode or median as a point estimate would lead to an unrealistically high number of main-sequence stars due to the fact that for a given input value of $\log g$ with a large uncertainty, a star will probabilistically be most likely on the main sequence. Since the {\it Kepler} target stars represent neither a volume nor a strictly magnitude-limited sample \citep[see, for example, the target selection criteria as described in][]{2010ApJ...713L.109B}, constructing a prior to characterize the most probable evolutionary state of a {\it Kepler} target star is not straightforward. The stellar classification in the KIC used a prior constructed from a volume-limited Hipparcos sample, which has been shown to underestimate the number of subgiants due to Malmquist bias \citep[see for example][]{2014ApJ...788L...9B}. Adopting the best-fit values ensures that the point estimates reported in the catalog account for some of the expected Malmquist bias in the {\it Kepler} sample, but we caution that some systematic biases likely remain in the catalog.




\subsection{Stellar Parameter Uncertainties}

\begin{figure}
\begin{center}
\includegraphics[width=4.2cm, trim=1cm 1.5cm 1cm 0]{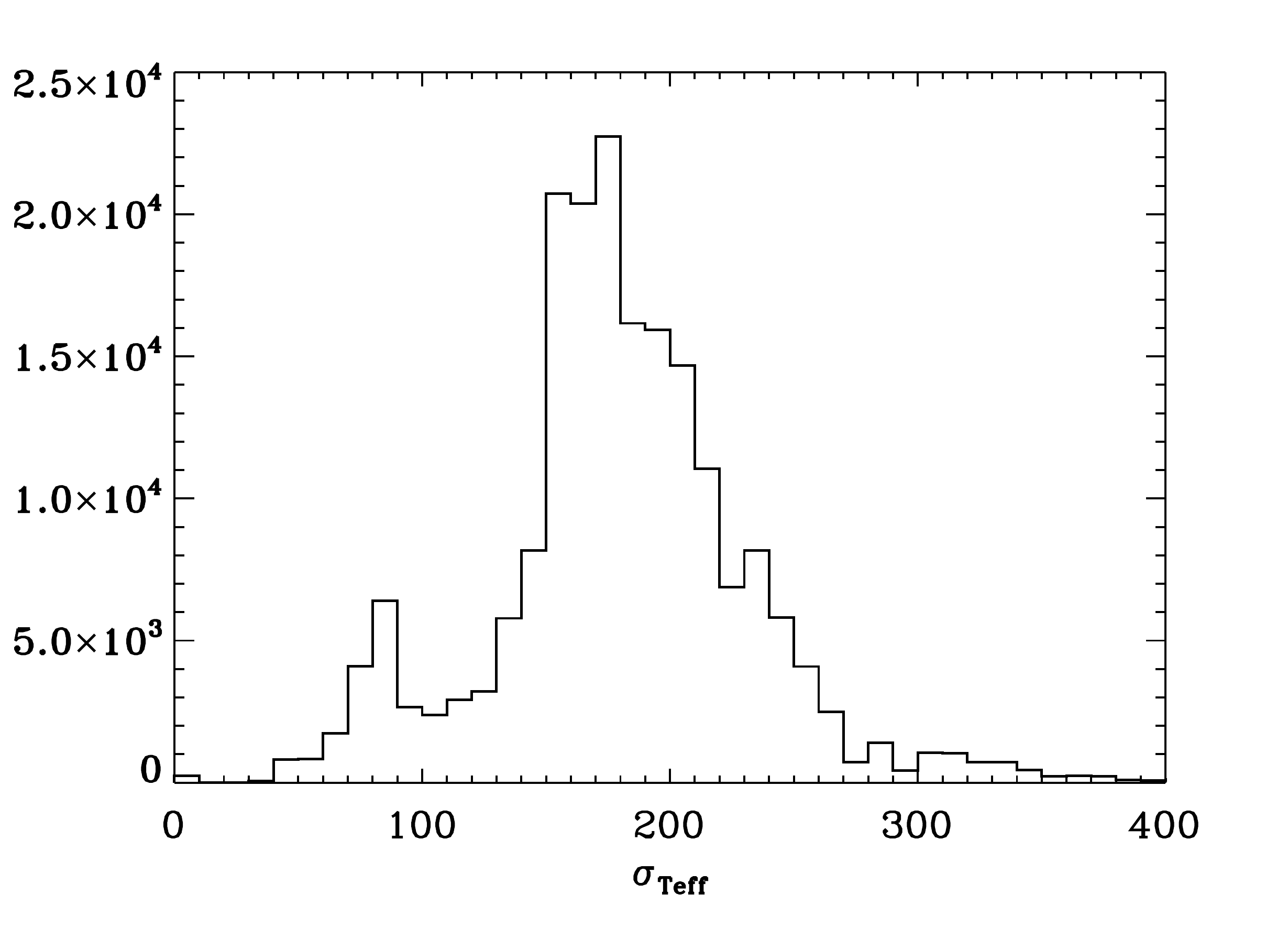}
\includegraphics[width=4.2cm, trim=1cm 1.5cm 1cm 0]{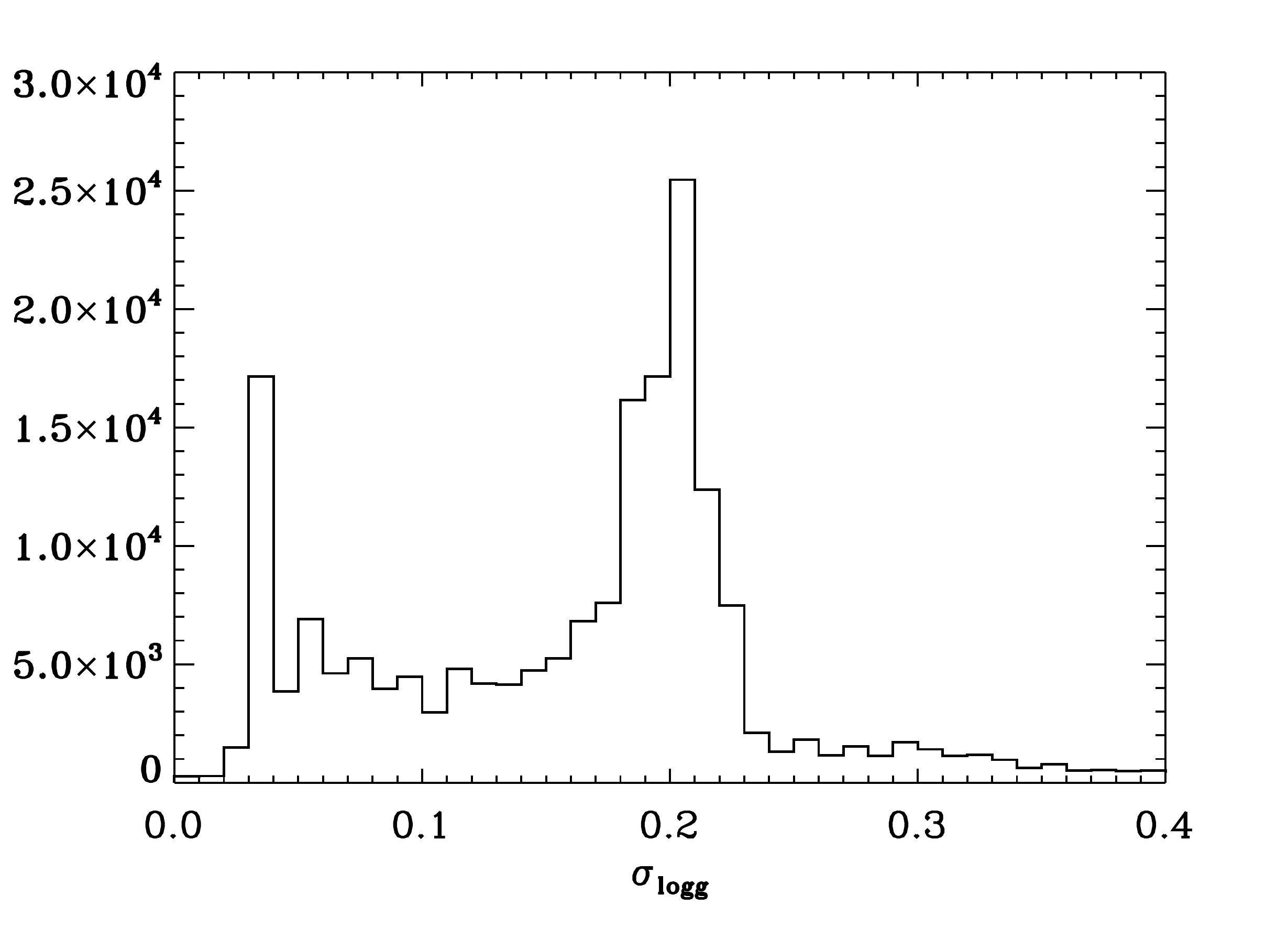}
\includegraphics[width=4.2cm, trim=1cm 1.5cm 1cm 0]{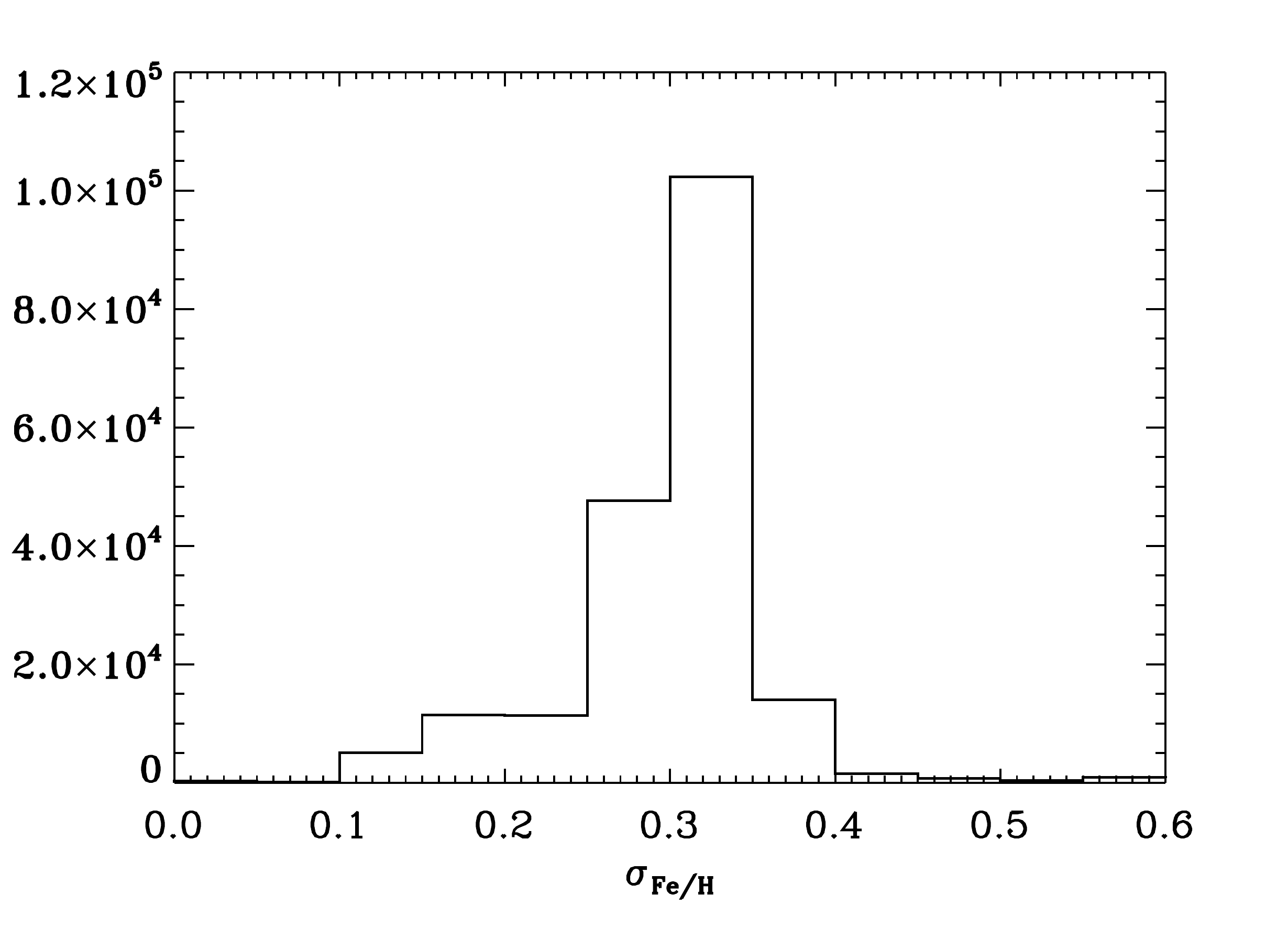}
\includegraphics[width=4.2cm, trim=1cm 1.5cm 1cm 0]{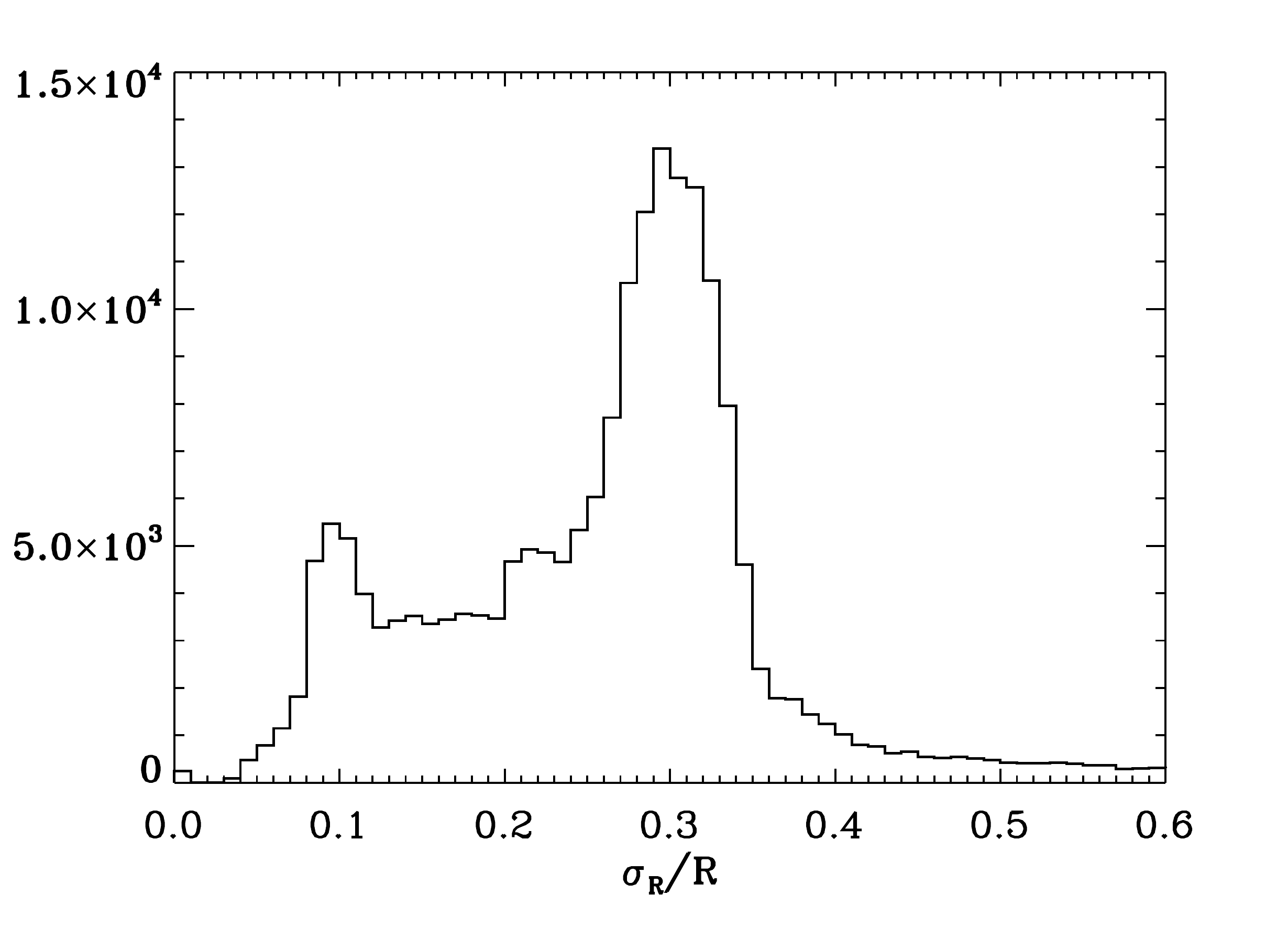}
\includegraphics[width=4.2cm, trim=1cm 1.5cm 1cm 0]{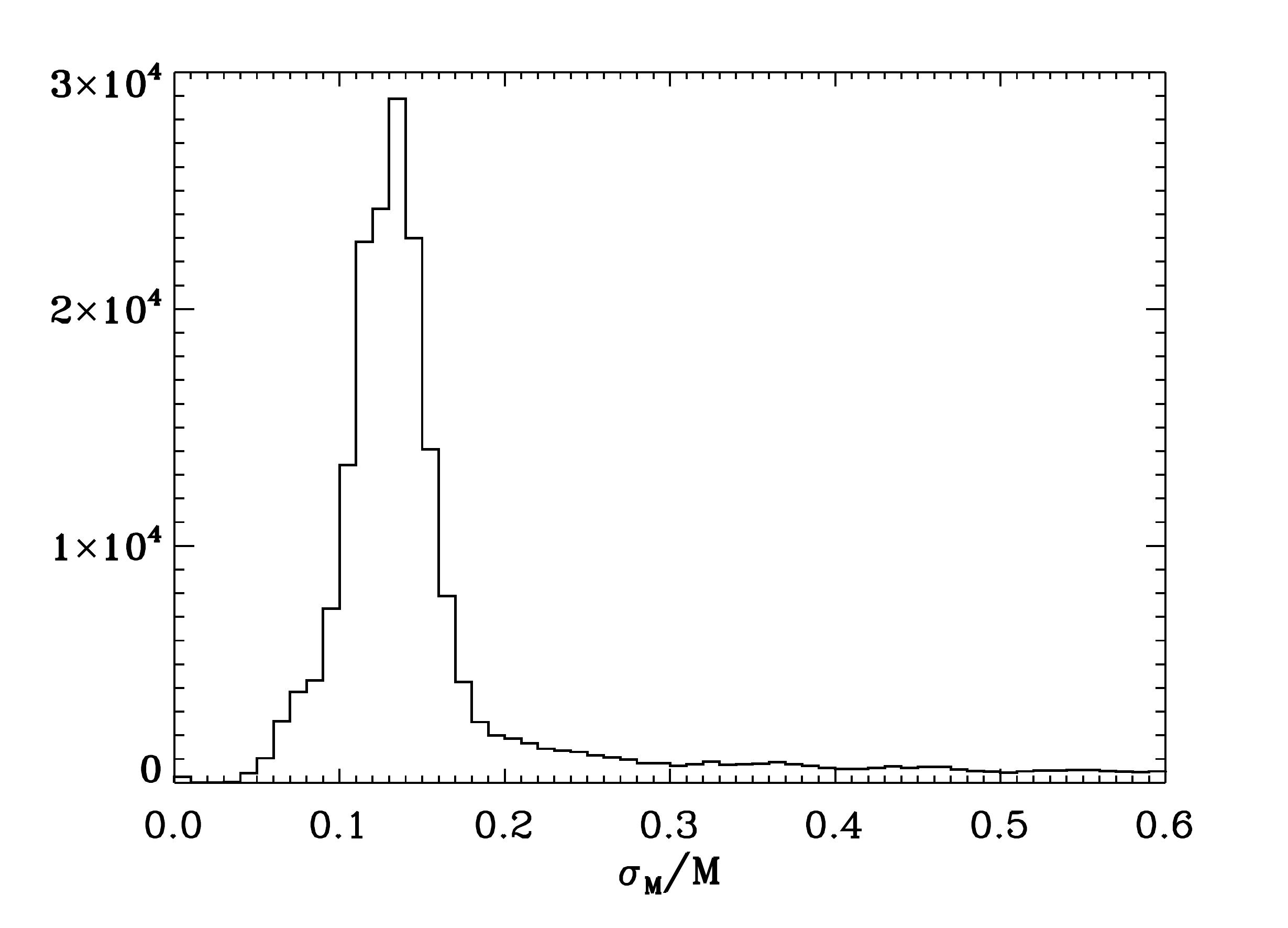}
\includegraphics[width=4.2cm, trim=1cm 1.5cm 1cm 0]{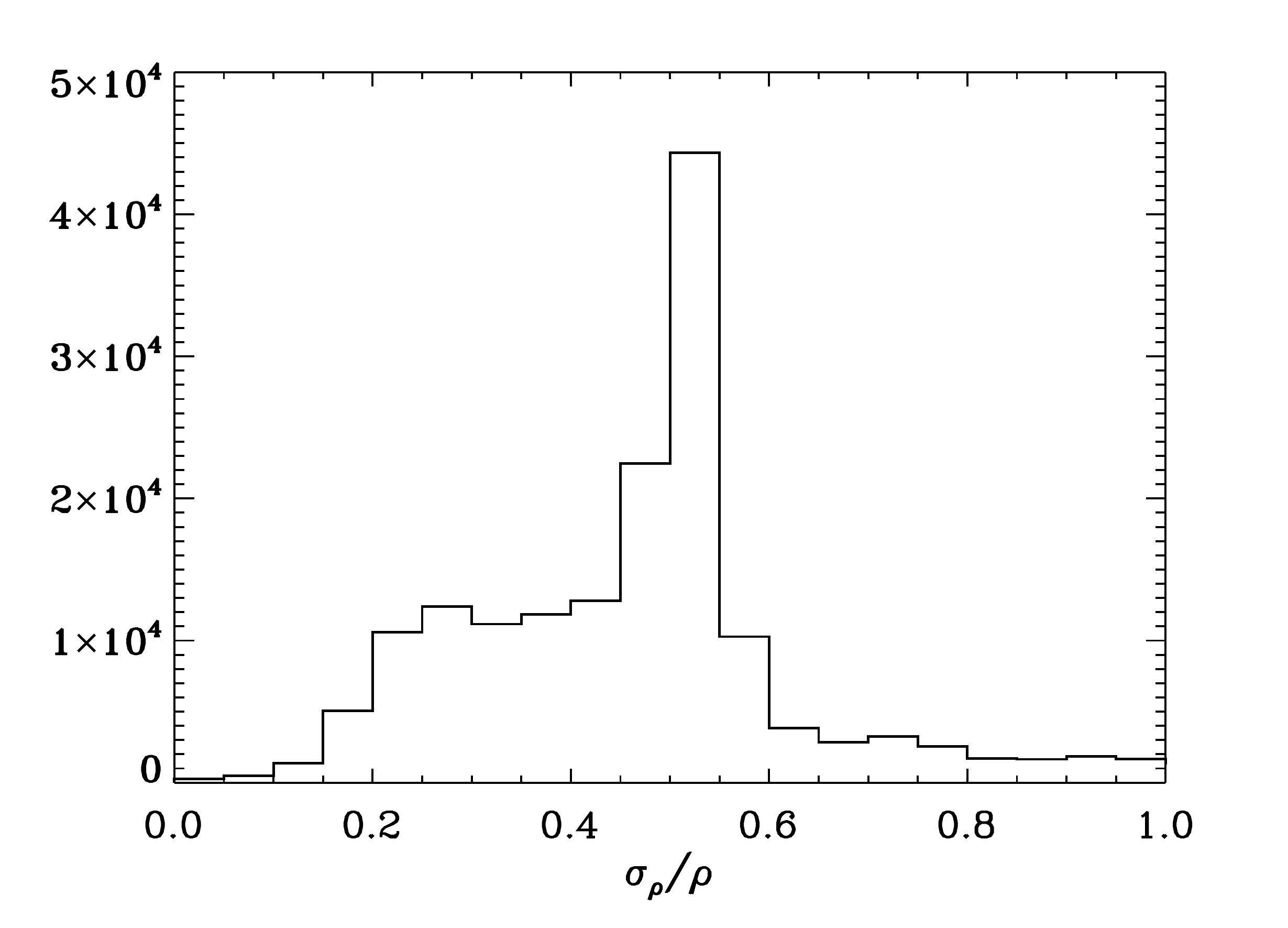}
\caption{Distribution of  absolute uncertainties in $T_{\rm eff}$, \logg, [Fe/H] and relative uncertainties in radius, mass and density for all stars in the sample.}
\label{histo_unc}
\end{center}
\end{figure}

Uncertainties for each reported stellar parameter are calculated from the 1-$\sigma$ interval around the best fit (Figure ~\ref{discrete_post}). Figure~\ref{histo_unc} shows the distribution of fractional uncertainties over all targets for various stellar parameters. We notice a bimodal distribution for surface gravity and radius, which is due to the two main provenances of the surface gravity input values from seismology and from the KIC with associated uncertainties of 0.03 dex and 0.40 dex, respectively. We observe a similar bi-modality with peaks at $\sim$80K and $\sim$\,150K for effective temperatures based on spectroscopic and photometric input values. While the bi-modality in radius and gravity was also present in the H14 catalog, the bi-modality in $T_{\rm eff}$ is new and reflects the increase in the number of stars that now have spectroscopic observations.

The typical reported uncertainties in the catalog are $\sim$\,27\% in radius, $\sim$\,17\% in mass and $\sim$\,51\% in density. We note that the uncertainties are on average smaller (e.g.\ $\sim$\,27\% versus $\sim$\,40\% in radius) compared to H14, which is mostly due to the volume weighting of each isochrone point which was not taken into account in the Q1-16 catalog. An additional factor for the reduced uncertainties are the increased number of stars with \logg\ input values derived from spectroscopy or {   Flicker}, which considerably increases the precision of the derived radii.

\subsection{Distances and Extinctions}

In addition to the stellar properties reported in H14 we also report estimates of distances and extinction in the $V$ band ($A_{V}$). For each model, we calculated a distance and extinction using observed apparent magnitudes, galactic coordinates of a given target, absolute magnitudes given by the model and the 3D extinction map by \citet{2005AJ....130..659A}. For apparent magnitudes we used $g$-band magnitudes when available, and 2MASS $J$-band otherwise. We adopted the extinction law from \citet{1989ApJ...345..245C} with $A_{J}/A_{V}=0.288$ and $A_{g}$/$A_{V}=1.234$ {\bf to convert between extinction values} in different passbands. The posteriors for distance and $A_{V}$ were then derived using the same methodology as for other parameters. We emphasize that the methodology described above assumes that the adopted reddening map is exact, which is unlikely to be the case. Hence, the uncertainties for the derived distances and extinction values are likely underestimated, and both may suffer from systematic errors compared to other extinction maps available in the literature (see also Section 5.2). 

Following the delivery of the {   DR25 stellar properties catalog on the NASA exoplanet archive,} we discovered a coding error which caused the extinction relations to be swapped, i.e. $A_{g}$/$A_{V}$ was applied to $J$-band measurements and $A_{J}$/$A_{V}$ was applied to $g$-band measurements, respectively. Since most distances were derived from $J$-band, this resulted in a systematic underestimation of reported distances by on average $\sim$\,20\% for typical solar-type stars, and up to $\sim$\,50\% for more distant red giant stars. Correspondingly, this also led to a systematic overestimation of $A_{V}$ values by up to $\sim$\,0.05\,mag. {   The online table was affected prior to 10 November 2016. After that date, the corrected distances and extinctions were updated and they} are reported in Table 4. {   Hence any distances and extinction values downloaded before this date should not be used. Similarly, the replicated posteriors (see Section 3.4) for these erroneous distances and extinction values downloaded prior to 15 December 2016 should not be used.}


\subsection{Stellar Replicated Posteriors}

While the discrete posteriors derived in Section 3.1 are valuable to inspect probability distributions and derive uncertainties for a given parameter, it is often desirable to use posterior samples to investigate parameter correlations and use posteriors in further analysis steps (e.g. transit fits). The classical tool to generate posterior samples is Markov Chain Monte Carlo (MCMC), which has previously been applied to stellar parameter inference using isochrones \citep{2015ApJ...809....7B,2015ApJ...804...64M,2015ascl.soft03010M}. Due to the significant computational effort involved in running MCMC on 190,000 stars, we chose an alternative approach to derive posterior samples by approximating discrete posterior distributions.

The method for approximating Discrete posteriors works as follows. The Discrete posteriors are based on a subset of $\sim$400,000 models from the grid of models used. Each model is a point on the isochrones and is described by a set of star parameters (i.e., $\teff$, [Fe/H], $\log g$, $M$, $R$, etc.). From the Discrete posterior, each individual model has some probability $x$. We scale the Discrete posterior by a factor $N_{\rm scale}$ so that the Discrete posterior values range from 0 to $N_{\rm scale}$. After a few tests $N_{\rm scale}$ was fixed to 50. We then draw a random model (from a uniformly random process) with a probability x from the Discrete posterior and replicate all its parameters $x \times N_{\rm scale}$ times. If $x \times N_{\rm scale}$<1, the model is not replicated. This process is repeated until the number of samples reaches the total number of samples desired, $N_{\rm sample}$, which for this delivery was fixed at 40,000. This value for $N_{\rm sample}$ was chosen as a compromise between achieving appropriate correlation lengths and keeping the file sizes to a reasonable value for each star. The posteriors obtained with this method are hereafter called ``Replicated Posteriors''. Importantly, Replicated posteriors conserve correlations between the parameters (similar to MCMC) because each set is drawn so as to correspond to a self-consistent model.

\begin{figure}
\begin{center}
\includegraphics[width=8.5cm]{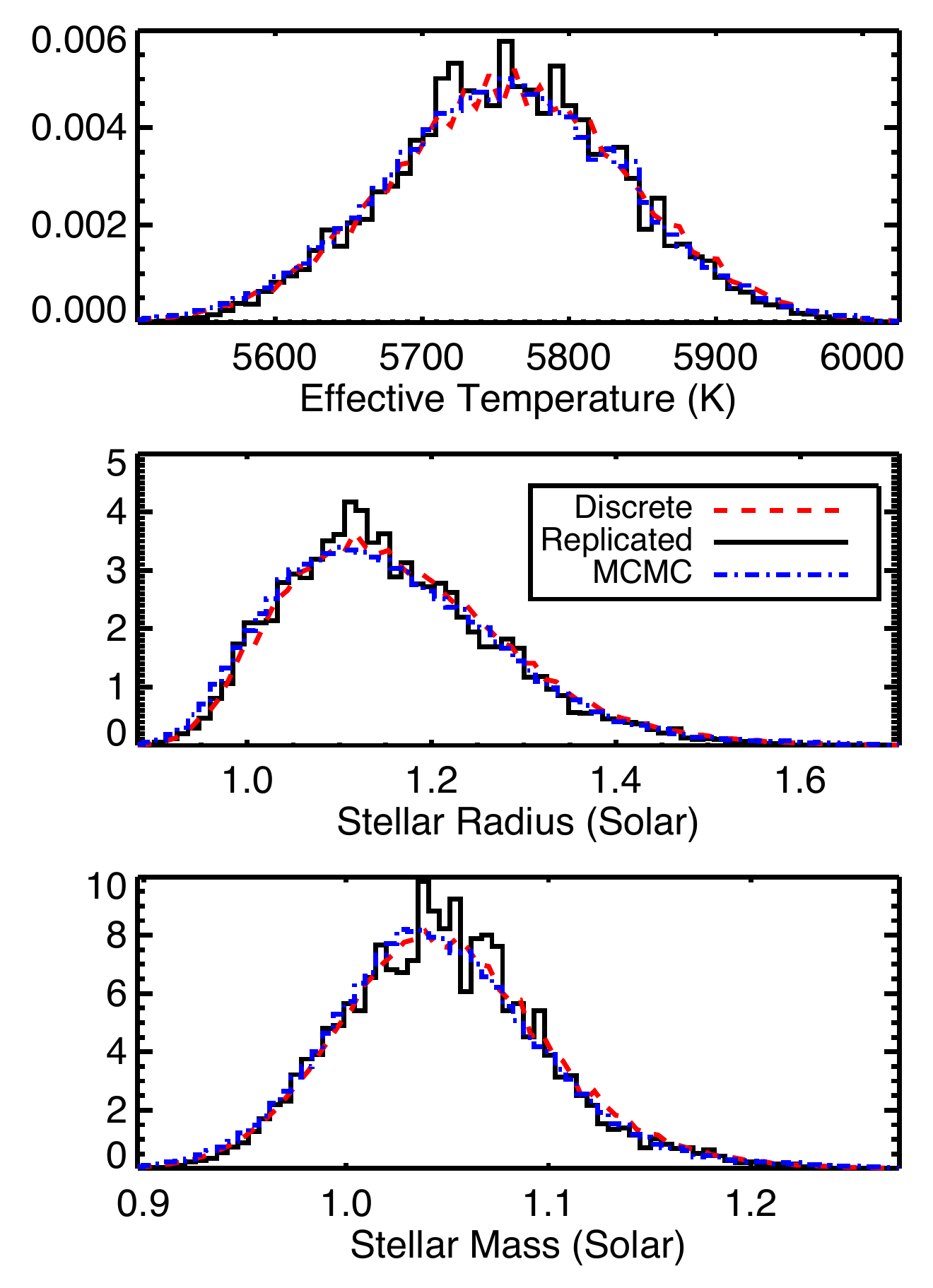}
\caption{Comparison of replicated posteriors (black solid lines), discrete posteriors (red dashed lines) and MCMC posteriors (blue dot-dashed lines) for the temperature, radius and mass of Kepler-452. The MCMC posteriors were taken from \citet{2015AJ....150...56J}.}
\label{kep452}
\end{center}
\end{figure}

To test the validity of the method, Figure ~\ref{kep452} compares the replicated posteriors for Kepler-452 (black solid line) to the discrete posterior (red dashed line) and posteriors derived from a full MCMC analysis by \citet{2015AJ....150...56J}. All three distributions agree well, demonstrating that the replicated posteriors provide a good approximation to MCMC methods (but with a factor of $\sim$\,10 faster computation time).

\begin{figure*}
\begin{center}
\includegraphics[width=16cm]{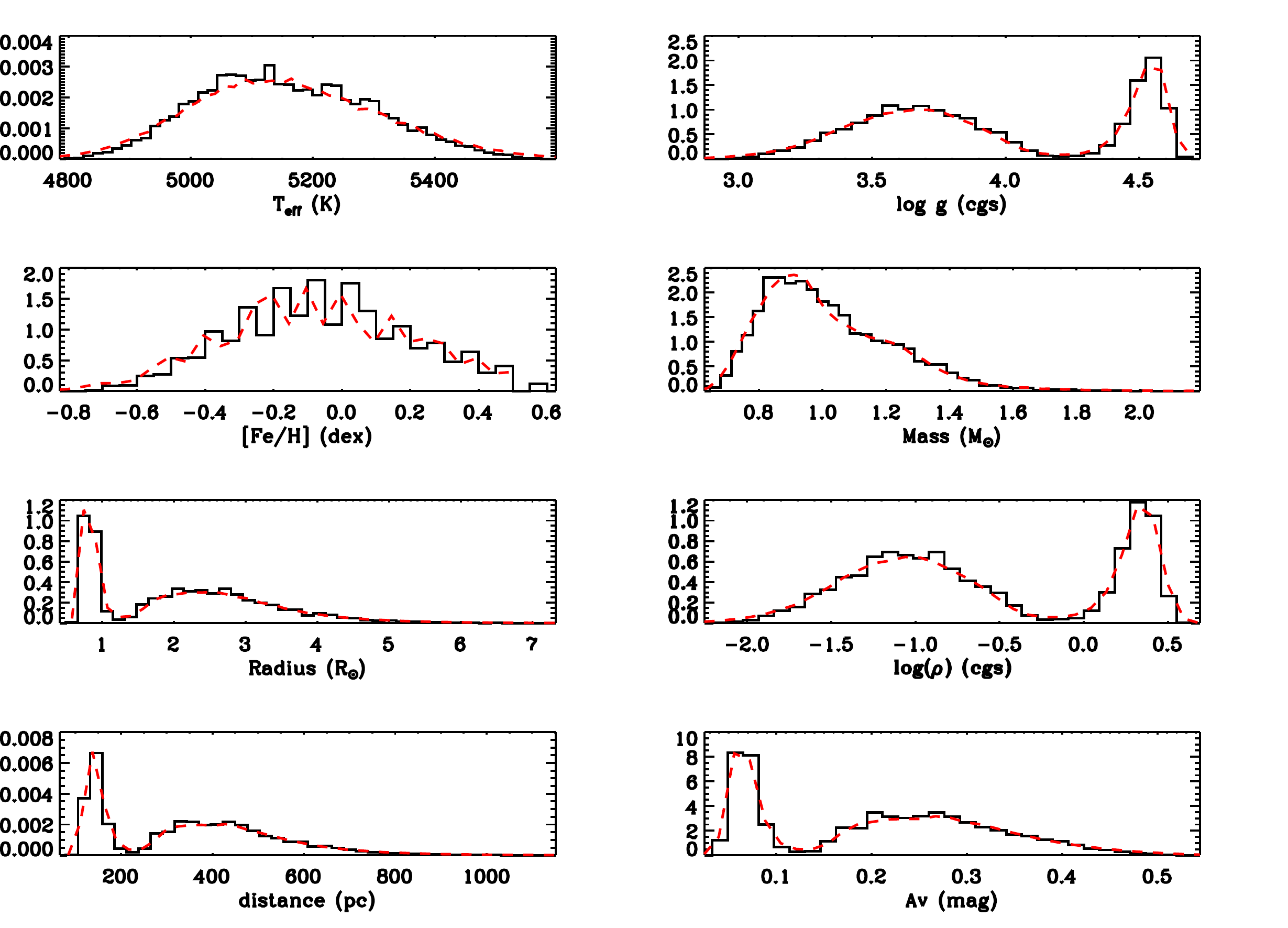}
\caption{Comparison of replicated posteriors (black solid lines) and discrete posteriors (red dashed line) for a typical solar-type star in the {\it Kepler} sample, KIC 757076 ($Kp=11.7$). The input values for $T_{\rm eff}$, $\log g$, and [Fe/H] are PHO1, KIC, and KIC respectively. }
\label{replicated}
\end{center}
\end{figure*}

Figure~\ref{replicated} shows an example comparison between Discrete posteriors and Replicated posteriors for a typical solar-type dwarf in the {\it Kepler} sample with a photometric input $\logg$. The Replicated posteriors again show good agreement with the Discrete posteriors, even in the case of bimodal distributions. We checked the results for different spectral types, which looked similar to this example.


\section{Final Catalog Description}

Applying the methodology described above to all stars in Table \ref{input} yielded best-fit values and 1-$\sigma$ confidence intervals for mass, radius, surface gravity, effective temperature, density, metallicity, distance and extinction for all 197,096 stars, which are listed in Table~\ref{tab4}. Each entry also gives the origin of the input values used for $T_{\rm eff}$, $\log g$, and [Fe/H] as described in Section 2.2. Finally, for each star we give the provenance for the parameters derived. While most of the stars have their output parameters derived from the isochrone fitting method of Section 3.1 (abbreviation DSEP), there are 235 stars where we used previously published values for cool dwarfs and stars falling off the isochrone grid (see Section 5.3 for more details). For this small sample of stars, distances and extinctions are not given and their provenance is BTSL since the parameters were estimated from polynomial fits to low-mass BT-Settl models \citep{2012RSPTA.370.2765A}. The abbreviation MULT corresponds to parameters derived from multiple evolutionary tracks and is given to a handful of stars. Using these flags, the reader can trace the reference(s) of the input values and the method used to derive the stellar parameters. We note that unlike previous catalog deliveries we did not override published solutions that provide better estimates for radii and masses (e.g. from asteroseismology) in order to homogeneously derive stellar properties (including distances) for all stars. This means that for some stars better estimates for radii and masses may be available in the literature.

The full catalog is available at the NASA exoplanet archive\footnote{{\url http://exoplanetarchive.ipac.caltech.edu/}} along with the Replicated posteriors obtained as described in Section 3.2. Note that the online catalog contains 200,038 entries. The difference between the number of stars with derived parameters (197,096) and the total number of {\it Kepler} targets during the whole mission comes from the fact that there are still 2,942 stars that are unclassified without any $T_{\rm eff}$, $\log g$, and [Fe/H] available. Among these stars 139 only had Q0 observations, 528 stars were only observed in Q17, 8 are flagged as a galaxy, and 516 stars do not have valid 2MASS photometry. We also note that 1800 of these unclassified stars are faint ($Kp > 16$).


\begin{table*}
\begin{center}
\caption{Output values of the DR25 stellar properties catalog with the updated distances and extinctions.}
\begin{tabular}{cccccccccc}
\hline
\hline
KIC & $T_{\rm eff}$ & $\log g$ & [Fe/H] & R & M & $\rho$ & d (kpc) & $A_V$ &$P_{\rm M, R, \rho}$  \\
\hline
757076 &  5160$^{+ 171}_{- 156}$& 3.580\,$\pm$\,0.232&-0.100$^{+0.300}_{-0.300}$
&   3.13$^{+  0.99}_{-  2.30}$ &   1.36$^{+ 0.20}_{- 0.48}$ &  
  0.06$^{+  1.81}_{-  0.04}$ &   0.52$^{+ 0.13}_{- 0.30}$& 
  0.32$^{+  0.08}_{-  0.24}$ &  DSEP \\
757099 &  5519$^{+ 182}_{- 149}$& 3.822\,$\pm$\,0.213&-0.220$^{+0.350}_{-0.250}$
&   2.11$^{+  0.67}_{-  1.25}$ &   1.08$^{+ 0.17}_{- 0.23}$ &  
  0.16$^{+  1.70}_{-  0.08}$ &   0.75$^{+ 0.17}_{- 0.34}$& 
  0.43$^{+  0.06}_{-  0.24}$ &  DSEP \\
757137 &  4706$^{+  74}_{- 102}$& 2.374\,$\pm$\,0.027&-0.100$^{+0.200}_{-0.300}$
&  15.45$^{+  3.54}_{-  3.93}$ &   2.06$^{+ 1.16}_{- 0.95}$ &  
  0.00$^{+  0.00}_{-  0.00}$ &   0.66$^{+ 0.13}_{- 0.14}$& 
  0.39$^{+  0.06}_{-  0.10}$ &  DSEP \\
757280 &  6543$^{+ 162}_{- 194}$& 4.082\,$\pm$\,0.172&-0.240$^{+0.250}_{-0.300}$
&   1.64$^{+  0.48}_{-  0.48}$ &   1.18$^{+ 0.21}_{- 0.16}$ &  
  0.38$^{+  0.57}_{-  0.17}$ &   0.49$^{+ 0.11}_{- 0.12}$& 
  0.30$^{+  0.08}_{-  0.09}$ &  DSEP \\
757450 &  5332$^{+ 106}_{-  96}$& 4.500\,$\pm$\,0.036&-0.080$^{+0.150}_{-0.150}$
&   0.84$^{+  0.05}_{-  0.05}$ &   0.82$^{+ 0.06}_{- 0.04}$ &  
  1.93$^{+  0.35}_{-  0.24}$ &   0.73$^{+ 0.04}_{- 0.04}$& 
  0.43$^{+  0.02}_{-  0.02}$ &  DSEP \\
891901 &  6323$^{+ 158}_{- 205}$& 4.418\,$\pm$\,0.232&-0.080$^{+0.250}_{-0.300}$
&   1.09$^{+  0.38}_{-  0.13}$ &   1.14$^{+ 0.16}_{- 0.15}$ &  
  1.23$^{+  0.37}_{-  0.67}$ &   0.59$^{+ 0.15}_{- 0.06}$& 
  0.36$^{+  0.09}_{-  0.04}$ &  DSEP \\
891916 &  5602$^{+ 167}_{- 151}$& 4.587\,$\pm$\,0.119&-0.580$^{+0.300}_{-0.300}$
&   0.74$^{+  0.14}_{-  0.07}$ &   0.77$^{+ 0.09}_{- 0.06}$ &  
  2.68$^{+  0.57}_{-  0.91}$ &   0.63$^{+ 0.10}_{- 0.05}$& 
  0.38$^{+  0.06}_{-  0.03}$ &  DSEP \\
892010 &  4729$^{+  70}_{- 182}$& 2.168\,$\pm$\,0.030& 0.070$^{+0.250}_{-0.450}$
&  26.09$^{+  0.51}_{-  9.62}$ &   3.65$^{+ 0.07}_{- 2.27}$ &  
  0.00$^{+  0.00}_{-  0.00}$ &   3.20$^{+ 0.05}_{- 0.97}$& 
  0.58$^{+  0.00}_{-  0.00}$ &  DSEP \\
892107 &  5080$^{+ 138}_{- 138}$& 3.354\,$\pm$\,0.248&-0.080$^{+0.250}_{-0.300}$
&   4.29$^{+  1.30}_{-  1.79}$ &   1.52$^{+ 0.23}_{- 0.54}$ &  
  0.03$^{+  0.09}_{-  0.01}$ &   0.90$^{+ 0.19}_{- 0.29}$& 
  0.48$^{+  0.05}_{-  0.16}$ &  DSEP \\
892195 &  5522$^{+ 194}_{- 155}$& 3.984\,$\pm$\,0.170&-0.060$^{+0.300}_{-0.250}$
&   1.67$^{+  0.50}_{-  0.75}$ &   0.98$^{+ 0.11}_{- 0.12}$ &  
  0.30$^{+  1.45}_{-  0.13}$ &   0.81$^{+ 0.16}_{- 0.30}$& 
  0.45$^{+  0.06}_{-  0.18}$ &  DSEP \\
892203 &  5947$^{+ 193}_{- 193}$& 4.080\,$\pm$\,0.147&-0.120$^{+0.300}_{-0.300}$
&   1.54$^{+  0.42}_{-  0.56}$ &   1.03$^{+ 0.16}_{- 0.15}$ &  
  0.40$^{+  0.97}_{-  0.16}$ &   0.81$^{+ 0.15}_{- 0.22}$& 
  0.45$^{+  0.05}_{-  0.13}$ &  DSEP \\
892376 &  3973$^{+ 124}_{- 152}$& 4.656\,$\pm$\,0.022& 0.140$^{+0.250}_{-0.300}$
&   0.60$^{+  0.03}_{-  0.07}$ &   0.60$^{+ 0.04}_{- 0.07}$ &  
  3.85$^{+  1.14}_{-  0.41}$ &   0.14$^{+ 0.01}_{- 0.02}$& 
  0.06$^{+  0.01}_{-  0.02}$ &  DSEP \\
892667 &  6609$^{+ 159}_{- 227}$& 4.105\,$\pm$\,0.164&-0.260$^{+0.250}_{-0.300}$
&   1.65$^{+  0.48}_{-  0.52}$ &   1.28$^{+ 0.17}_{- 0.24}$ &  
  0.40$^{+  0.59}_{-  0.18}$ &   0.86$^{+ 0.18}_{- 0.19}$& 
  0.47$^{+  0.05}_{-  0.10}$ &  DSEP \\
892675 &  6316$^{+ 181}_{- 227}$& 4.038\,$\pm$\,0.144&-0.240$^{+0.250}_{-0.300}$
&   1.69$^{+  0.46}_{-  0.56}$ &   1.13$^{+ 0.19}_{- 0.19}$ &  
  0.33$^{+  0.81}_{-  0.14}$ &   0.92$^{+ 0.17}_{- 0.25}$& 
  0.49$^{+  0.05}_{-  0.12}$ &  DSEP \\
892678 &  6137$^{+ 167}_{- 186}$& 3.936\,$\pm$\,0.143&-0.260$^{+0.300}_{-0.300}$
&   1.89$^{+  0.44}_{-  0.82}$ &   1.12$^{+ 0.17}_{- 0.20}$ &  
  0.23$^{+  0.93}_{-  0.10}$ &   0.55$^{+ 0.10}_{- 0.19}$& 
  0.34$^{+  0.06}_{-  0.15}$ &  DSEP \\
  ...\\
\hline
\end{tabular}
\end{center}
\label{tab4}
\end{table*}%




\section{Discussion}

\subsection{Quality Control Tests}

\subsubsection{Comparison of Input and Output Values}




The first quality control test was to compare the input and output values for a well characterized sample of stars that have asteroseismic gravities or spectroscopic effective temperatures. Large deviations between input and output \teff\ or \logg\ values may indicate potential misclassifications due to problems with the adopted input values or the isochrone fitting methodology. 

\begin{figure}
\begin{center}
\includegraphics[width=9cm]{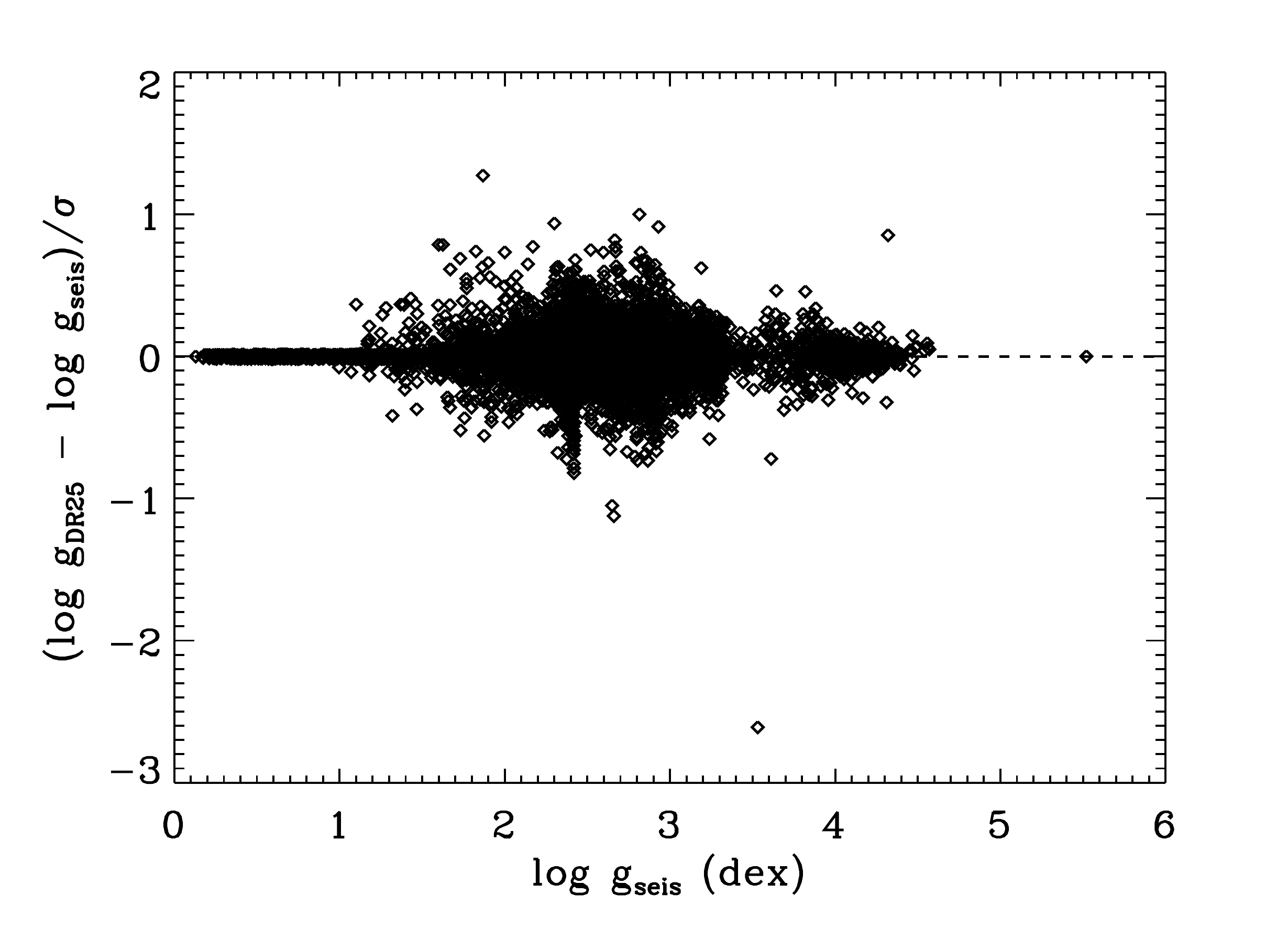}
\includegraphics[width=9cm]{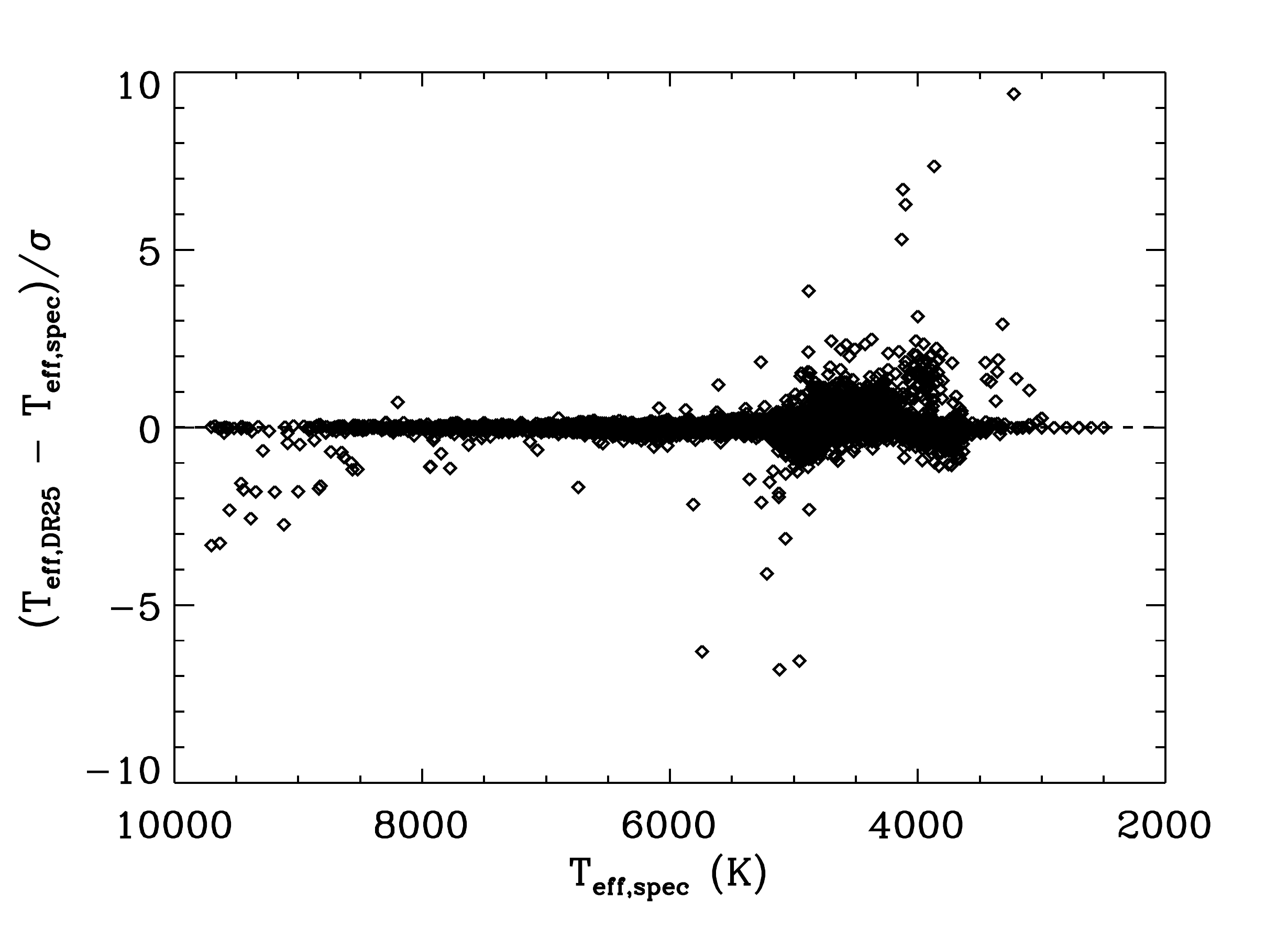}
\caption{Top panel: differences between input and output $\log g$ values in units of $\sigma$ for stars with asteroseismic input values for $\log g$. The adopted typical uncertainty for asteroseismic $\log g$ values is 0.03\,dex. Bottom panel: Same as top panel but for stars with spectroscopic $T_{\rm eff}$. The adopted uncertainty is 2\%.
}
\label{complogg}
\end{center}
\end{figure}

Asteroseismic gravities are available for 16,947 stars (red giants and dwarfs) while spectroscopic temperatures were obtained for 14,813 stars. Figure~\ref{complogg} shows the difference between the input values of $\logg$ and $T_{\rm eff}$ and the DR25 values for the subsample of aforementioned stars. We see that for $\log g$ most of the output values agree with the seismic values within 1$\sigma$ and 5 stars disagree by more than 1$\sigma$. The largest disagreement concerns stars with $\log g$ values between 2 and 3, i.e. red giants including red clump stars. This can be explained by the fact that the DSEP models do not include helium-burning red giant models, as pointed out by H14.
 
The  effective temperature comparison (bottom panel of Figure~\ref{complogg}) shows that in most cases the values provided in the catalog agree with the spectroscopic input values within 1$\sigma$. A large number of stars with $T_{\rm eff}$ between 3500K and 5500K disagree by more than 1$\sigma$. These stars are again mostly red giants. {   We note that 11} stars disagree by more than 5\,$\sigma$. Among them, three stars (KIC 8714886, 10536147, and 10797526) have $T_{\rm eff} >$\,15,000\,K, well beyond our grid of models {   (and out of the plot)}, so we reported their DR24 stellar parameters for which the effective temperatures are close to 16,000\,K. The remaining nine stars have input values that are slightly off the model grid, thus the code converges to the parameter space that is significantly different than the input values. Three of these stars (KIC 2585447, 3968716, and 8559125) are new red giants with seismic $\log g$ and spectroscopic $T_{\rm eff}$. The first two stars are flagged in \citet{2016ApJ...827...50M} as a possible blend. This means that either the oscillation detection comes from another close-by star or that the blend has an impact on the estimate of the effective temperature in the spectroscopic analysis. KIC 8559125 is not a misclassified red giant anymore as it was removed from the list after the delivery of the DR25 catalog as explained in Section 5.3. The last five stars (KIC 3335176, 3346584, 4078024, 4263398, 8710336) have seismic and/or spectroscopic input values but are slightly off the grid, which explains the large difference between the input and the output values.
 
\subsubsection{Comparison to Previous Catalogs}

\begin{figure*}
\begin{center}
\includegraphics[width=17cm]{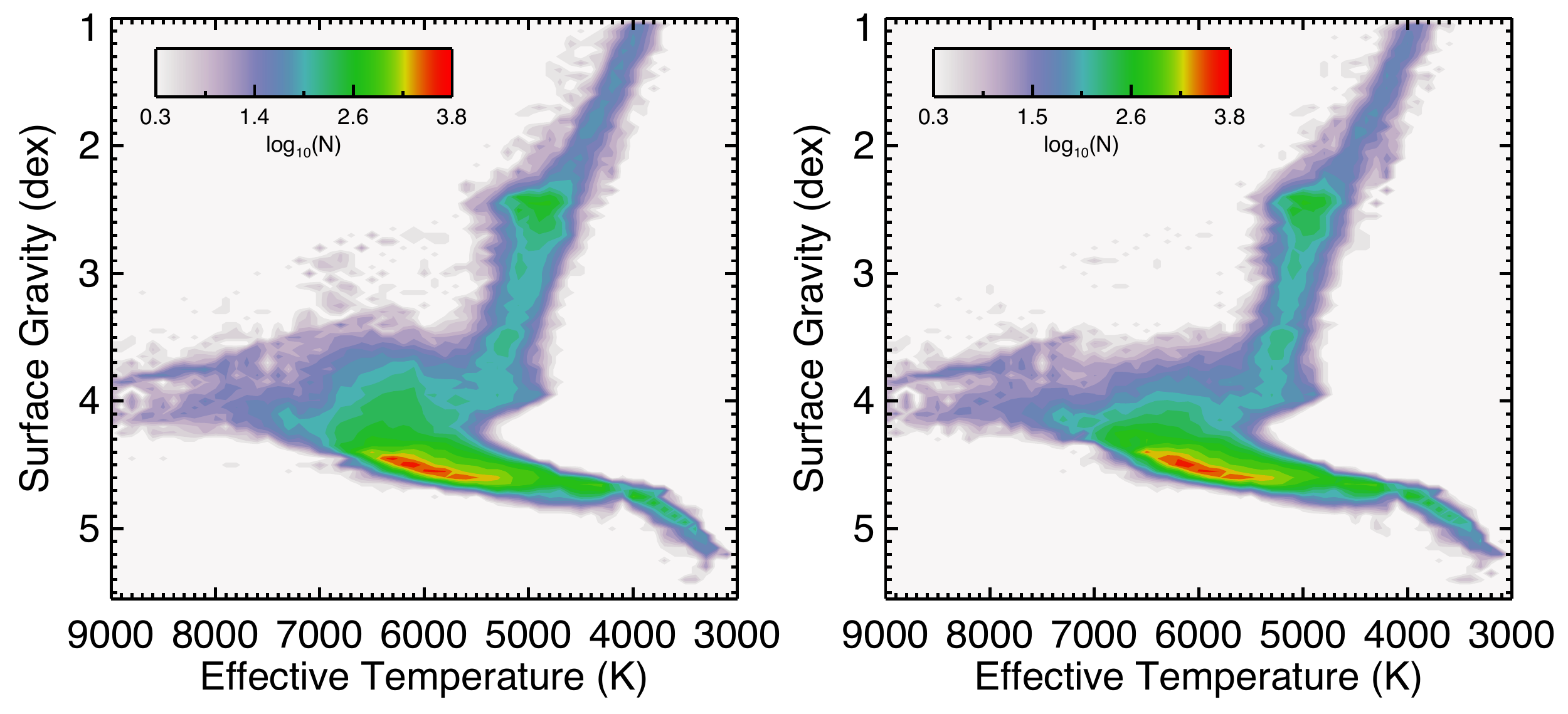}
\caption{Surface gravity versus effective temperature for all classified stars in this catalog (DR25, left panel) and the previous catalog (DR24, right panel). Color denotes the logarithmic number density of stars.}
\label{Fig3}
\end{center}
\end{figure*}

Figure~\ref{Fig3} shows the surface gravity versus temperature distribution for DR25 (left panel) and DR24 (right panel). It is evident that the DR25 catalog contains a significantly larger fraction of subgiants, mostly due to the inclusion of the LAMOST and Flicker surface gravities. Using the equations (8) and (9) from \citet{2016ApJS..224....2H}, we computed the number of subgiants and found that DR25 contains 15,893 subgiants compared to 11,078 in DR24, a 43.5\% increase. 
While these updates generally only affect the brighter {\it Kepler} targets ($Kp \lesssim 13$), this indicates that the DR25 catalog should be less prone to the systematic underestimation of radii for solar-type dwarfs than previous catalogs.

\begin{figure*}
\begin{center}
\includegraphics[width=5.9cm, trim=2cm 0.5cm 2cm 0]{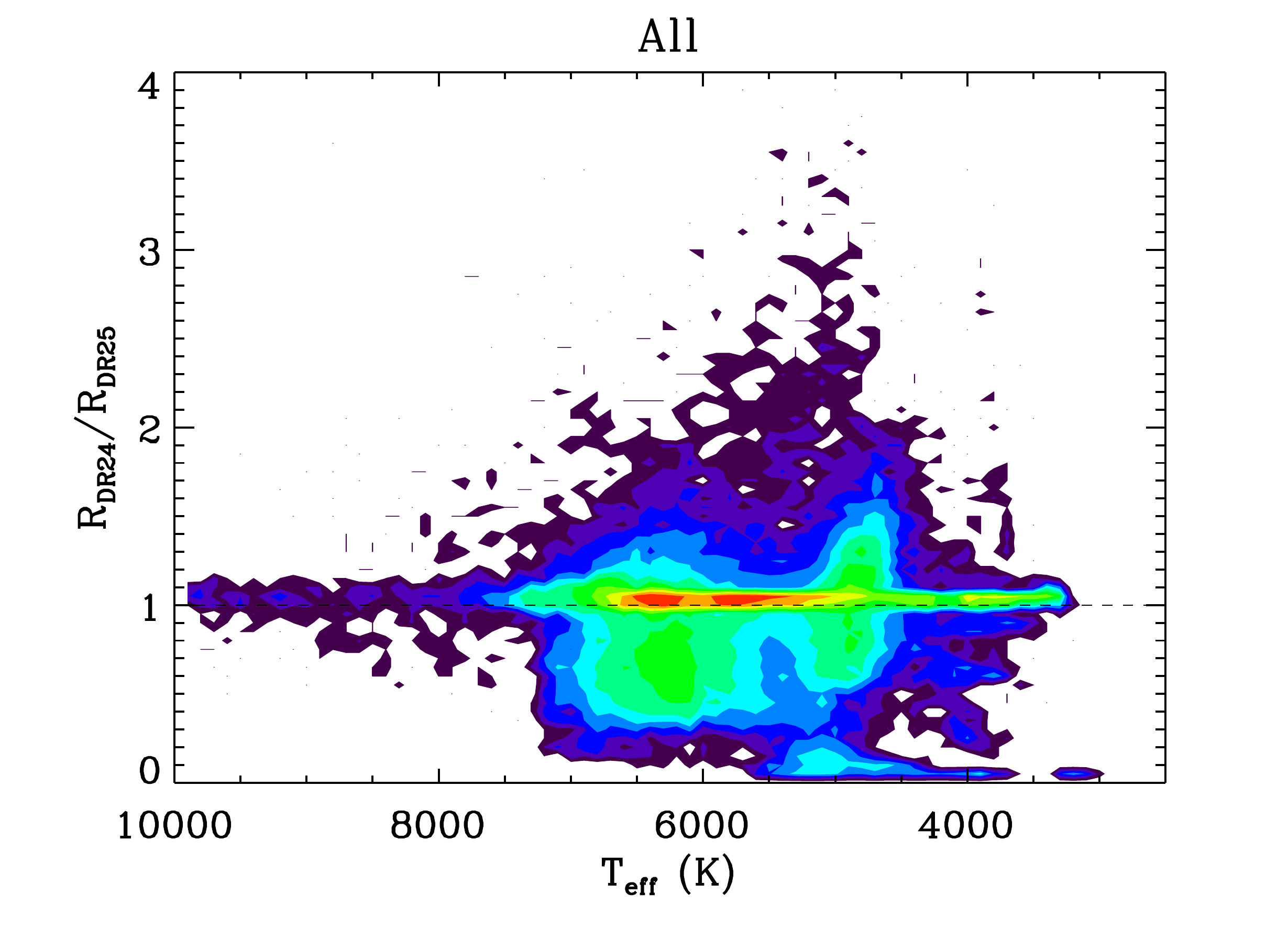}
\includegraphics[width=5.9cm, trim=2cm 0.5cm 2cm 0]{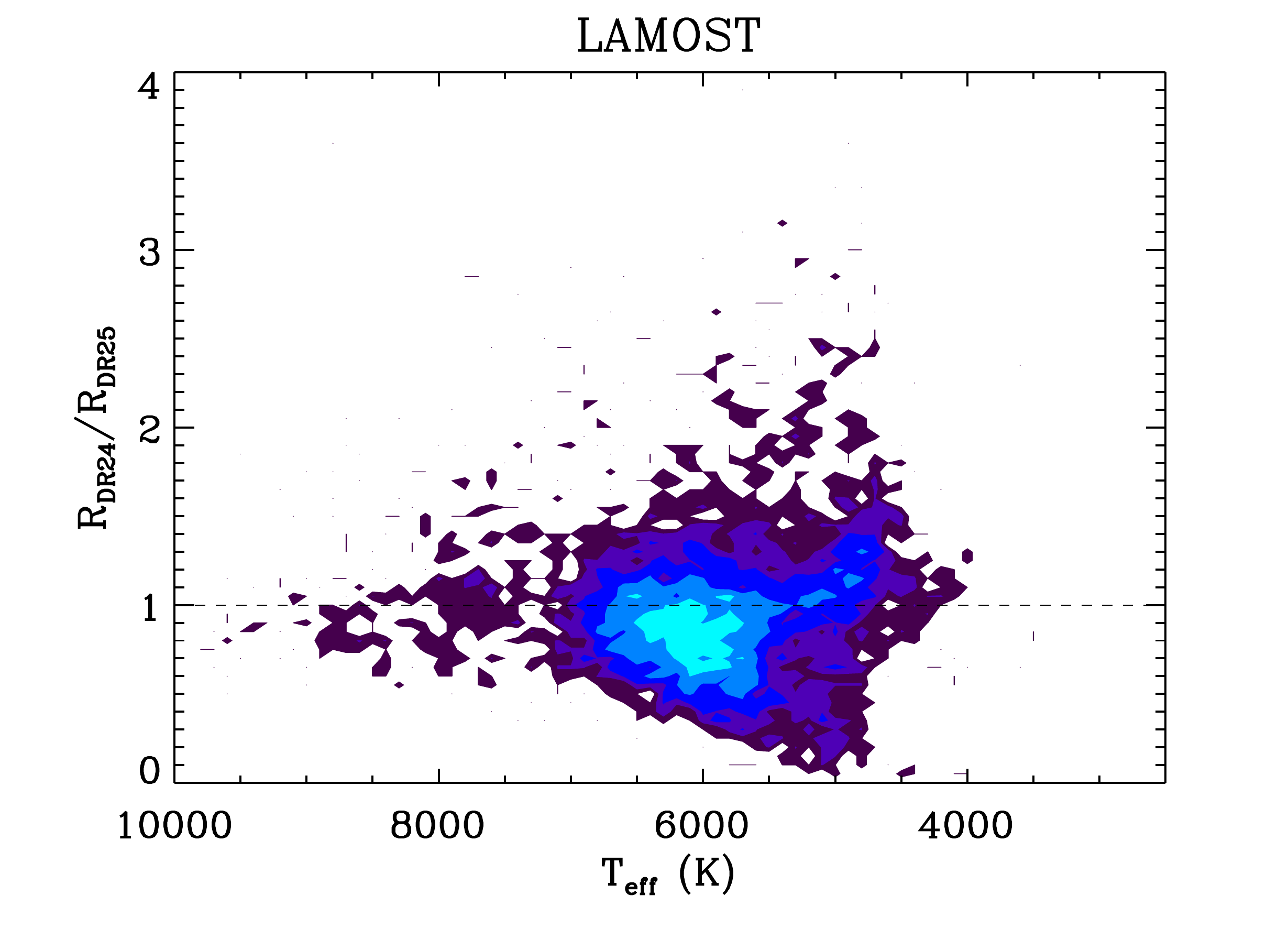}
\includegraphics[width=5.9cm, trim=2cm 0.5cm 2cm 0]{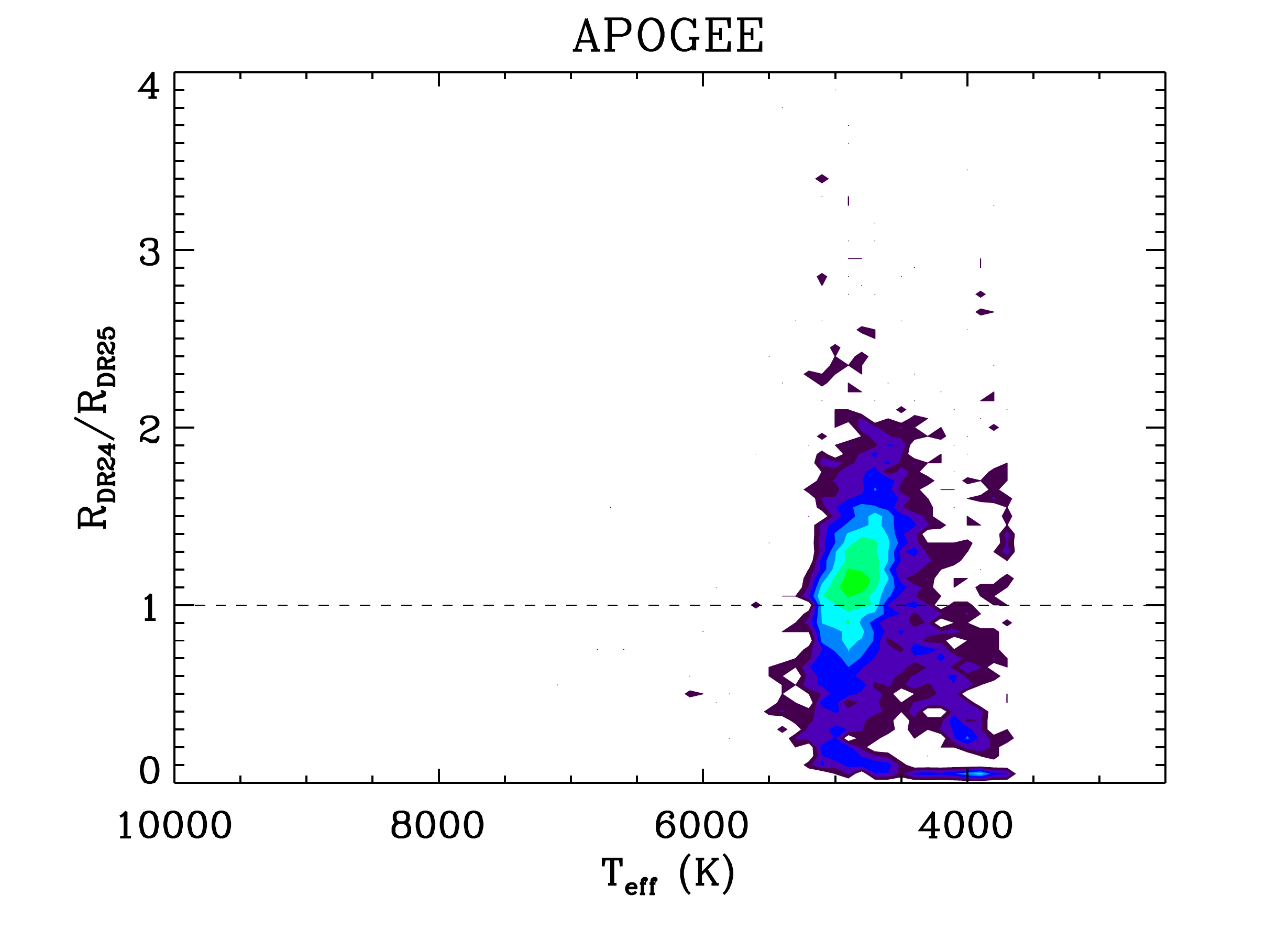}
\includegraphics[width=5.9cm, trim=2cm 0.5cm 2cm 0]{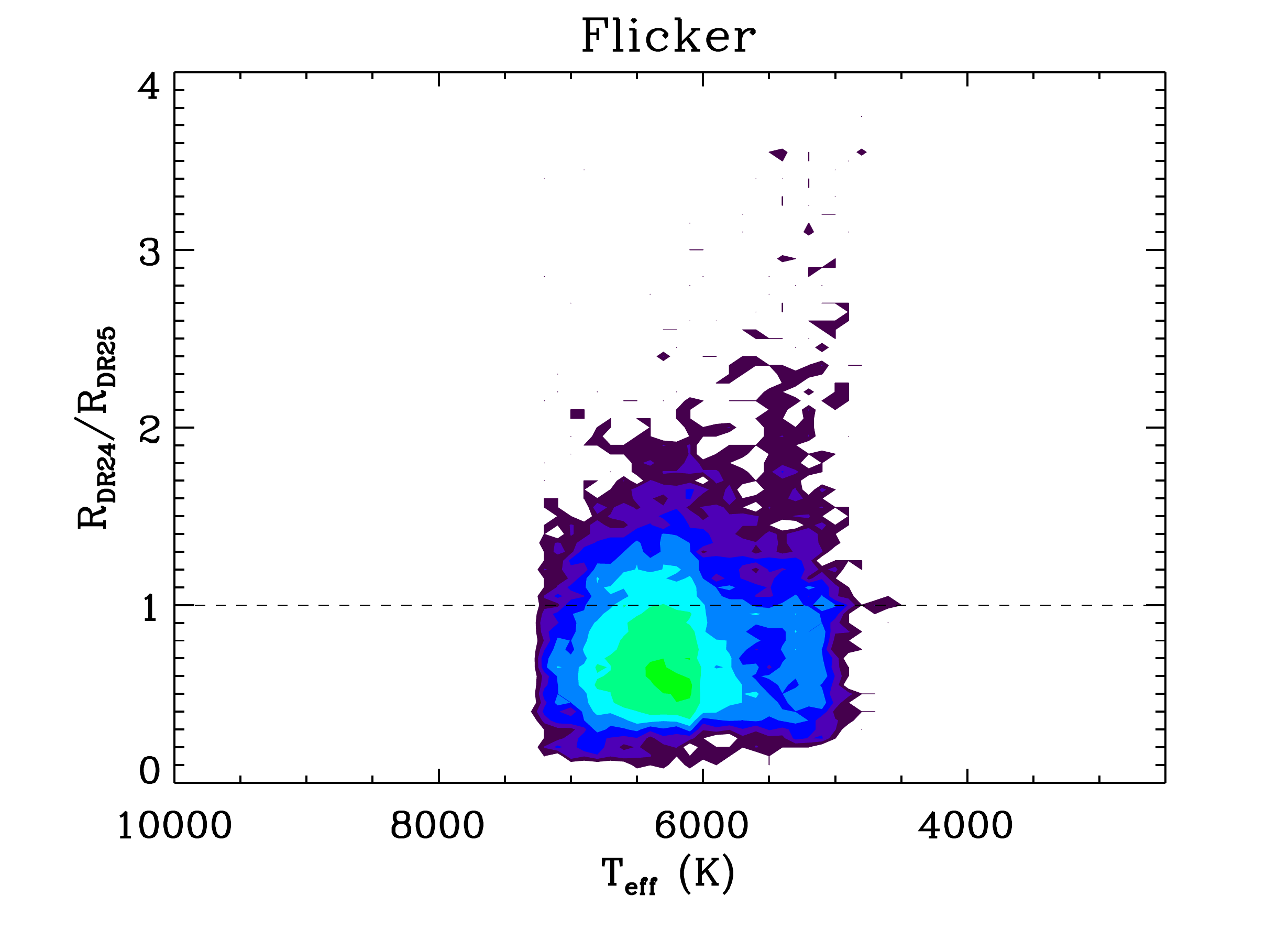}
\includegraphics[width=5.9cm, trim=2cm 0.5cm 2cm 0]{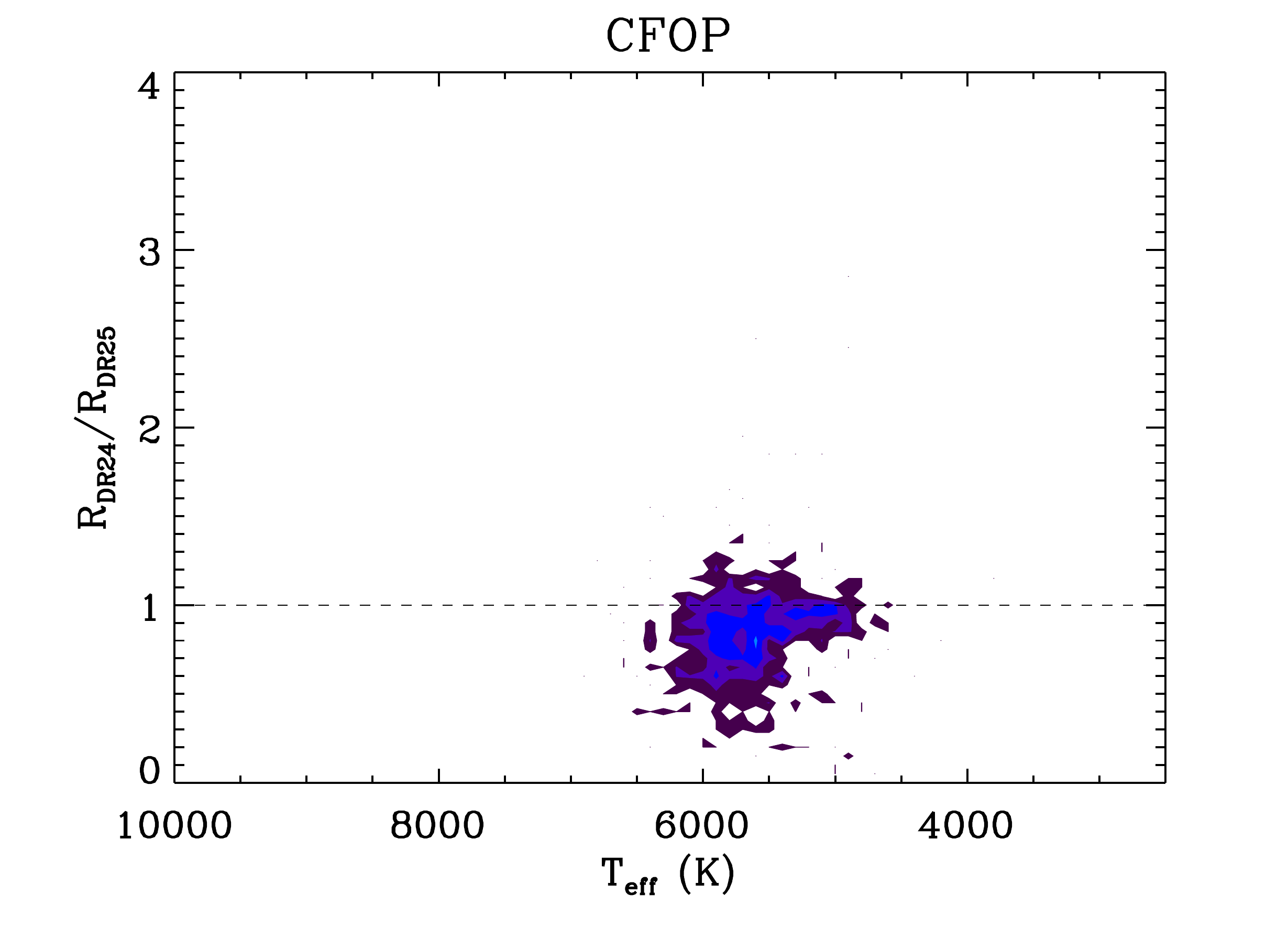}
\includegraphics[width=5.9cm, trim=2cm 0.5cm 2cm 0]{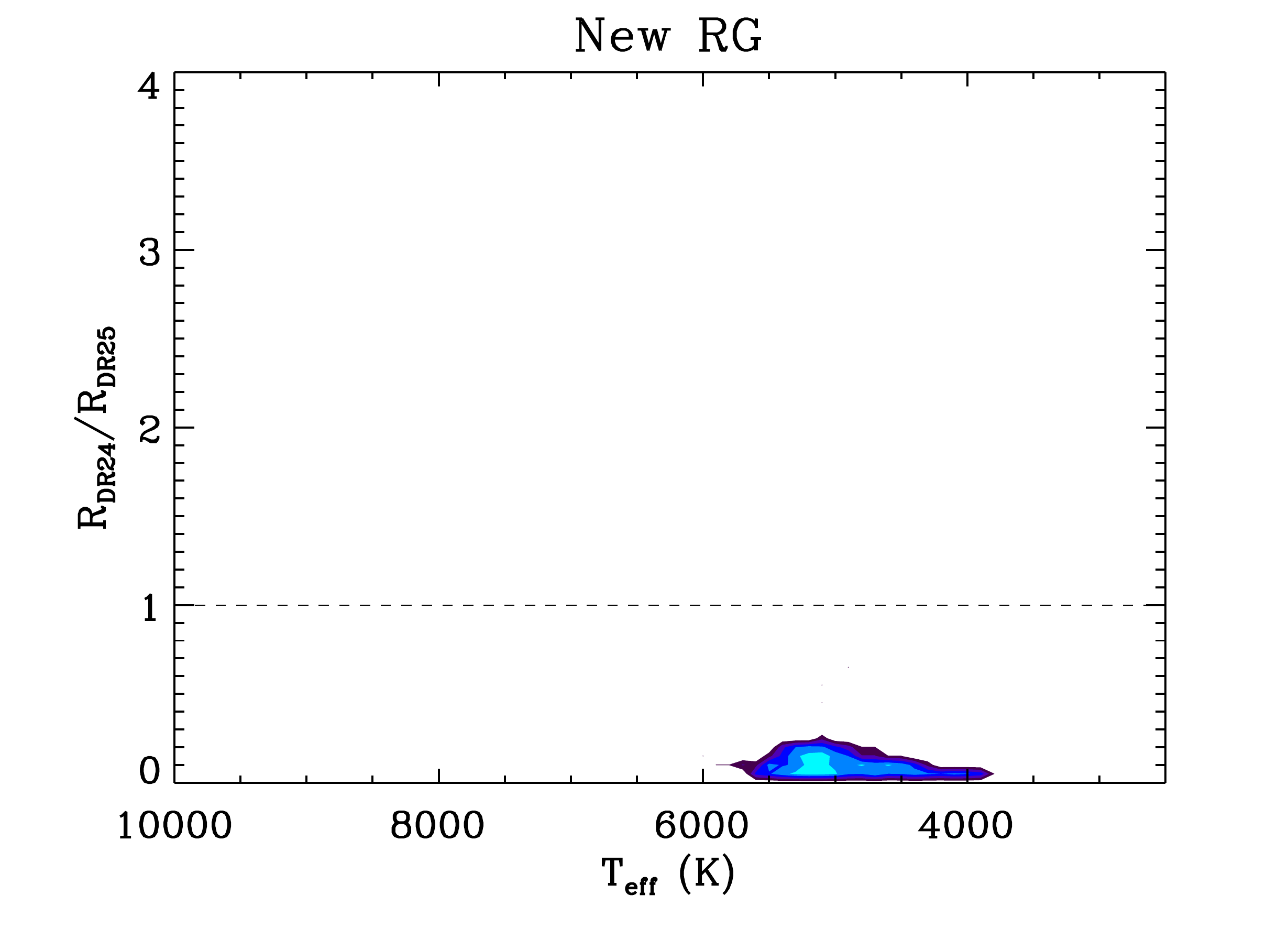}
\caption{Ratio of radii from DR24 and DR25 for full sample (top left panel), the LAMOST sample (top middle panel), the APOGEE sample (top right panel), the sample with Flicker \logg (bottom left panel), the sample of stars with CFOP spectroscopy (bottom middle panel), and the sample of new red giants (bottom right panel). Color denotes the number density of stars.}
\label{compR}
\end{center}
\end{figure*}

\begin{figure*}[htbp]
\begin{center}
\includegraphics[width=5.9cm, trim=2cm 0.5cm 2cm 0]{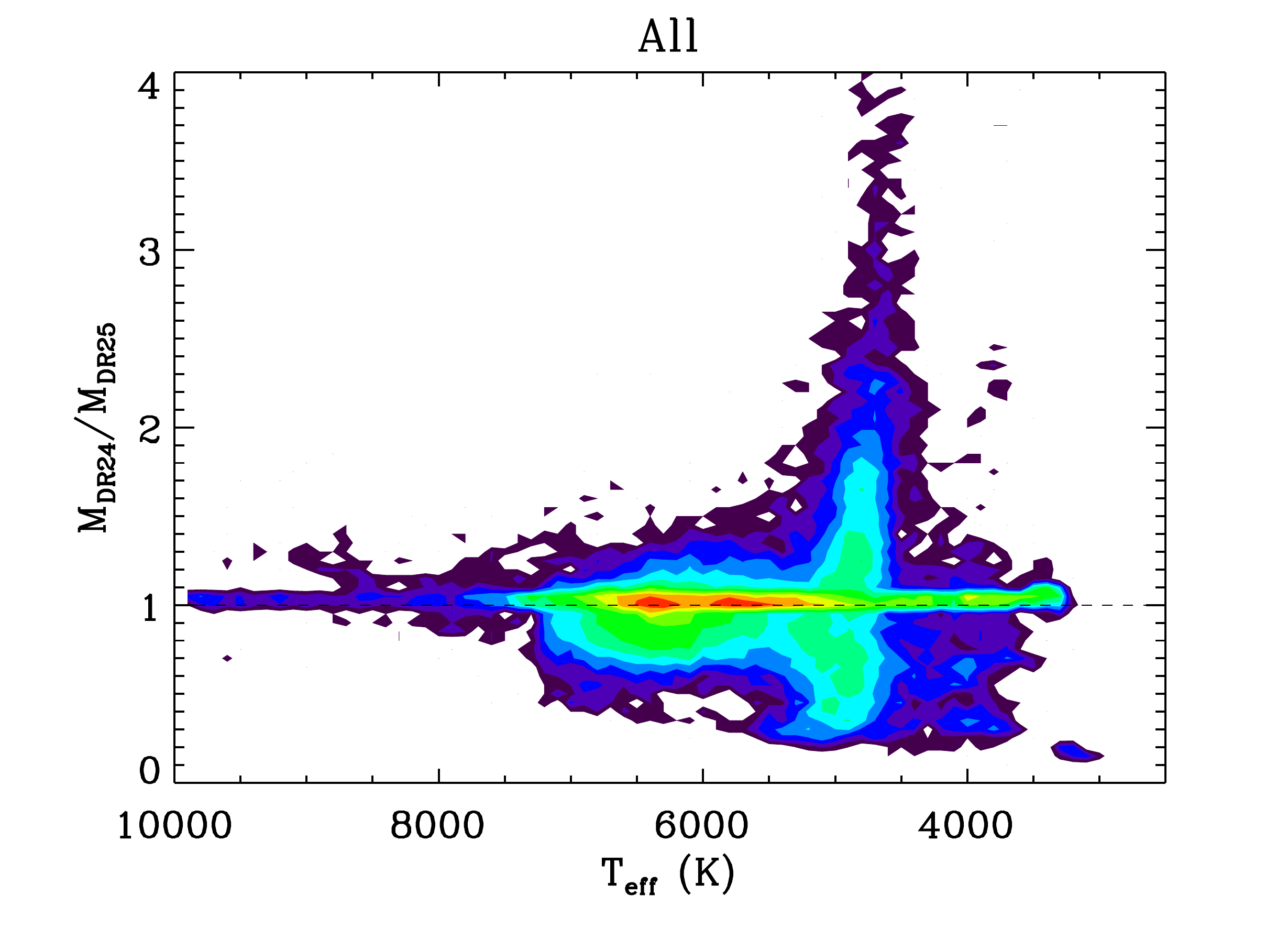}
\includegraphics[width=5.9cm, trim=2cm 0.5cm 2cm 0]{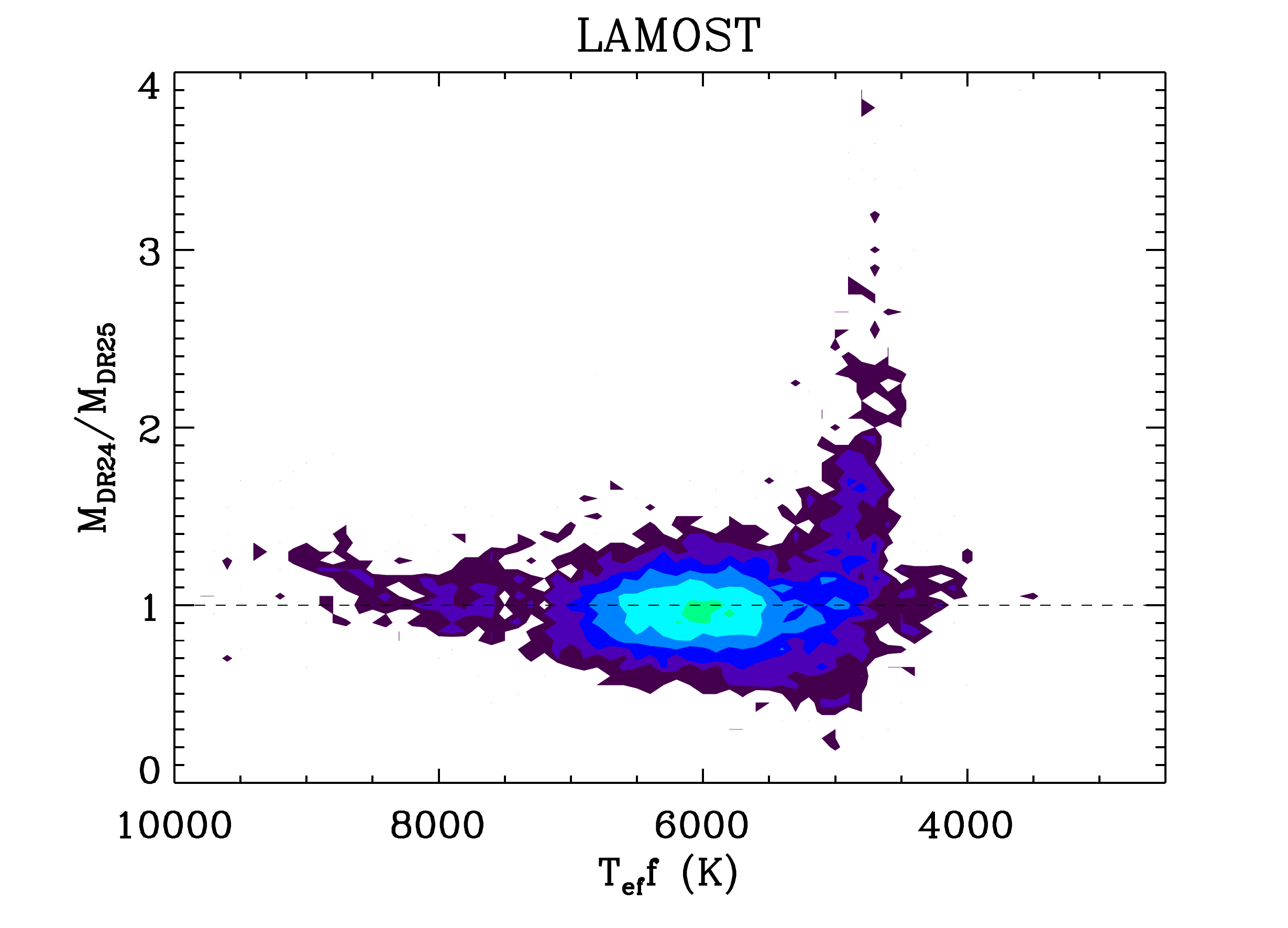}
\includegraphics[width=5.9cm, trim=2cm 0.5cm 2cm 0]{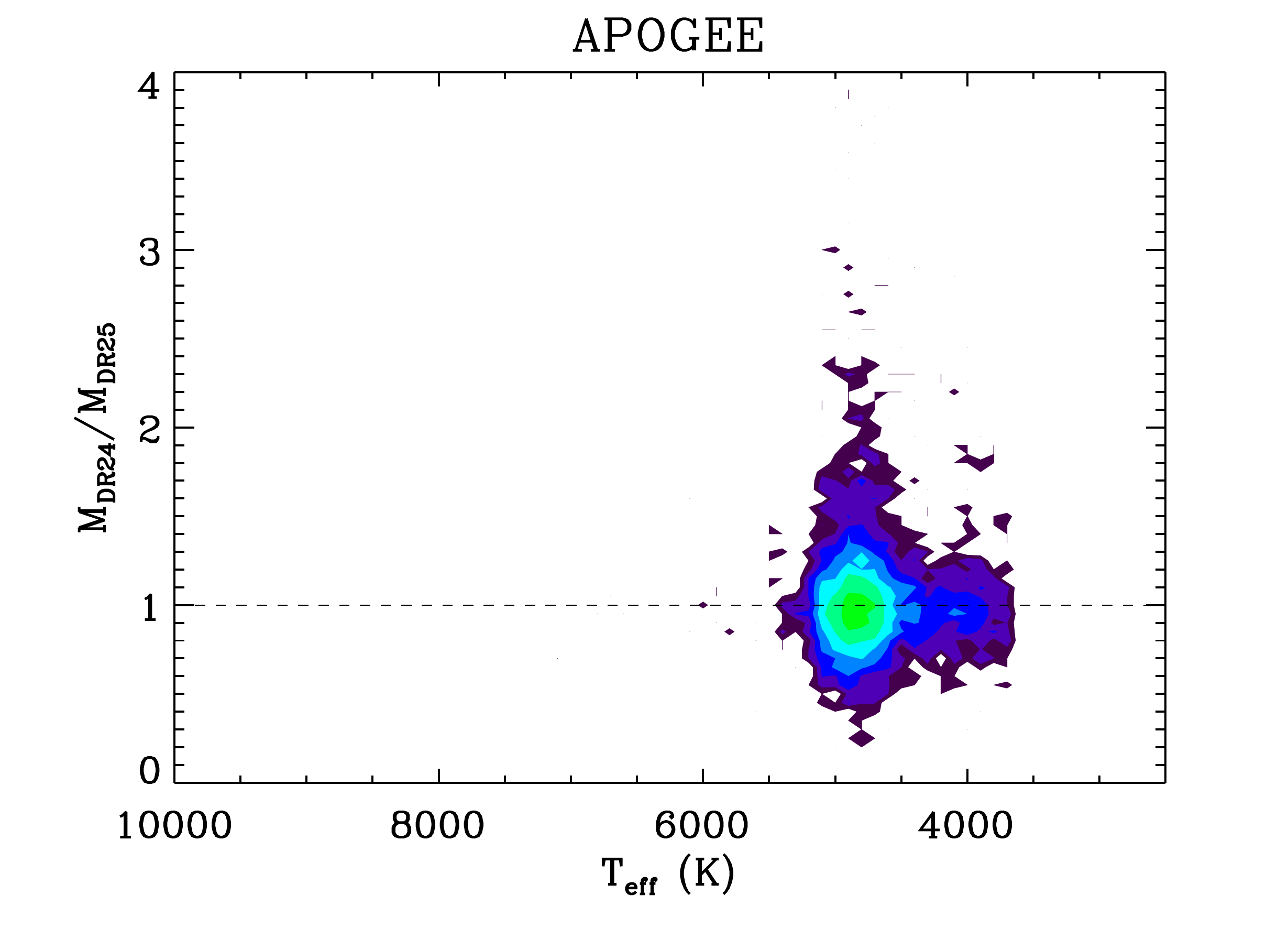}
\includegraphics[width=5.9cm, trim=2cm 0.5cm 2cm 0]{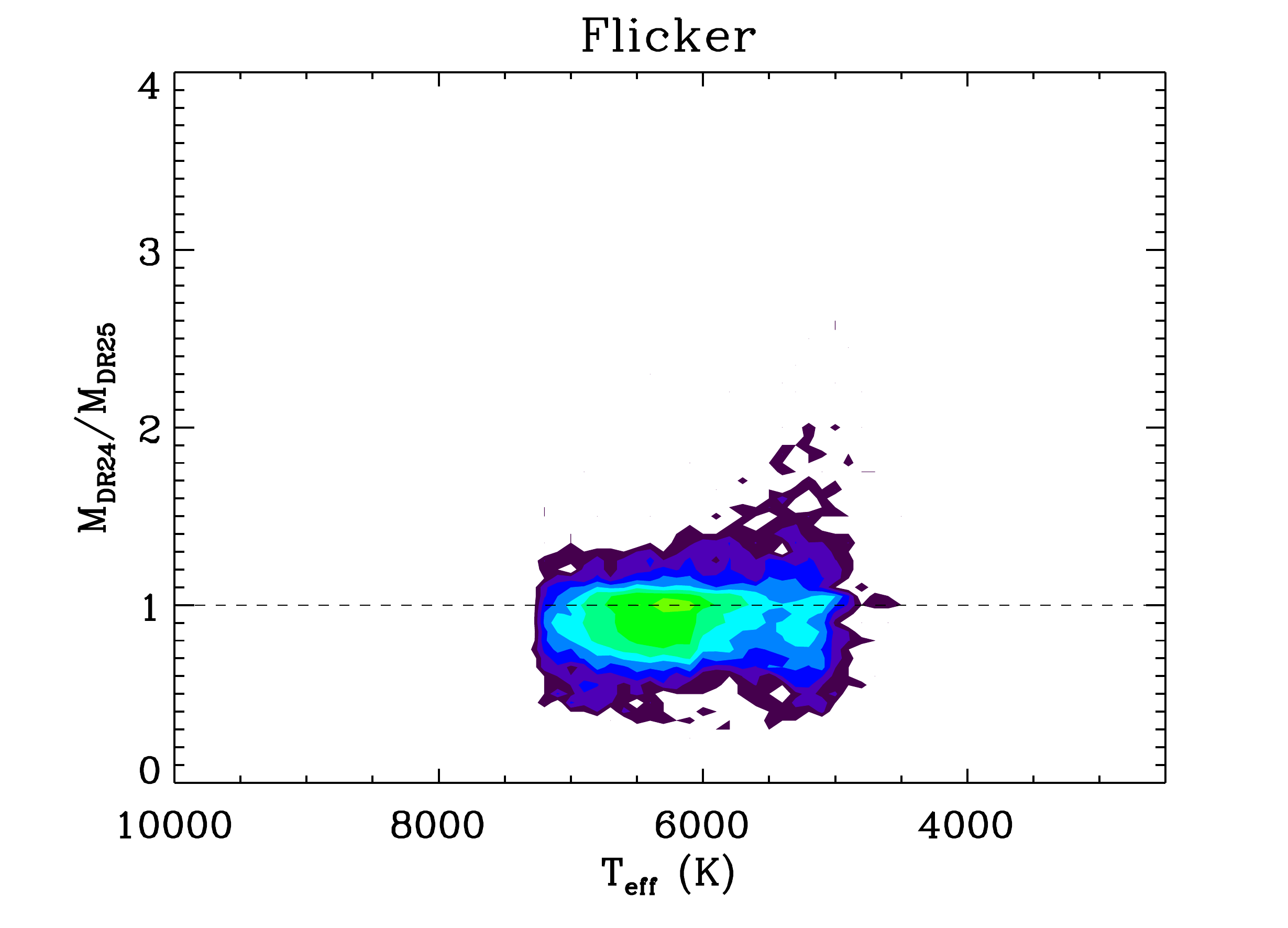}
\includegraphics[width=5.9cm, trim=2cm 0.5cm 2cm 0]{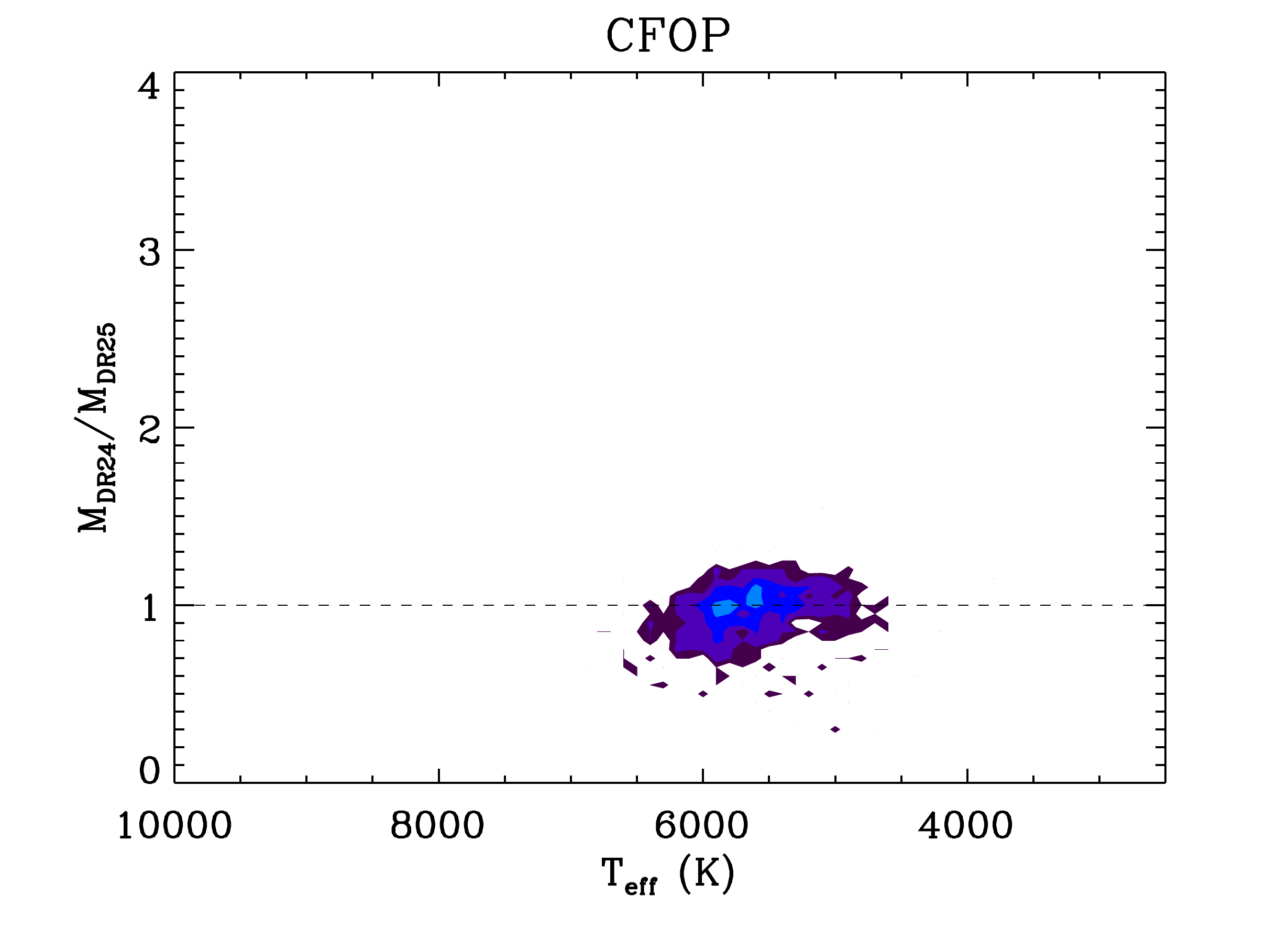}
\includegraphics[width=5.9cm, trim=2cm 0.5cm 2cm 0]{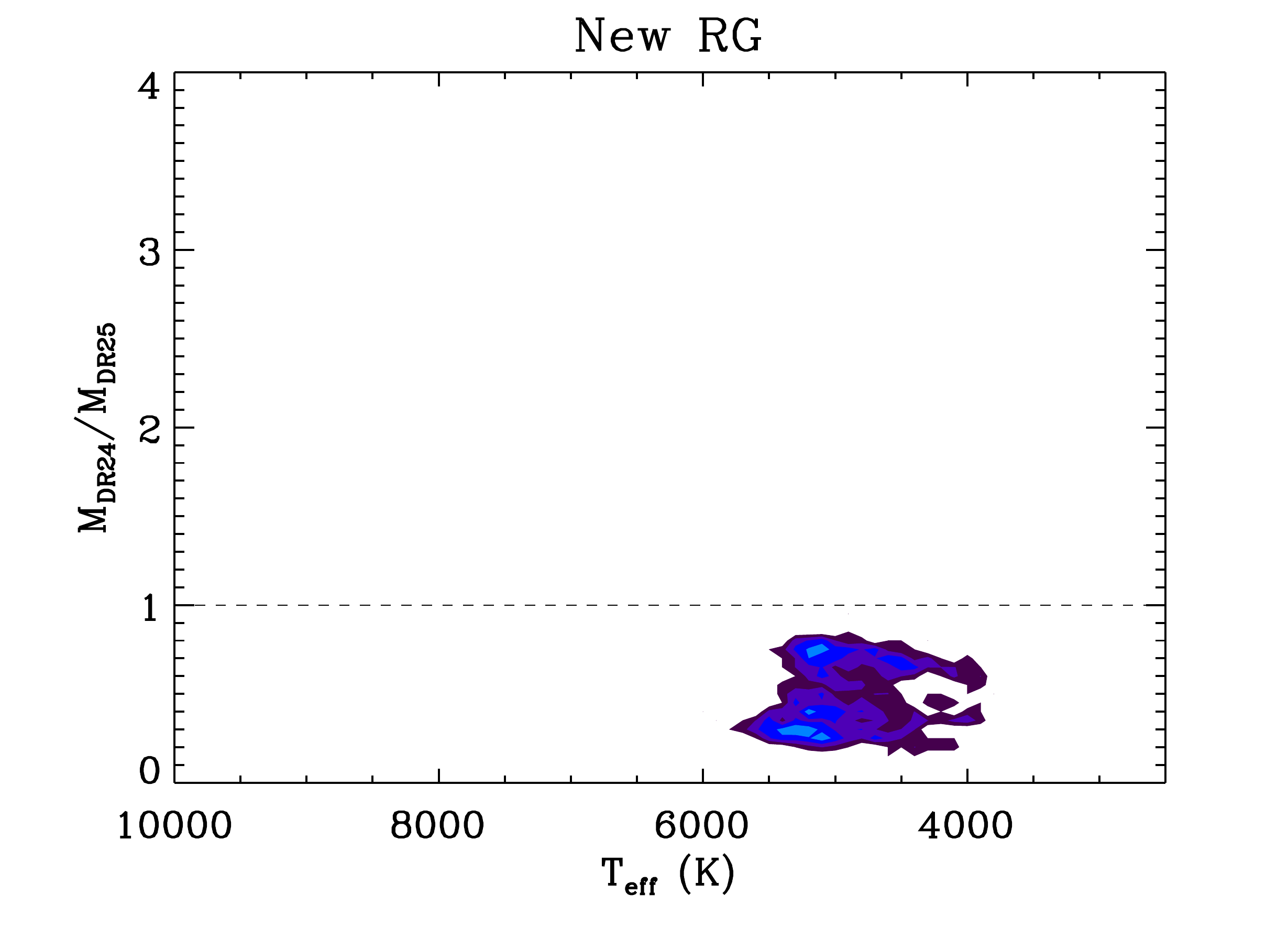}
\caption{Same as Figure \ref{compR} but for stellar mass.}
\label{compM}
\end{center}
\end{figure*}

Figures~\ref{compR} and \ref{compM}  show the ratios of DR24 to DR25 radii and masses. These plots represent the logarithm of the number density of stars for different effective temperatures and gravity bins. Figures \ref{compR} and \ref{compM} are included for all stars (upper left) as well as the samples highlighted in Section 2.2.

Figure~\ref{compR} shows that the highest density of stars is close to the $R_{\rm DR24}/R_{\rm DR25}$=1 line, which means their radii did not change. The stars with the most significant changes in the stellar parameters correspond to stars with new input values, as expected. Stars with LAMOST and Flicker inputs have a larger number density of stars slightly below the ratio equals 1 line, which means that these stars have become larger (up to a factor of 2). Stars with APOGEE inputs are on both sides of the 1 line with higher number density above 1 (i.e. smaller radii in the DR25). Finally, the new red giants have a radius ratio close to 0, corresponding to a large increase of the size of the star from a dwarf to a red giant. Some of these cases are also present in the APOGEE sample. 


The mass comparison also shows that the highest number density of stars is close to 1:1 line. Stars with Flicker and CFOP inputs see their masses change by less than a factor of 2. The new inputs from LAMOST and APOGEE show a similar behaviour except that they also lead to smaller masses for stars with $\teff \sim 5000$\,K.

In both Figures \ref{compR} and \ref{compM} we see a group of  cool stars ($T_{\rm eff} <$\,3250\,K) which systematically fall below the 1:1 line. These are stars that were erroneously classified as giants in the H14 catalog and corrected using the dwarf classifications by \citet{2012ApJ...753...90M} in the DR25 catalog, as further explained in Section 5.3. 

After completion of DR25 catalog, \citet[][hereafter G16]{2016MNRAS.457.2877G} published the revised properties of 4216 M dwarfs observed by {\it Kepler}. A total of 699 stars in G16 are not included in the DR25 catalog since they were only observed during Q17 and {   neither had  KIC values available nor spectroscopic inputs. For 68 stars spectroscopic parameters were also published by \citet{2016A&A...594A..39F}}. For the stars in common between G16 and DR25, the two temperature scales are close for cool stars below 3500K, although the temperatures from G16 are on average 200K hotter for 63 stars. Above 3500K, the temperatures from G16 are cooler compared to the DR25 values with differences larger than 200K (up to 2000K) for 487 stars. We found that 54 stars in G16 are {   classified} as red giants in the DR25. A small sample of these stars (16) were classified as red giants from seismology so the detection of oscillations does not agree with the dwarf classification of G16. A majority of the stars with DR25 temperatures hotter than 4000K have a $T_{\rm eff}$ provenance from the KIC and PHO54. Given that the analysis by G16 was specifically tailored towards cool dwarfs some of these stars may be misclassified in the DR25 catalog, and hence the classifications by G16 should be preferred over the DR25 catalog. We list these potentially misclassified stars in Appendix C, Table~\ref{tab:Mdwarfs}.

\subsubsection{Effects on Planet Host Star Parameters}

\begin{figure}[htbp]
\begin{center}
\includegraphics[width=9cm]{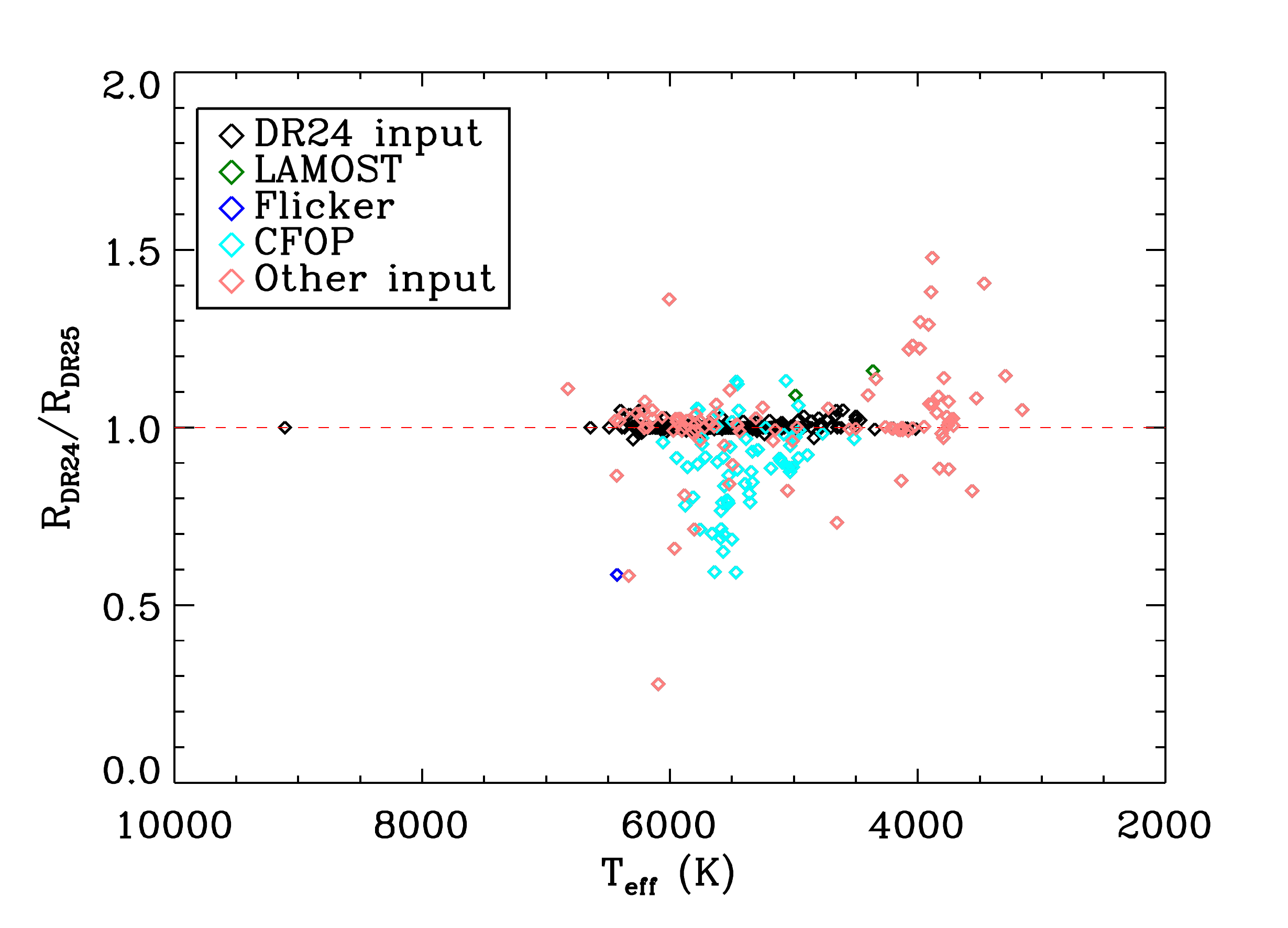}
\includegraphics[width=9cm]{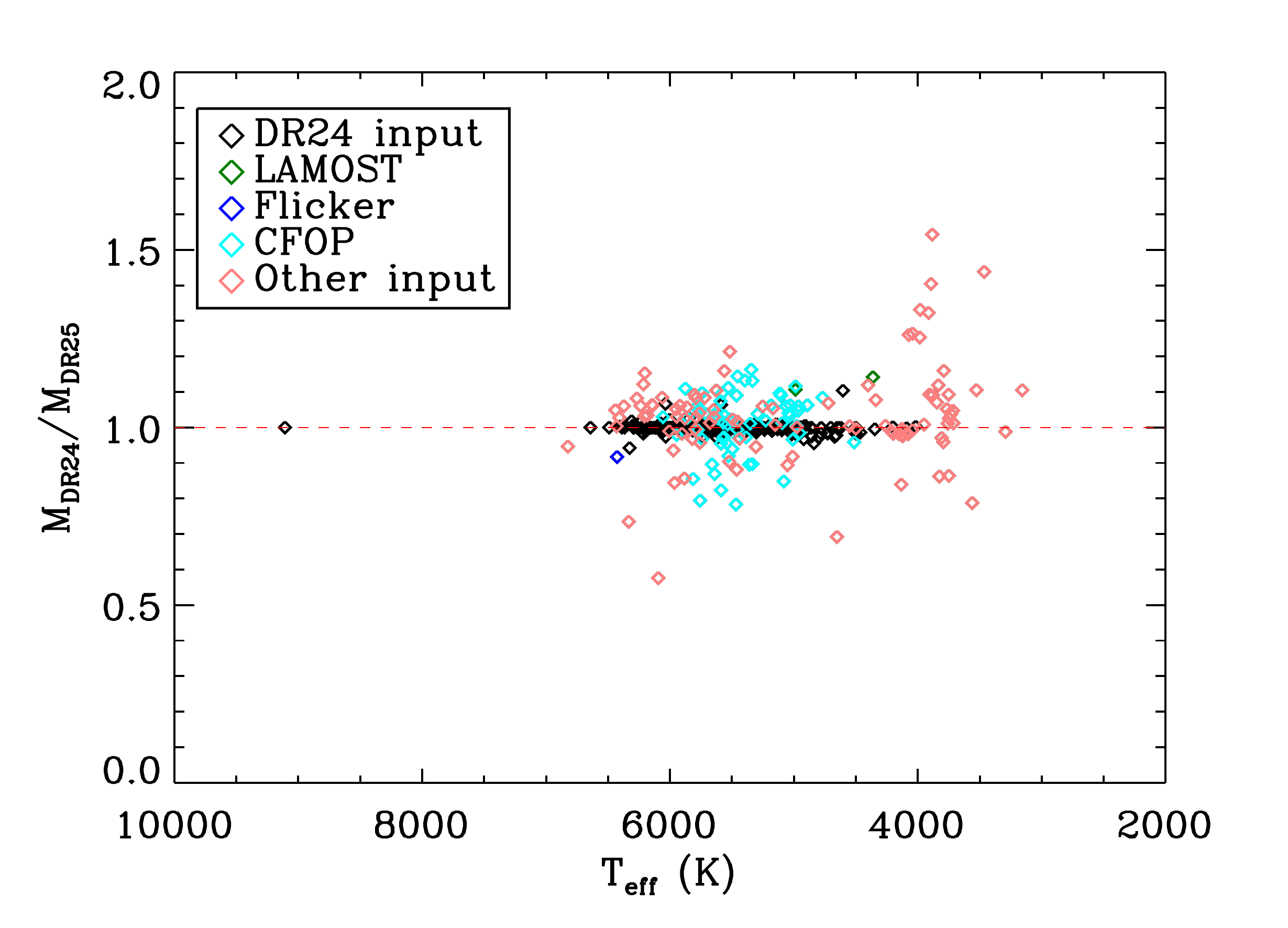}
\caption{Comparison of radii and masses of planet host stars showing the different subsamples where we used either new inputs values or the same input values as in the previous catalog.}
\label{compRM_planet}
\end{center}
\end{figure}

As a final test, we looked in particular at planet host stars parameters as they directly impact the size inferred for the planets. Figure~\ref{compRM_planet} compares the radii and masses of the planet host stars computed in this work with the DR24 catalog. It is comforting to see that stars where we used the same inputs as the DR24 catalog (black diamonds in the figure) fall on or are very close to the line $R_{Q1-17}$/$R_{\rm new}$=1, indicating that the radii of these stars changed by a few percent at most. The small change can be explained by the updated isochrone grid that was used in this work.

As expected, the largest changes affected stars with new input values. Many stars with new CFOP parameters have a different evolutionary stage. Indeed, we aforementioned that a fraction of stars moved from main sequence stars to more evolved subgiants. This explains the number of stars that now have a larger radius than the previous catalog (cyan symbols). This is also the case for the star with the Flicker input (blue symbol) and some of the individual new inputs (pink symbols).


For stars cooler than ~4500K, we notice that a significant number of host stars become smaller and less massive. Specifically, for eight host stars the spectroscopic classification by \citet{2014ApJ...784...45R} was subsequently shown to lead to systematically overestimated effective temperatures and radii, and hence led to biased estimates in the DR24 catalog. To correct this, we adopted the inputs from H14 for these stars for the DR25 catalog.


The following is a list of specific host stars with significant changes in their stellar parameters:

1) The radii of the K-dwarfs KIC 5640085 (KOI-448 and Kepler-148) and KIC 10027323 (KOI-1596 and Kepler-309) decreased by $\sim\,40-50$\% due to the correction of the spectroscopic input values from \citet{2014ApJ...784...45R}, as discussed above. The input parameters were reversed back to the H14 catalog, which were based on \citet{2012ApJ...750L..37M}. 

2) KIC 7529266 (KOI-680, Kepler-635) is a solar-type star ($\sim$\,6000K) and shows that the largest change in radius ($R_{\rm Q1-17}/R_{\rm new}$\,$\sim$\,0.3). We adopted updated input values from \citet{2015AA...575A..71A}, which lists a $\logg$ of 3.5\,dex compared to 4.35\,dex in the KIC where the $\log g$, leading to a large increase in radius. It is not surprising to see this change given that the original KIC had known shortcomings regarding the classification of subgiants. 

3) KIC 8733898 (KOI-2842, Kepler-446), with $T_{\rm eff}\sim$3500K and $R_{\rm Q1-17}$/$R_{\rm new}\sim$\,1.4, had its input values changed from \citet{2013ApJ...767...95D} to \citet{2015ApJ...801...18M}, leading to a smaller radius. The spectroscopic input should be more reliable than the photometric classification in the previous catalog.



\subsection{Distances and extinction}

\begin{figure}[htbp]
\begin{center}
\includegraphics[width=6.5cm, angle=90]{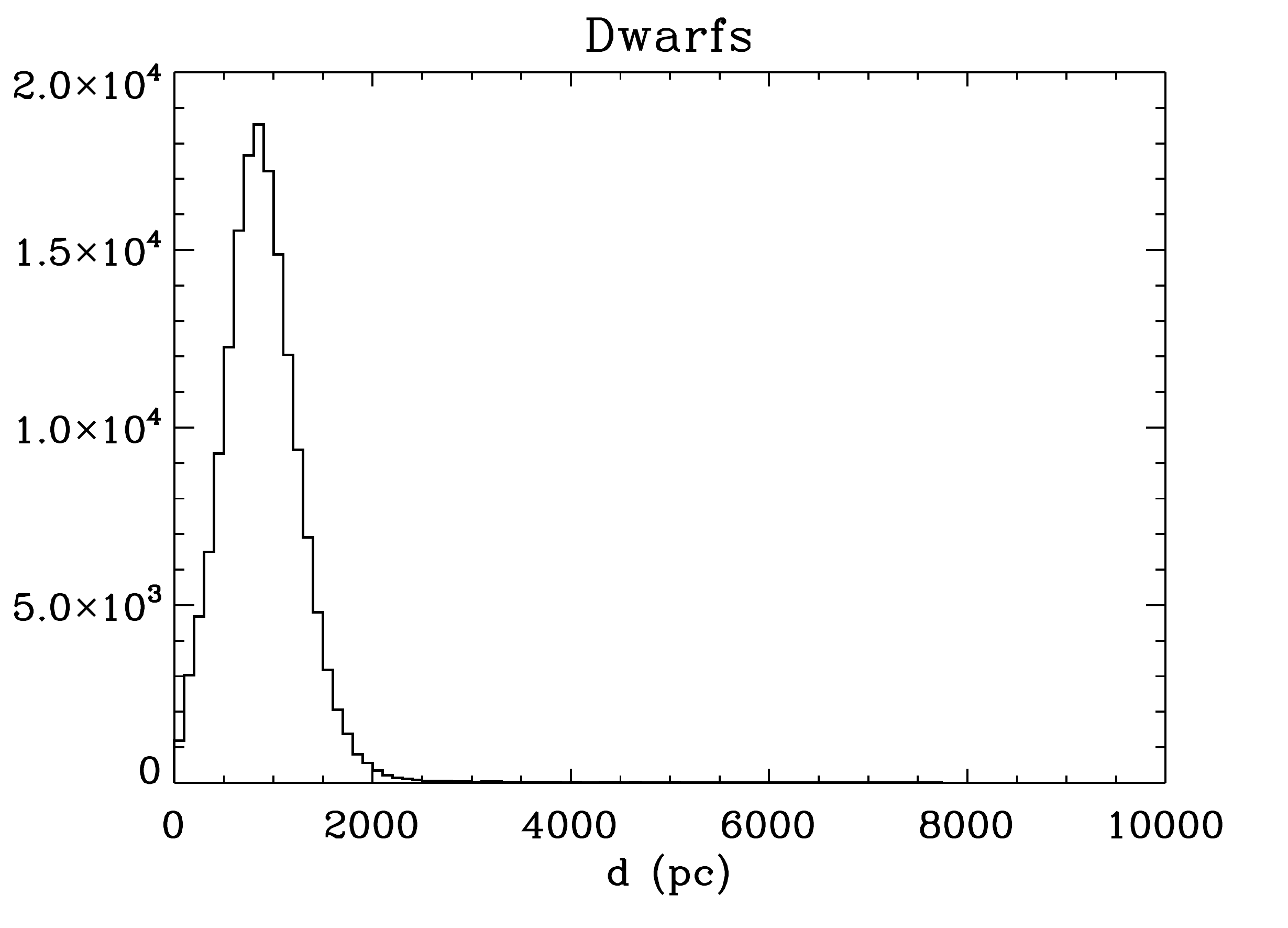}
\includegraphics[width=6.5cm, angle=90]{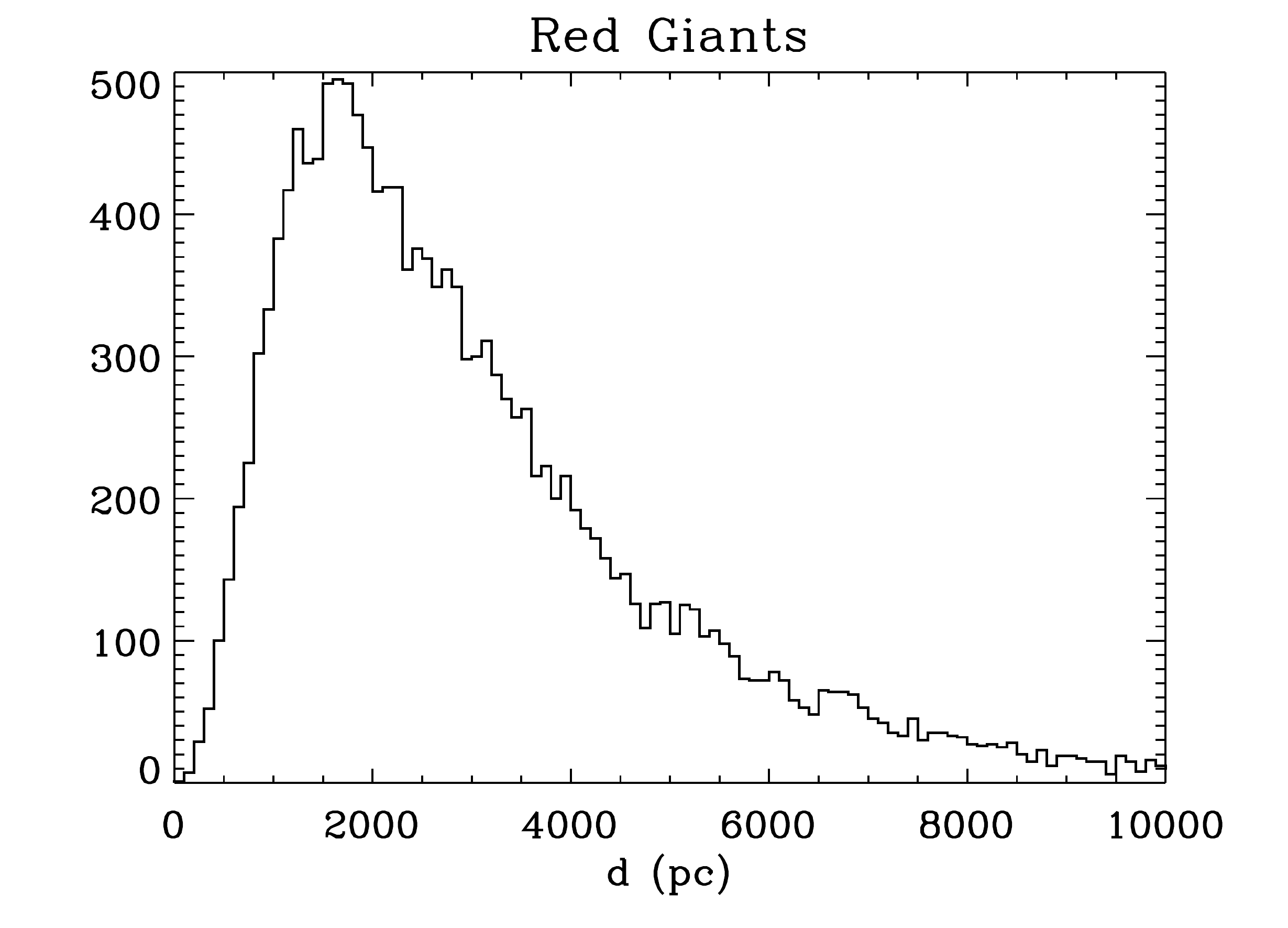}
\caption{Distribution of distances for dwarfs (top panel) and red giants (bottom panel) in the DR25 catalog.}
\label{histDW_RG}
\end{center}
\end{figure}

In addition to stellar parameters, the DR25 catalog also includes distances and extinction values for $\sim$\,196,850 stars (see Section 5.3 for more details). 
Figure~\ref{histDW_RG} shows the distribution of distances for dwarfs (left panel) and for red giants (right panel) observed by {\it Kepler}. As expected, red giants observed by \kep\ are on average more distant than dwarfs.

We also compared our catalog distances to \citet[][hereafter R14]{2014MNRAS.445.2758R}, who combined asteroseismology with APOGEE spectra to derive distances and extinctions for a sample of $\sim$\,2000 \kep\ red giants. The comparison showed that the catalog distances are systematically larger by up to \,50\%, which is due to the fact that our model grid does not include He-core burning models for low-mass stars, and hence giants are preferentially fitted to higher-mass, more luminous and hence more distant models. This bias was already pointed out in H14, and should be kept in mind when using catalog results for red giant stars. We emphasize that this distance bias is not expected to be relevant for dwarfs and subgiant stars, which form the majority of the {\it Kepler} target sample. 

Finally, a comparison of extinction values to \citet[][]{2014MNRAS.445.2758R} showed that the catalog values for giants are systematically larger by $\sim$0.1--0.3\,mag on average, similar to the results found for the KIC (see Figure 17 of R14). This is most likely due to the simplified 3D reddening model adopted in this work and the KIC compared to the method adopted by R14, which derives reddening values by comparing synthetic to observed photometry on a star-by-star basis. Since this method is only effective if \teff-\logg-\feh\ can be derived independently from photometry, it cannot be applied to the full {\it Kepler} sample at this point.

\subsection{Catalog Shortcomings}

While this paper provides important improvements over previous {\it Kepler} stellar properties catalogs, several shortcomings remain. In particular:

\begin{itemize}
\item For stars with input values that fall off the Dartmouth isochrone grid (e.g. very cool dwarfs) we adopted the input and output values from H14. There are also 3 stars where we adopted the published values (KIC 5807616, 5868793 and 10001893). Indeed these three stars fall out of the grid because they are too hot with a temperature above 25,000K. These stars do not have distances and extinction values. The provenance for the mass, radius and density is MULT as they come from a different method.

\item Unlike in previous deliveries we did not override catalog values with published solutions that provide better estimates for radii and masses (e.g. from asteroseismology) in order to homogeneously derive posterior distributions (including distances) for all stars. This means that for some stars better estimates for radii and masses may be available in the literature. 

\item Similar to H14, the adopted isochrone grid does not include He-core burning models for low-mass stars and hence derived properties for red giants (such as radius, mass, and distances) will be systematically biased towards higher-mass stars (and more distant for red giants). Users are strongly encouraged to adopt values from dedicated Kepler red-giant classification programs such as the APOKASC \citep[e.g.][]{2014ApJS..215...19P} or SAGA \citep{2014ApJ...787..110C} surveys for such stars, or use the {\bf provided} \teff, \logg\ and \feh\ values in this catalog as input for deriving more accurate stellar properties.

\item The new catalog also includes several corrections that were pointed out by the community since the release of the H14 catalog. Due to a coding error, every star in the Q1-16 catalog with input $T_{\rm eff}$\,<\,3250K was automatically classified as a dwarf using BT-Settl models even if the input $T_{\rm eff}$ indicated that it was a giant. To correct this, we revisited all dwarfs that have been classified using BT-Settl models and verified their evolutionary state using the \citet{2012ApJ...753...90M} spectroscopic classifications. When this was verified, we adopted the Q1-16 BT-Settl solution. These stars do not have distances and extinction values. The provenance for the mass, radius and density is BTSL.

\item The number of misclassified red giants reported in \citet{2016ApJ...827...50M} is of 854 while in this delivery the misclassified red giants represent 835 stars. Between the delivery of the catalog and the finalization of the misclassified red giants some stars were dropped due to the pollution from nearby known red giants while others were added. Hence, there is a discrepancy of 51 stars. 

\item For the vast majority of targets the input classifications assumed that all the stars are single systems, which can lead to biased stellar parameters if the targets are in fact multiple star systems. While we expect that this effect is small compared to the typical uncertainties in the derived stellar properties, future catalog releases will attempt to take into account information from various high-resolution imaging programs \citep[e.g.][]{2012AJ....144...42A,2014AJ....148...78D,2014AA...566A.103L,2016AJ....152...18B,2016arXiv161202392F,2016AJ....152....8K} for stellar classifications.

\end{itemize}

\section{Summary}

The DR25 {\it Kepler} stellar properties catalog includes improved stellar properties for over 28,800 stars, including spectroscopic surveys (CFOP, APOGEE, LAMOST), $\log g$ values derived from stellar granulation ({   Flicker}), and new asteroseismic reclassifications of more than 800 stars \citep{2016ApJ...827...50M}. We also added 311 stars that were targeted during the last quarter observed by {\it Kepler}, Q17. Finally, 317 stars which had so far not been classified were included in this catalog using spectroscopic classifications from LAMOST and APOGEE. This leads to a total number of stars in the {\it Kepler} DR25 catalog to 197,096, including 4085 planet(-candidate) host stars. The DR25 stellar properties catalog has been used for the final Transiting Planet Search/Data Validation (TPS/DV) by the {\it Kepler} Mission, and is available at the NASA Exoplanet Archive (http://exoplanetarchive.ipac.caltech.edu) and the Mikulski Archive for Space Telescopes (MAST, http://archive.stsci.edu/kepler/stellar17/search.php). We note that there are still $\sim$3000 unclassified stars that do not have reliable colors and were not analysed in this work. A major addition compared to the DR24 catalog is the delivery of the posterior samples for all stellar parameters for $\sim$\,196,850 stars.

The catalog was constructed with similar methodology to H14, using input data from different techniques such as asteroseismology, spectroscopy, photometry, or {   Flicker}. The effective temperature, surface gravity and metallicity were then conditioned on  a grid of isochrones to provide posterior distributions of all parameters. While the input values still come from a variety of sources, the updated methodology allowed an in principle homogeneous estimation of all derived quantities such as mass, radius, density, distance, and extinction. The update of the methodology from the H14 catalog also led to slightly smaller and more realistic uncertainties associated with the stellar parameters. However, we emphasize that there are still a number of significant shortcomings in the catalog, as described in Section 5.3. We also note that distances and extinctions listed in this paper are systematically different from the values in the original delivery to the NASA Exoplanet archive due to the coding error explained in Section 3.3. All other stellar properties are unaffected, but we recommend to use the corrected distances listed in this paper for scientific investigations of the Kepler sample.


Even though the DR25 catalog forms the basis for the final Transiting Planet Search in the \kep\ mission close out, the improvement in the characterization of all \kep\ targets will continue to develop over the coming years. Indeed, since the delivery of the catalog additional observations and analyses have been performed for  {\it Kepler} targets. For example, \citet{2016MNRAS.457.2877G} obtained spectra for more than 3,000 dwarfs providing more accurate $T_{\rm eff}$ and $\log g$. More recently, \citet{2016MNRAS.tmp.1200Y} used asteroseismology to re-classify more than 1,500 subgiants in DR25 as red giants. Finally, the most important update of the {\it Kepler} stellar properties catalog can be expected with the advent of high-precision parallaxes by the ESA Gaia mission \citep{2005ASPC..338....3P}, {   for which the first data release has been announced \citep{2016A&A...595A...2G}. These parallaxes} will at last provide an efficient tool to precisely determine the evolutionary states of nearly all targets observed by \kep.  


\acknowledgments
The authors would like to thank Michael Haas and Eric Gaidos for useful discussions. SM would like to thank R.~A. Garc\'ia and the CEA Saclay (France) for their computing resources. 
SM and DH acknowledge support by the National Aeronautics 
and Space Administration under Grant NNX14AB92G issued through the Kepler Participating Scientist Program, and DH acknowledges support by the Australian Research Council's Discovery Projects funding scheme (project number DE140101364).
FB is supported by NASA through Hubble Fellowship grant \#HST-HF2-51335 awarded by the Space Telescope Science Institute, which is operated by the Association of Universities for Research in Astronomy, Inc., for NASA, under contract NAS5-26555.

\bibliographystyle{apj} 

\bibliography{apj-jour,/Users/Savita/Documents/BIBLIO_sav}


\appendix

\section{Provenances of input parameters}

We followed the same scheme introduced by H14 to numerically cross-link literature sources of input parameters to a given provenance (see Section 6.5 in H14). Table~\ref{tab:refs} lists the complete references for all input sources used in the DR25 catalog. As an example, a \logg\ provenance of AST10 indicates that the input \logg\ value was derived from asteroseismology and taken from \citet{2014ApJS..210....1C}.

\begin{table*}[b]
\begin{footnotesize}
\begin{center}
\caption{Reference Key}
\begin{tabular}{c l l}
\hline
Key & Reference & Methods \\
\hline
 0 & \citet{2011AJ....142..112B}&Photometry				    		\\	    
 1 & \citet{2012ApJS..199...30P}	&Photometry				    		\\
 2 & \citet{2013ApJ...767...95D}&Photometry				    		\\
 3 & \citet{2012Natur.486..375B}			&Spectroscopy  			            \\
 4 & \citet{2011AA...534A.125U}		&Spectroscopy  			            \\
 5 & \citet{2012ApJ...750L..37M}			&Spectroscopy  			            \\
 6 & \citet{2012MNRAS.423..122B}&Spectroscopy/Asteroseismology      \\
 7 & \citet{2012AA...543A.160T} &Spectroscopy/Asteroseismology      \\
 8 & \citet{2013ApJ...767..127H}&Spectroscopy/Asteroseismology      \\
 9 & \citet{2013ApJ...765L..41S}&Asteroseismology			    	\\
10 & \citet{2014ApJS..210....1C}&Asteroseismology			    	\\
11 & \citet{2011ApJ...743..143H}&Asteroseismology			    	\\
12 & \citet{2013ApJ...770...69P}			&Spectroscopy  			            \\
13 & \citet{2013MNRAS.434.1422M}&Spectroscopy  			            \\
14 & \citet{2012ApJ...753...90M}&Spectroscopy  			    		\\
15 & \citet{2013ApJ...770...43M}			&Spectroscopy  			    		\\
16 & \citet{2013ApJ...770...90G}			&Photometry				    		\\
17 & \citet{2013AA...555A.108M}		&Spectroscopy  			            \\
18 & \citet{2013ApJS..204...24B}			&Spectroscopy/Transits 			    \\    
19 & \citet{2013MNRAS.433.1262W}			&Spectroscopy/Asteroseismology		\\	  	  
20 & \citet{2010ApJ...710.1724B}			&Spectroscopy/Transits/EBs		    \\
21 & \citet{2010ApJ...713L..79K}&Spectroscopy/Transits/EBs	    	\\
22 & \citet{2010ApJ...713L.136D}			&Spectroscopy/Transits/EBs		    \\
23 &\citet{2010ApJ...724.1108J} 			&Spectroscopy/Transits/EBs		    \\
24 & \citet{2010Sci...330...51H}			&Spectroscopy/Transits/EBs		    \\
25 & \citet{2013ApJ...770..131L}			&Spectroscopy/Transits/EBs		    \\
26 & \citet{2011ApJS..197....9F}		&Spectroscopy/Transits/EBs		    \\
27 &\citet{2011ApJS..197...13E}				&Spectroscopy/Transits/EBs	    	\\
28 & \citet{2011Sci...333.1602D}			&Spectroscopy/Transits/EBs		    \\
29 & \citet{2011ApJS..197...14D}			&Spectroscopy/Transits/EBs		    \\
30 & \citet{2011ApJS..197....7C}			&Spectroscopy/Transits/EBs		    \\
31 & \citet{2011ApJ...743..200B}			&Spectroscopy/Transits/EBs		    \\
32 & \citet{2012Natur.482..195F}			&Spectroscopy/Transits/EBs		    \\
33 & \citet{2012MNRAS.421.2342S}		&Spectroscopy/Transits/EBs		    \\
34 & \citet{2012ApJ...750..114F}			&Spectroscopy/Transits/EBs		    \\
35 &\citet{2012ApJ...750..112L} 			&Spectroscopy/Transits/EBs		    \\
36 & \citet{2012Natur.481..475W}			&Spectroscopy/Transits/EBs		    \\
37 & \citet{2012Sci...337.1511O}			&Spectroscopy/Transits/EBs		    \\
38 & \citet{2011AA...533A..83B}			&Spectroscopy/Transits/EBs		    \\
39 &\citet{2011AA...528A..63S} 		&Spectroscopy/Transits/EBs		    \\
40 & \citet{2011AA...536A..70S}			&Spectroscopy/Transits/EBs		    \\
41 & \citet{2012ApJ...747..144M}		&Spectroscopy/Transits/EBs		    \\
42 & \citet{2012AA...538A..96B}			&Spectroscopy/Transits/EBs		    \\
43 & \citet{2012AJ....143..111J}			&Spectroscopy/Transits/EBs		    \\
44 & \citet{2012Sci...336.1133N}			&Spectroscopy/Transits/EBs		    \\
45 & \citet{2012ApJ...758...87O}			&Spectroscopy/Transits/EBs		    \\
46 & \citet{2013ApJ...773...98B}			&Spectroscopy/Transits/EBs		    \\
47 & \citet{2013Natur.499...55M}			&Spectroscopy/Transits/EBs		    \\
48 & \citet{2013Natur.494..452B}			&Spectroscopy/Transits/EBs		    \\
49 & \citet{2011Natur.480..496C}  		&Spectroscopy/Transits/EBs		    \\
50 & \citet{2010ApJ...725.1633H}			&Spectroscopy/Transits/EBs		    \\
51 & \citet{2013AA...554A.114H}			&Spectroscopy/Transits/EBs		    \\
52 & \citet{2013ApJ...771...26F}			&Spectroscopy/Transits/EBs		    \\
53 & \citet{2013ApJ...775...54S}			&Spectroscopy/Transits/EBs		    \\
54 & \citet{2014ApJS..211....2H}&Photometry/Asteroseismology	    \\	
\hline								
\end{tabular}
\label{tab:refs}
\end{center}

\end{footnotesize}
\end{table*}

\begin{table*}
\begin{footnotesize}
\begin{center}
\caption{Reference Key}
\begin{tabular}{c l l}
\hline
Key & Reference & Methods \\
\hline
55 & \citet{2014ApJS..215...19P}& Photometry/Asteroseismology/Spectroscopy\\
56 & \citet{2014ApJ...787..110C}& Photometry/Asteroseismology\\
57 & \citet{2013PNAS..11019273P}& Spectroscopy\\
58 & \citet{2014ApJ...784...45R}& Spectroscopy\\
59 & \citet{2014Natur.509..593B}& Spectroscopy\\
60 & \citet{2013ApJ...770...43M,2013ApJ...779..188M}& Spectroscopy\\
61 & \citet{2014ApJS..210...20M}& Spectroscopy\\
62 & \citet{2013Sci...340..587B}& Spectroscopy\\
63 & \citet{2013ApJ...775...54S}& Spectroscopy\\
64 & \citet{2013AA...557A..74G}& Spectroscopy/Transits\\
65 & \citet{2014AA...561A.103O}& Spectroscopy/Transits\\
66 & \citet{2014AA...564A..56D}& Spectroscopy/Transits\\
67 & \citet{2014AA...567A..14T}& Spectroscopy\\
68 & \citet{2015RAA....15.1095L}& Spectroscopy\\
69 & \citet{2015MNRAS.452.2127S}& Spectroscopy/Asteroseismology\\
70 & \citet{2014ApJS..213....5M}& Spectroscopy\\
71 &\citet{2016ApJ...827...50M} & Asteroseismology\\
72 & Chaplin et al. (in prep.)& Spectroscopy\\
73 & \citet{2016ApJ...818...43B} & Flicker\\
74 & \citet{2015ApJS..219...12A}& Spectroscopy\\
75 & \citet{2016AA...590A.112M}& Spectroscopy\\
76 & \citet{2015AA...575A..71A}& Spectroscopy\\
77 & \citet{2014AA...572A..93H}& Spectroscopy\\
78 & \citet{2014AA...571A..37S}& Spectroscopy\\
79 & \citet{2014ApJ...791...89D}& Spectroscopy\\
80 &\citet{2014ApJ...795...25K} & Spectroscopy\\
81 & \citet{2014ApJ...795..151E}& Spectroscopy\\
82 & \citet{2015AA...576A..11G}& Spectroscopy\\
83 & \citet{2014AA...570A.130S}& Spectroscopy\\
84 & \citet{2015AJ....149...55E}& Spectroscopy\\
85 &\citet{2015ApJ...800...99T} & Spectroscopy\\
86 & \citet{2015ApJ...801...18M}& Spectroscopy\\
87 & \citet{2015AA...577A.105L}& Spectroscopy\\
88 & \citet{2015AA...579A..55B}& Spectroscopy\\
89 & \citet{2012ApJ...745..120B}& Spectroscopy\\
90 & Furlan et al. (in prep.) & Spectroscopy\\
\hline								
\end{tabular}
\label{tab:refs}
\end{center}

\end{footnotesize}
\end{table*}

\clearpage

\section{Stellar replicated posteriors}


The DR25 catalog delivery includes replicated posteriors for each star obtained as described in Section 3.4. The files contain 40,000 samples of self-consistent stellar parameters, together with the logarithm of total likelihood and the isochrone weights, corresponding to volume of each model in mass, metallicity, and age. Age, mass and metallicity priors are not listed as we used uniform priors for these quantities. 

\begin{figure*}[htbp]
\begin{center}
\includegraphics[width=8cm]{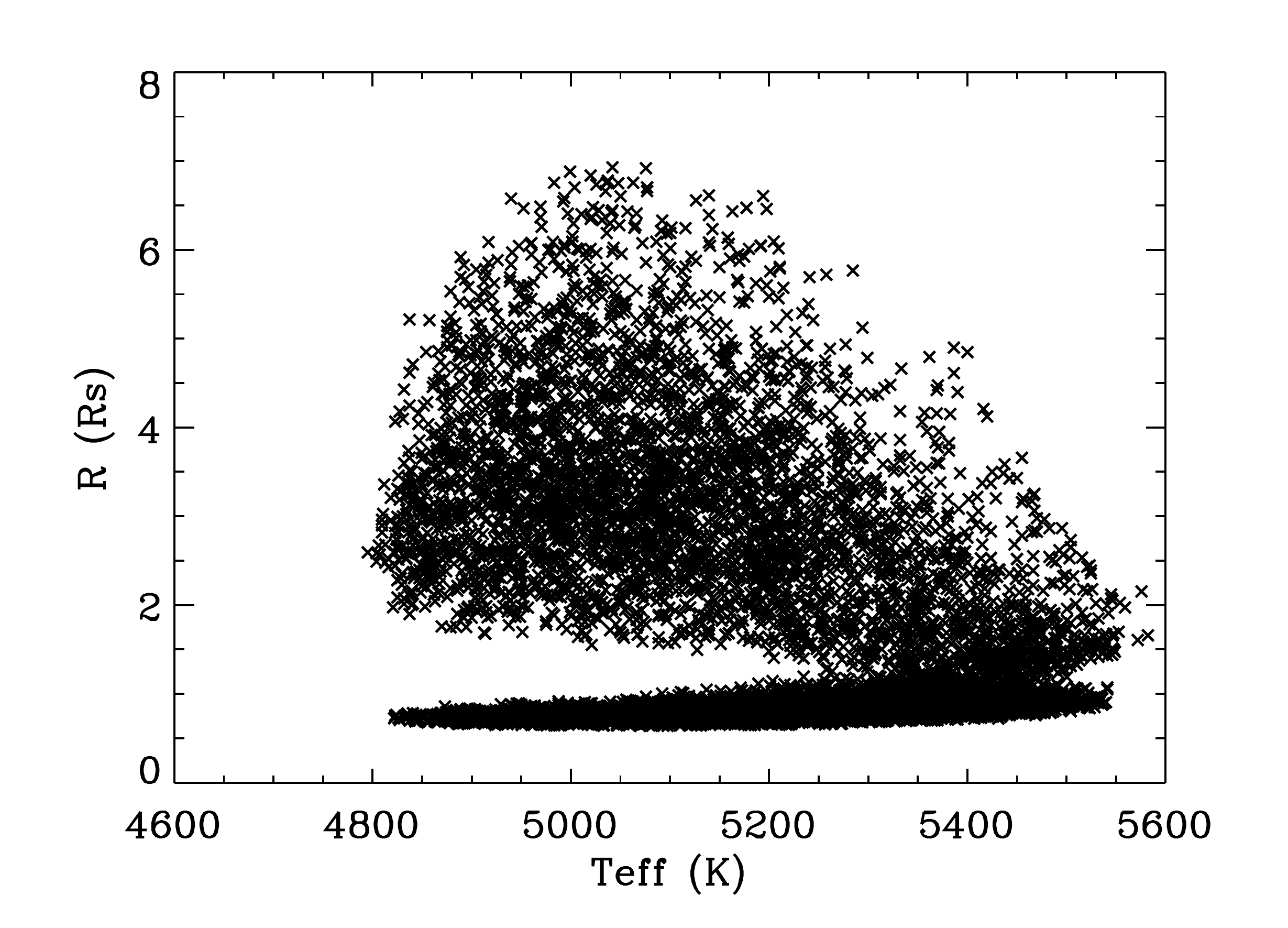}
\includegraphics[width=8cm]{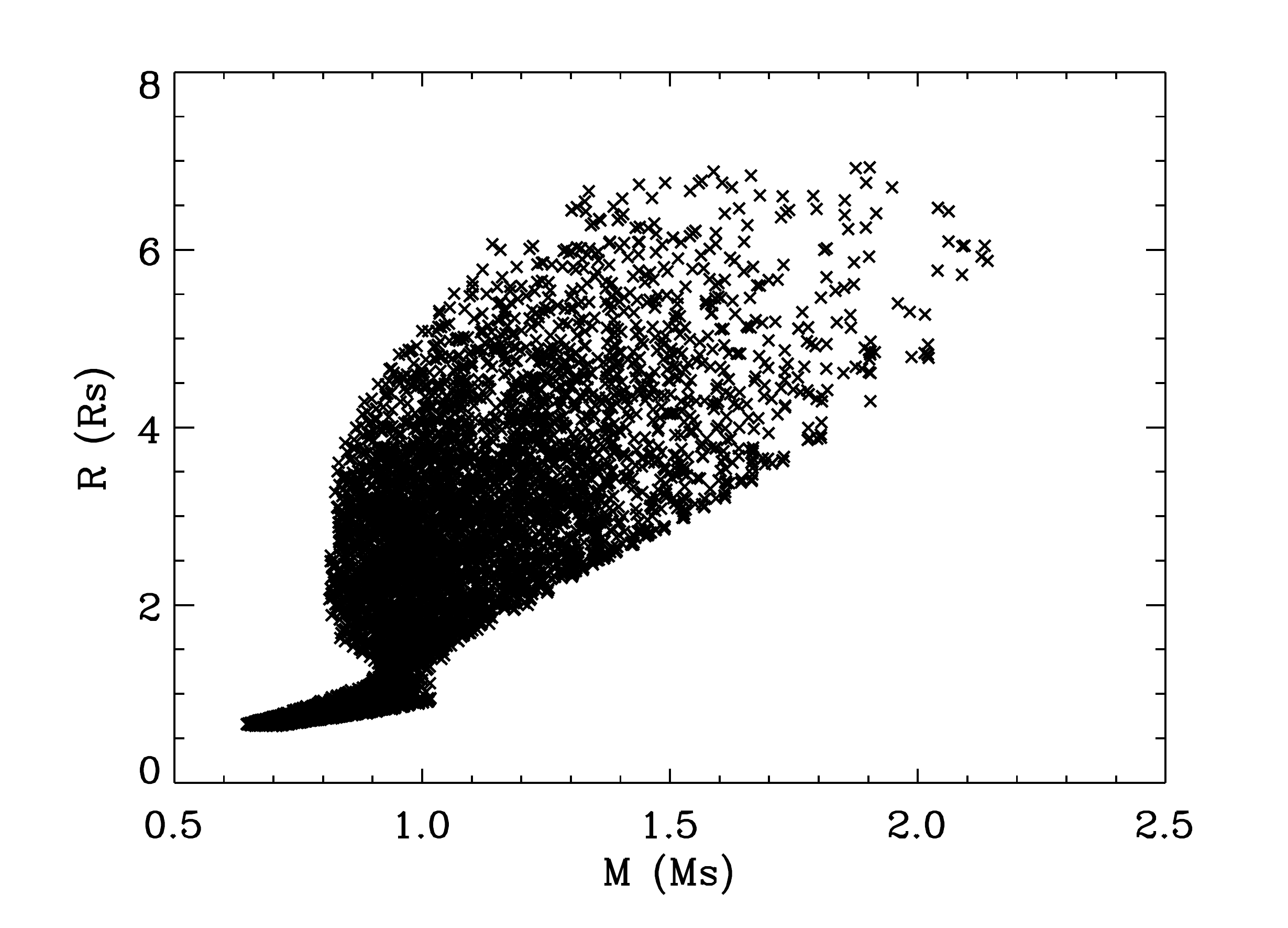}
\includegraphics[width=8cm]{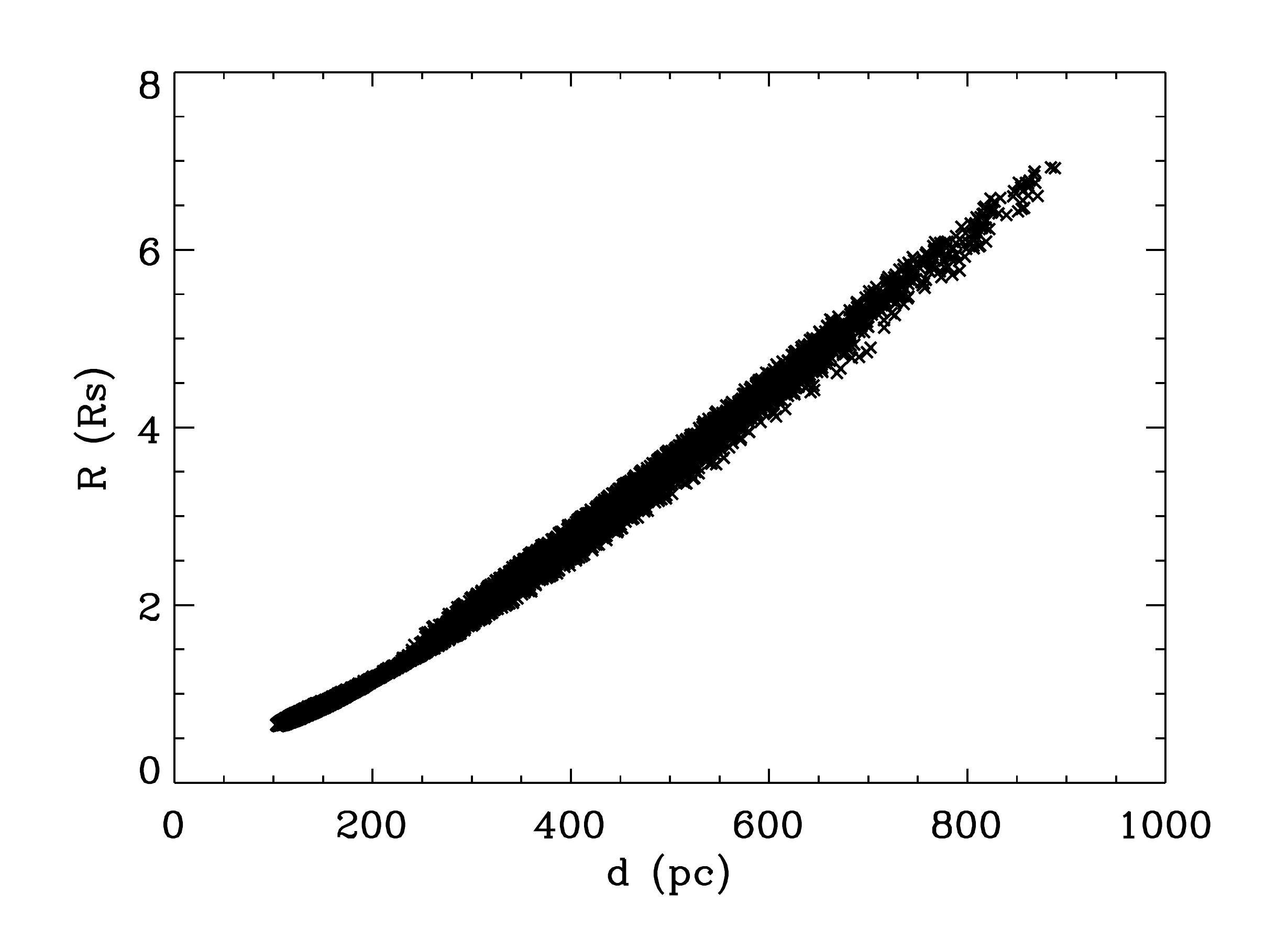}
\caption{Example of correlations between posterior samples for KIC 757076: radius versus \teff\ (top left), radius versus mass (top right) and radius versus distance (bottom).}
\label{Fig_corr}
\end{center}
\end{figure*}

Replicated posteriors files have the generic name: ``kplr<kepler id> dr25-stellarposterior.txt'' and contain 10 space-separated columns for each star:
$T_{\rm eff}$, \logg, [Fe/H], Mass, Radius, $\log \rho$, distance, Av, log(likelihood), log(weights). Each row corresponds to a set of self-consistent stellar properties, and hence can be used to produce marginalized distributions or explore parameter correlations. Figure~\ref{Fig_corr} shows an example of parameter correlations for  KIC 757076, which has a best-fit $T_{\rm eff}=5160^{+171}_{-156}$\,K and $\log g=3.58^{+0.93}_{-0.23}$\,dex in DR25. 

\section{Possible misclassified M dwarfs}

As discussed in Section 5.1.2, 54 targets which are classified as giants in the DR25 catalog have been classified as cool dwarfs by G16. Since these targets are potentially interesting for planet searches, we list them in Table~\ref{tab:Mdwarfs} together with the listed \teff\, \logg\, and R values in both catalogs. We note that follow-up spectroscopy will be needed to unambiguously determine the evolutionary state for these stars.

\begin{table}
\begin{center}
\caption{Possible M dwarfs according to Gaidos et al. (2016).}
\begin{tabular}{ccccccccc}
\hline
\hline
KIC & $T_{\rm eff}^{*}$ (K) & $\log g^*$ & R$^*$ (R$_\odot$) & $T_{\rm eff}$ (K)  & $\log g$  & R (R$_\odot$) & $P_{T_{\rm eff}}$ & $P_{logg}$\\
\hline
  1575570 &  3429 & 4.89  &0.36 & 3370 & 0.46 & 175.79  & KIC0 & KIC0\\
  3629762 & 3241 & 5.04 & 0.25 & 3279 & 0.16 & 151.95  & KIC0  & KIC0\\
  4454364 & 3586 & 4.87 & 0.38 & 4102 & 1.61 &  25.69  & PHO2  &AST71\\
  4466520 & 3385 & 4.89 & 0.36 & 3500 & 0.66 & 147.76  & KIC0  & KIC0\\
  4473475 & 3449 & 4.92 & 0.34 & 4477 & 2.33 &  10.90  & PHO2  &AST71\\
  4732678 & 3963 & 4.66 & 0.62 & 3683 & 0.73 &  92.23  &PHO54  &AST54\\
  5122206 & 3359 & 4.96 & 0.30 & 3400 & 0.50 & 179.17  & KIC0  & KIC0\\
  5446961 & 3676 & 4.68 & 0.59  &4487 & 2.93 &   5.62  & KIC0  &AST71\\
  5471005 & 3965 & 4.59 & 0.72  &4297 & 1.78 &  26.80  &PHO54  &AST54\\
...\\
\hline								
\end{tabular}
\label{tab:Mdwarfs}
\flushleft Notes: $T_{\rm eff}^{*}$, $\log g^*$, and $R^*$ are the effective temperature, surface gravity, and radius from G16. $T_{\rm eff}$, $\log g$, and R are the effective temperature, surface gravity, and radius from DR25.  $P_{T_{\rm eff}}$ and $P_{logg}$ are the provenances of $\teff$ and $\logg$ in DR25 as described in Table~\ref{tab:refs}.
\end{center}
\end{table}

\end{document}